[a4paper,superscriptaddress,floatfix]{revtex4}

\usepackage{graphicx}
\usepackage{subcaption}

\usepackage[utf8]{inputenc}
\usepackage[T1]{fontenc}
\usepackage[english]{babel}
\usepackage{blindtext}

\usepackage{graphicx}

\usepackage{caption}
\usepackage{slashed}
\usepackage{ragged2e}

\usepackage{pstricks}
\usepackage{color}
\usepackage{hepnames}

\usepackage{amsmath}
\usepackage{amsfonts}
\usepackage{amssymb}
\usepackage{fancyhdr}
\usepackage{setspace}
\usepackage{bm}
\usepackage{units}
\usepackage{lineno}
\usepackage{multirow}

\usepackage[final]{feynmp-auto}
\makeatletter
\def\endfmffile{%
	\fmfcmd{\p@rcent\space the end.^^J%
			end.^^J%
			endinput;}%
	\if@fmfio
		\immediate\closeout\@outfmf
	\fi
	\ifnum\pdfshellescape=\@ne
		\immediate\write18{mpost \thefmffile}%
	\fi}

\usepackage{hyperref}
\usepackage{doi}
\usepackage{url}

\usepackage{flafter}
\usepackage{placeins}
\usepackage{float}

\usepackage{lmodern}

\usepackage{tabularx}

\usepackage{pifont}

\usepackage{enumitem}

\newcommand{\explain}[2]{\underbrace{#1}_{\footnotesize\raggedright \textrm{#2}}}

\newcommand{\cba}          {\ensuremath{\cos(\beta - \alpha)}\xspace}
\newcommand{\tb}           {\ensuremath{\tan\beta}\xspace}
\newcommand{\tanb}         {\ensuremath{\tan\beta}\xspace}
\newcommand{\mA}           {\ensuremath{m_{A}}\xspace}

\newcommand{\cf}       {c.f.\xspace}

\newcommand{\Magellan}{\texttt{Magellan}\xspace}

\newcommand{\pandas}{\texttt{pandas}\xspace}
\newcommand{\DataFrame}{\texttt{DataFrame}\xspace}

\newcommand{\bokeh}{\texttt{bokeh}\xspace}
\newcommand{\holoviews}{\texttt{holoviews}\xspace}
\newcommand{\matplotlib}{\texttt{matplotlib}\xspace}

\newcommand{\TTPS}{\texttt{T3PS}\xspace}
\newcommand{\HiggsSignals}{\texttt{HiggsSignals}\xspace}
\newcommand{\HiggsBounds}{\texttt{HiggsBounds}\xspace}
\newcommand{\THDMC}{\texttt{2HDMC}\xspace}
\newcommand{\SusHi}{\texttt{SusHi}\xspace}

\newcommand{\GeV}{\ensuremath{\mathrm{GeV}}\xspace}

\newcommand{\MCMCnPoints}{4,259,823}

\newcommand{\alphan}{127.934}
\newcommand{\alphaS}{0.119}
\newcommand{\alphaE}{137.035997}
\newcommand{\mtop}  {172.5}
\newcommand{\mhiggs}{125.09}

\begin{document}


\title{LHC data interpretation within the 2HDM type II via a new analysis toolkit}

\author{E. Accomando}
\email[E-mail: ]{e.accomando@soton.ac.uk}
\affiliation{School of Physics \& Astronomy, University of Southampton, Highfield, Southampton SO17 1BJ, UK}
\affiliation{Particle Physics Department, Rutherford Appleton Laboratory, Chilton, Didcot, Oxon OX11 0QX, UK}

\author{D. Englert}
\email[E-mail: ]{david.englert@soton.ac.uk, d.englert@qmul.ac.uk}
\affiliation{School of Physics \& Astronomy, University of Southampton, Highfield, Southampton SO17 1BJ, UK}
\affiliation{Particle Physics Research Centre, School of Physics and Astronomy, \\
Queen Mary University of London, Mile End Road, London E1 4NS, UK}

\author{J. Hays}
\email[E-mail: ]{j.hays@qmul.ac.uk}
\affiliation{Particle Physics Research Centre, School of Physics and Astronomy, \\ 
Queen Mary University of London, Mile End Road, London E1 4NS, UK}

\author{S. Moretti}
\email[E-mail: ]{s.moretti@soton.ac.uk}
\affiliation{School of Physics \& Astronomy, University of Southampton, Highfield, Southampton SO17 1BJ, UK}
\affiliation{Particle Physics Department, Rutherford Appleton Laboratory, Chilton, Didcot, Oxon OX11 0QX, UK}

\begin{abstract}
{
We review the status of the 2-Higgs Doublet Model (2HDM) Type-II, in the light of
the current experimental results and various theoretical consistency conditions.
Compared to the existing literature, in this paper we apply for the first time a new method that can improve the standard procedure for setting bounds on the 2HDM parameter space, as no experimental evidence has been found so far. Our new numerical framework, called \Magellan, and statistical techniques can be applied to any BSM scenarios. Here, we take as testing ground the 2HDM, particularly as it is physically interesting and moreover characterised by a far from trivial multi-dimensional parameter space where the effectiveness of the new methods can be proved. \Magellan uses a Markov Chain Monte Carlo technique for scanning the parameter space and leverages the use of data processing and visualisation methods, allowing the user to perform inference on the model in a complete and efficient way. The novelty of the proposed method is that the parameter space of any BSM theory can still be projected onto any bi-dimensional plane but one can map whichever portion of this sub-space into any other bi-dimensional plane thus having a full control of the whole parameter space at once. The \Magellan’s website interactive dashboards can be accessed via a public link. Through this website, the user can explore the full parameter space and exploit the phenomenological features of the model with ease. 
}
\end{abstract}

\maketitle

\section{Introduction}

The discovery of a Higgs boson at the Large Hadron Collider (LHC) has been a triumph for
particle physics~\cite{ATLAS-Higgs,CMS-Higgs}, revealing that the masses of the
fundamental particles in Nature are indeed generated through the Higgs mechanism of
(spontaneous) Electro-Weak Symmetry Breaking (EWSB). This particle eventually revealed
itself to have properties close to those of the Standard Model (SM) Higgs state. However, even if technically possible, it is rather unnatural thinking that the discovered state would ultimately complete the particle physics scenario. Such a light Higgs state leaves in fact the hierarchy problem unresolved, that is, the great disparity between the Higgs mass itself (125 GeV) and the Planck scale (of order 10$^{19}$ GeV). Under the assumption that the discovered Higgs state is of a fundamental nature, i.e., not a composite state, in order to surpass the hierarchy problem, one has to invoke Beyond the SM (BSM) scenarios that inevitably involve an enlarged Higgs sector. One could have any number of singlet Higgs fields and/or Higgs doublets. 

\noindent
In this paper, we consider the presence of a second Higgs doublet, thereby introducing a generic 2-Higgs Doublet Model (2HDM). The presence of a second Higgs doublet naturally arises in many models of BSM physics, ranging from the most recent ones addressing for example the generation of neutrino masses and the $g-2$ anomaly to the more traditional ones, like the Supersymmetric theories. Just to cite a few of them, a class of axion models~\cite{Celis2014185,Axion1987}, which can explain the lack of observed CP violation in the strong sector, and certain realisations of composite Higgs models with pseudo-Nambu-Goldstone bosons~\cite{Mrazek20111,Bertuzzo2013,Agashe2005165,DeCurtis:2018iqd,DeCurtis:2018zvh} can both give rise to an effective low-energy theory with two Higgs doublets. The additional source of CP violation present in this type of enlarged (pseudo)scalar sector could further provide an explanation of the matter-antimatter asymmetry. Particular realisations of the 2HDM also have the appealing features of being able to explain the neutrino mass generation~\cite{Neutrino-mass}, to provide a candidate for dark matter \cite{Ko2014} or to accommodate the muon $g-2$ anomaly \cite{muong2_lepton,muong2_limiting,Wang2015}. From a more traditional perspective, it is well known that the Minimal Supersymmetric Standard Model (MSSM) requires the existence of two doublets to generate the mass of both up-type and down-type quarks and charged leptons. In this case, the Yukawa couplings should have Type-II values. The representative model chosen in this paper, the 2HDM Type-II, would therefore coincide with the MSSM\cite{HiggsHunters,DjouadiMSSM} in the sparticle decoupling limit where the SUSY scale is assumed to be much higher than the EW one. This coincidence does not spoil the pure generality of the chosen 2HDM representation of the scalar sector, embedded in the wide  variety of models concisely recalled above.

\noindent
From the experimental point of view, the additional four states of a generic
2HDM~\cite{HiggsHunters,Branco2012} provide a range of observables through
which all the above theoretical models could in principle be tested, or at least have a first evidence of their validity through their scalar sector. Hence, it is worthwhile investigating in detail the scope of the LHC in discovering the new Higgs bosons described by the 2HDM generic representation. 

\noindent
There exists a vast literature on the phenomenological analyses setting bounds on the 2HDM parameter space, as no experimental evidence has been found so far. The last two decades have seen the implementation and development of the global fits, which collect the data coming from different experiments and make rigorous statistical analyses to extract limits on BSM theories. The package GFitter \cite{Baak_2012} was a pioneer in releasing a global electroweak fit to constrain New Physics predicted by a variety of models, including the 2HDM. Other toolkits are published in the literature, with their main focus centred on SUSY and its numerous variants (the 2HDM Type II is one of them in the decoupling limit). A global analysis of SUSY is provided by SFitter \cite{Lafaye:2004cn}, SuperBayeS \cite{Austri_2006, Strege_2012, Strege_2013}, Fittino \cite{Bechtle_2006, Bechtle_2010, Bechtle_2012}, Lilith \cite{Kraml_2019, Bernon_2015} and MasterCode \cite{Buchmueller_2009, Buchmueller_2011, Buchmueller_2012_2, Buchmueller_2012_6, Buchmueller_2014}. A much wider range of BSM theories is covered by the package Gambit \cite{Athron_2017_11, Athron_2017_12}, a global fitting software framework characterised by theory flexibility and straightforward extension to new observables and external interfaces. Bounds on the MSSM are addressed by Gambit in Ref. \cite{Athron_2017_12} and limits on the 2HDM are derived in Ref. \cite{Rajec_2020}, specifically.

The standard procedure, generally adopted by the global fitting packages, makes use of all relevant experimental data and theoretical arguments that can constrain the model. These constraints can be categorised into the following three main sources: measurements of the discovered 125 \GeV Higgs boson properties (i.e. production and decay signal strengths), direct and indirect searches for the extra Higgs bosons present in the model and, finally, theory considerations based on perturbativity, unitarity, triviality and vacuum stability. The statistical analysis is then performed, with the likelihood function expressing the plausibilities of different parameter values for the given sample(s) of data. As the 2HDM parameter space is six-dimensional, the standard way of extracting bounds is projecting the full parameter space onto bi-dimensional planes, defined by any two model parameters. In doing so, the statistical procedure is maximising the (log)likelihood on all the other four remaining parameters. 

\noindent
In this paper, we apply for the first time a new method that can improve this standard procedure. Our new tools and statistical techniques can be applied to any BSM scenarios. Here, we take as testing ground the 2HDM, particularly as it is physically interesting and  moreover characterised by a multi-dimensional parameter space. This latter feature allows us to prove the effectiveness and efficiency of our new approach in a far from trivial setup. First of all, we perform an efficient scanning of the parameter space through a Markov Chain Monte Carlo (MCMC) approach with \TTPS~\cite{T3PS}. After this first step, instead of projecting the parameter space onto bi-dimensional planes by maximising the (log)likelihood over the other remaining parameters, we keep the full punctual information on all the model parameters, simultaneously. We introduce effective data processing and visualisation methods based on \pandas~\cite{pandas}, \matplotlib~\cite{matplotlib}, \bokeh~\cite{bokeh} and \holoviews~\cite{holoviews}. With the help of these packages and the wrapper framework called \Magellan, we can still project the parameter space onto any bi-dimensional plane but we can map any portion of this sub-space into any other bi-dimensional plane thus having a full control of the whole parameter space at once. This constitutes the novelty of our method. The code \Magellan is not published. However, its website interactive dashboards can be accessed via the link given in Ref.~\cite{Magellan-Web}. We have moreover created an open-access repository in Zenodo. There, we have published the datasets generated by \Magellan. The MCMC scan of the 2HDM Type II parameter space is stored in a HDF5 file and instructions are given to load the dataset as a pandas data-frame. Through the \Magellan website, the user can explore the complete parameter space at once and exploit the phenomenological features of the model with ease. 

\noindent
To envisage the wider use of the proposed toolkit, we highlight that applications of \Magellan to BSM theories other than the 2HDM presented in this paper are already well documented. The first one addresses the analysis of the Higgs boson pair production in six different channels. In this case, the toolkit has been used to extract the excluded regions in the parameter space of the EWK-singlet model and the hMSSM model. This analysis has been published by ATLAS \cite{Aad_2020}. The second application concerns the analysis of the extra CP-even Higgs decaying into two light Higgses at the LHC within the 2HDM Type II. This analysis is already public on the Magellan website. There, also the analyses of both the heavy CP-even and CP-odd Higgs decaying into tau pairs are published. These latter studies represent extensions of the main analysis carried out in this paper which, for illustrative purposes, is focussed on the associated production $pp\rightarrow Zh\rightarrow 4f$ within the Type-II 2HDM.

\noindent
The plan of the paper is as follows. In Section 2, we describe the 2HDM Type-II, taken as prototypical example to illustrate the described approach. In Section 3, the scanning procedure is specified. Section 4 enumerates the theoretical and experimental constraints that are taken into account during the scan. Section 5 shows how data interpretation is facilitated by these new tools. Finally, in Section 6, we conclude.

\section{The 2HDM}

In this section we give a brief introduction to the 2HDM, with a focus on the
aspects relevant to our analysis. Extensive reviews of the 2HDM can be found in
Refs.~\cite{HiggsHunters,Branco2012,DjouadiMSSM}.

\noindent
An important feature of the model is the number of degrees of freedom (d.o.f)
of the fields, which we can be enumerated before and after the spontaneous  breaking of the EW symmetry due to the shape of the Higgs potential. Initially, we have two complex doublets, $\Phi_{1}$ and $\Phi_{2}$, giving 8 d.o.f. in total. After EWSB, the spectrum contains two CP-even scalars $h$ and $H$, one pseudo-scalar $A$ and two charged Higgs bosons $H^{\pm}$ (i.e.,  5 d.o.f.). The Goldstone bosons of the theory will then become the longitudinal components of the weak $W^{\pm}$ and $Z$ bosons (3 d.o.f). Hence, the total d.o.f. number is unchanged.

\noindent
The most general renormalisable (i.e.,  quartic) scalar potential of two doublets
can be written as
\begin{equation}
\begin{aligned}
	\mathcal{V}_{gen}
	&=
	m_{11}^{2} \Phi_{1}^{\dag} \Phi_{1} 
	+ m_{22}^{2} \Phi_{2}^{\dag} \Phi_{2}
	- \left[ m_{12}^{2} \Phi_{1}^{\dag} \Phi_{2} + \textrm{h.c} \right] + \\
	&+ \frac{1}{2} \lambda_{1} \left( \Phi_{1}^{\dag} \Phi_{1} \right)^{2}
	+ \frac{1}{2} \lambda_{2} \left( \Phi_{2}^{\dag} \Phi_{2} \right)^{2}
	+ \lambda_{3} \left( \Phi_{1}^{\dag} \Phi_{1} \right) \left( \Phi_{2}^{\dag} \Phi_{2} \right)
	+ \lambda_{4} \left( \Phi_{1}^{\dag} \Phi_{2} \right) \left( \Phi_{2}^{\dag}
	\Phi_{1} \right) + \\
	&+ \left\{ \frac{1}{2} \lambda_{5} \left( \Phi_{1}^{\dag} \Phi_{2} \right)^{2} + \left[ \lambda_{6} \left( \Phi_{1}^{\dag} \Phi_{1} \right) + \lambda_{7} \left( \Phi_{2}^{\dag} \Phi_{2} \right) \right] \Phi_{1}^{\dag} \Phi_{2} \right\},
\end{aligned}
\end{equation}
\noindent
where $m_{11}^{2}$, $m_{22}^{2}$, $m_{12}^{2}$ are the mass squared parameters
and $\lambda_{i}$ ($i = 1, ..., 7$) are dimensionless quantities describing the
coupling of the order-4 interactions. Six parameters are real ($m_{11}^{2}$,
$m_{22}^{2}$, $\lambda_{i}$ with $i = 1, ..., 4$) and four are a priori complex
($m_{12}^{2}$ and $\lambda_{i}$ with $i = 5, ..., 7$). Therefore, in general,
the model has 14 free parameters. Under appropriate constraints, this number
can however be reduced.

\noindent
The potential is explicitly CP-conserving if and only if there exists a basis choice for the scalar fields in which $m_{12}^{2}$, $\lambda_{5}$, $\lambda_{6}$ and $\lambda_{7}$ are real. Notice that, even in this case, the vacuum could still break CP spontaneously. The spontaneous CP-violation of the vacuum takes place if and only if the scalar potential is explicitly CP-conserving, but there is no basis in which the scalars are real
\cite{Haber2015}. In the following, we assume that both the scalar potential and the vacuum are CP-conserving. Consequently, by requiring CP-conservation, one looses four d.o.f. reducing the number of free parameters down to 10. After EWSB, each scalar doublet acquires a Vacuum Expectation Value (VEV) that can be parametrised as follows:
\begin{equation}
  \langle \Phi_{1} \rangle =
	\frac{v}{\sqrt{2}}
  \begin{pmatrix}  
	0 \\
	\cos\beta
  \end{pmatrix}
\quad
\quad
  \langle \Phi_{2} \rangle =
	\frac{v}{\sqrt{2}}
  \begin{pmatrix}  
	0 \\
	\sin\beta
  \end{pmatrix},
\end{equation}
\noindent
where the angle $\beta$ determines the ratio of the two doublet VEVs, $v_1$ and $v_2$, through the definition of $\tan\beta = v_{2} / v_{1}$. The angle $\beta$ is an additional parameter that adds to the free parameters defining the scalar potential. 
{
\begin{table}[t!]
\centering
\begin{tabular}{|c|c|c|c|c|c|c|c|c|c|}
   \hline
	\multirow{2}{*}{Model} & \multicolumn{3}{c|}{$h$} & \multicolumn{3}{c|}{$H$} & \multicolumn{3}{c|}{$A$} \tabularnewline
	\cline{2-10}
	 &
	$u$ & $d$ & $l$ &
	$u$ & $d$ & $l$ &
	$u$ & $d$ & $l$
	\tabularnewline
   \hline
	{Type-I}  &
	$\phantom{-}\frac{\cos\alpha}{\sin\beta}$ & $\phantom{-}\frac{\cos\alpha}{\sin\beta}$ & $\phantom{-}\frac{\cos\alpha}{\sin\beta}$ & 
	$\phantom{-}\frac{\sin\alpha}{\sin\beta}$ & $\phantom{-}\frac{\sin\alpha}{\sin\beta}$ & $\phantom{-}\frac{\sin\alpha}{\sin\beta}$ & 
	$\phantom{-}\cot\beta                   $ & $         - \cot\beta                   $ & $         - \cot\beta                $         
	\tabularnewline
   \hline
	{Type-II}  &
	$\phantom{-}\frac{\cos\alpha}{\sin\beta}$ & $          -\frac{\sin\alpha}{\cos\beta} $ & $          -\frac{\sin\alpha}{\sin\beta} $ & 
	$\phantom{-}\frac{\sin\alpha}{\sin\beta}$ & $\phantom{-}\frac{\cos\alpha}{\sin\beta} $ & $\phantom{-}\frac{\cos\alpha}{\sin\beta} $ & 
	$            \cot\beta                  $ & $           \tan\beta                    $ & $           \tan\beta                    $   
	\tabularnewline
   \hline
	{Type-III}  &
	$\phantom{-}\frac{\cos\alpha}{\sin\beta}$ & $\phantom{-}\frac{\cos\alpha}{\sin\beta}$  & $           -\frac{\sin\alpha}{\sin\beta}$ & 
	$\phantom{-}\frac{\sin\alpha}{\sin\beta}$ & $\phantom{-}\frac{\sin\alpha}{\sin\beta} $ & $\phantom{-}\frac{\cos\alpha}{\cos\beta} $ & 
	$          -\cot\beta                   $ & $\phantom{-}\cot\beta                    $ & $           -\tan\beta                   $   
	\tabularnewline
   \hline
	{Type-IV}  &
	$\phantom{-}\frac{\cos\alpha}{\sin\beta}$ & $           -\frac{\sin\alpha}{\cos\beta}$ & $\phantom{-}\frac{\cos\alpha}{\sin\beta} $ & 
	$\phantom{-}\frac{\sin\alpha}{\sin\beta}$ & $\phantom{-}\frac{\cos\alpha}{\cos\beta} $ & $\phantom{-}\frac{\sin\alpha}{\sin\beta} $ & 
	$           -\cot\beta                  $ & $           -\tan\beta                   $ & $\phantom{-}\cot\beta                    $   
	\tabularnewline
   \hline
\end{tabular}
\caption{
Couplings of the neutral Higgs bosons to fermions, normalised to the corresponding SM
value ($m_{f}/v$) in the 2HDM Type-I, II, III and IV.}
\label{tab:interaction}
\end{table}
}
\noindent
In general, the Yukawa matrices corresponding to the two  doublets are not simultaneously
diagonalisable, which can pose a problem, as the off-diagonal elements lead to tree-level Higgs mediated Flavour Changing Neutral Currents (FCNCs) on which severe experimental bounds exist. The Glashow-Weinberg-Paschos (GWP) theorem~\cite{GWP1,GWP2} states that this type of FCNCs is absent  if at most one Higgs multiplet is responsible for providing mass to fermions of a given electric charge. This GWP condition can be enforced by a discrete 
$\mathbb{Z}_2$-symmetry ($\Phi_1 \rightarrow +\Phi_1$ and $\Phi_2\rightarrow -\Phi_2$) on the doublets, in which case the absence of FCNCs is natural. The soft $\mathbb{Z}_2$ breaking condition relies on the existence of a basis where $\lambda_6 = \lambda_7$ = 0. Therefore, one looses two additional d.o.f. reducing the number of free parameters down to 9. Finally, $m_{11}^{2}$ and $m_{22}^{2}$ can be expressed as a function of the other parameters, owing to the fact that the scalar potential is in a local minimum when computed in the VEVs. So, globally, with restrictions to CP-conservation and soft $\mathbb{Z}_2$-symmetry breaking, there remain 7 free parameters in the 2HDM.

\noindent
Under the above conditions, there are several alternative basis in which the 2HDM can be described: the \textit{general parametrisation} (as given above in terms of $m^2_{ij}$ and $\lambda_i$s), the \textit{Higgs basis}, where one of the doublets gets zero VEV, and the \textit{physical basis}, where one uses the physical masses of the scalars. However, in  the light of the discovery of the $125$ GeV Higgs boson, herein the $h$ state, it is customary to parametrise the theory using the \textit{hybrid basis}~\cite{Haber2015}, where the parameters provide a convenient choice to give a direct control on both the CP-even and CP-odd Higgs masses, the $hVV$ couplings ($V$ = $W^\pm , Z$), the $Aq\bar{q}$ vertices and the Higgs quartic couplings. The parameters in this basis are:
\begin{equation}
	\explain{
	m_{h}, \quad
	m_{H}}
	{CP-even Higgs masses}, \quad
	\explain{\cos(\beta - \alpha)}{\text{\shortstack{ determines the \\ $g_{hVV}$ \& $g_{HVV}$ couplings }}}
	, \quad
	\explain{
	\tan\beta}
	{ratio of the vevs}
	, \quad
	\explain{
	Z_{4}, \quad
	Z_{5}, \quad
	Z_{7},}
	{\text{\shortstack{Higgs self-coupling \\ parameters}}}
\end{equation}
\noindent
with $m_H\ge m_h$, $0\leq\beta\leq\pi /2$ and $0\le\sin (\beta -\alpha )\leq1$. The remaining (pseudo)scalar masses can be expressed in terms of the quartic scalar couplings in the Higgs basis:
\begin{equation}
	m_A^2 = m_H^2\sin^2(\beta -\alpha) + m_h^2\cos^2(\beta -\alpha) - Z_5v_1^2,
\end{equation}
\begin{equation}
	m_{H^\pm}^2 = m_A^2 -{1\over 2}(Z_4 - Z_5)v^2.
\end{equation}
\noindent
In the hybrid basis, by swapping the self-couplings $Z_4$ and $Z_5$ with the scalar masses given above, the 7 free parameters can be recast into four physical masses and 3 parameters that are related to the couplings of the scalars to gauge bosons, fermions and scalars themselves, respectively:
\begin{equation}
	m_h, ~ m_H, ~ m_A, ~ m_{H^\pm}, ~ \cos (\beta - \alpha ), ~ \tan (\beta ), ~ Z_7.
\end{equation}
\noindent
In the above list, $Z_7$ enters only the triple and quartic scalar interactions. Finally, as $m_h$ has been measured with excellent accuracy at the LHC, the number of d.o.f comes down to 6, globally.

\noindent
Beside the (pseudo)scalar fields, also fermions are required to have a definite charge under the discrete $\mathbb{Z}_2$-symmetry. The different assignments of the $\mathbb{Z}_2$-charge in the fermion sector give rise to the four different types of 2HDM. 
The couplings of the neutral Higgses to fermions, normalised to the corresponding SM value
($m_{f}/v$, henceforth, denoted by $\kappa_{hqq}$ for the case of the SM-like Higgs state coupling to a quark $q$, where $q=d,u$), can be found in Tab.~\ref{tab:interaction}. 

\noindent
As intimated, in the remainder of this paper, we will concentrate on the 2HDM Type-II. There are two limiting scenarios, giving rise to two distinct regions in the 
$(\cos (\beta - \alpha), \tan \beta)$ parameter plane \cite{Ferreira2014}. They can be understood by examining the behaviour of $\kappa_{hqq}$ as a function of the angles $\alpha$ and $\beta$. Taking the limits $\beta - \alpha\rightarrow \frac{\pi}{2}$ (upper lines in the upcoming figure) and $\beta + \alpha \rightarrow \frac{\pi}{2}$ (lower
lines in the upcoming figure), the couplings become (recall  Tab.~\ref{tab:interaction}):
{
\begin{equation}
\begin{aligned}
	\kappa_{hdd}
	=
	- \frac{\sin \alpha}{ \cos \beta}
	&=
	\sin (\beta  - \alpha) - \cos( \beta - \alpha) \tan \beta 
	\xrightarrow[\beta - \alpha = \frac{\pi}{2}]{}
	1 ~ \textrm{(middle-region),}
	\\ 
	&=
	- \sin (\beta + \alpha) + \cos(\beta + \alpha ) \tan \beta
	\xrightarrow[\beta + \alpha = \frac{\pi}{2}]{}
	- 1 ~ \textrm{(right-arm),}
	\\
	\kappa_{huu}
	=
	\frac{\cos \alpha}{ \sin \beta}
	&=
	\sin (\beta  - \alpha) + \cos( \beta - \alpha) \cot \beta
	\xrightarrow[\beta - \alpha = \frac{\pi}{2}]{}
	1 ~ \textrm{(middle-region),}
	\\
	&=
	\sin (\beta + \alpha) + \cos(\beta + \alpha ) \cot \beta 
	\xrightarrow[\beta + \alpha = \frac{\pi}{2}]{}
	1 ~ \textrm{(right-arm).}
\end{aligned}
\end{equation}
}

\begin{figure}[t!]
\begin{center}
\includegraphics[width=0.99\linewidth]{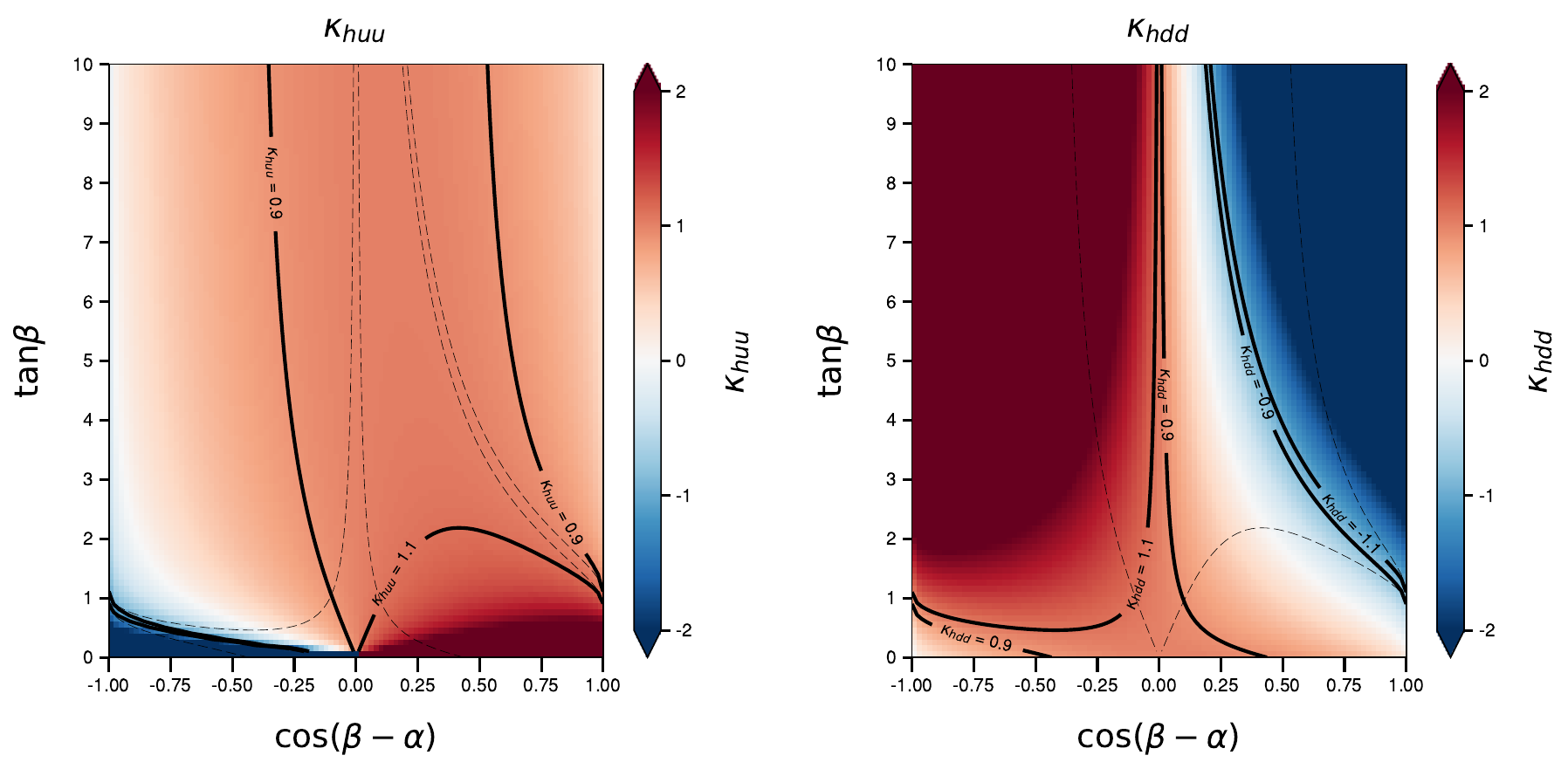}
\caption{\label{fig:couplings}
Light CP-even Higgs couplings to the up-type (left) and down-type (right) quarks, normalised to the corresponding
SM value, in the ($\cos(\beta - \alpha), \tan \beta$) plane.}
\end{center}
\end{figure}
\noindent
The dependence of $\kappa_{hdd}$ and $\kappa_{huu}$ on $\cos (\beta - \alpha )$ and 
$\tan (\beta )$ is illustrated in Fig.~\ref{fig:couplings}. The $\beta - \alpha \rightarrow \frac{\pi}{2}$ case corresponds to the ``middle-region'', which is the SM-limit of the theory. In the right-hand side plot of Fig.~\ref{fig:couplings}, this domain is identified by the contour region where $0.9 \le\kappa_{hdd}\le 1.1$, that is assuming a 10\% discrepancy from the SM couplings. The $\beta + \alpha\rightarrow \frac{\pi}{2}$ case corresponds to the ''right-arm'', where one gets an opposite sign for the coupling between the SM-like Higgs $h$ and the down-type quarks, relative to the SM value. This is called the wrong-sign Yukawa coupling scenario. In the right-hand side plot of Fig.~\ref{fig:couplings}, this region is represented by the narrow arm (or tongue) where the
coupling is negative and again has a 10\% displacement from the SM value: $-1.1 \le \kappa_{hdd}\le -0.9$. Both the alignment and the wrong-sign regions are well within the O(10\%) discrepancy from the corresponding SM value allowed for the coupling of the SM-like Higgs to the up-type quarks, $\kappa_{huu}$, as shown in the left-hand plot of Fig.~\ref{fig:couplings}.

\begin{figure}[t]
\begin{center}
\includegraphics[width=0.45\linewidth]{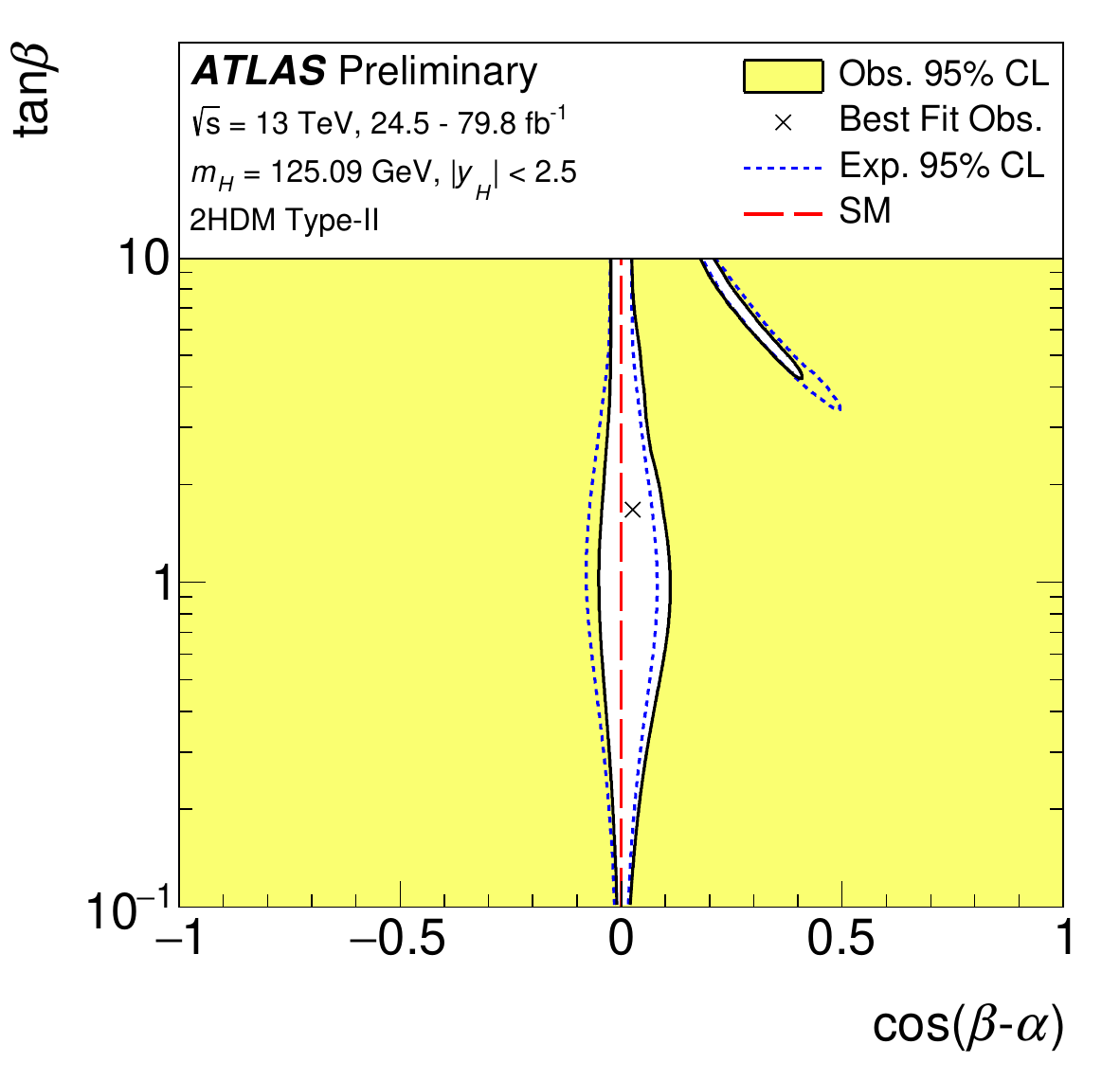}
\includegraphics[width=0.45\linewidth]{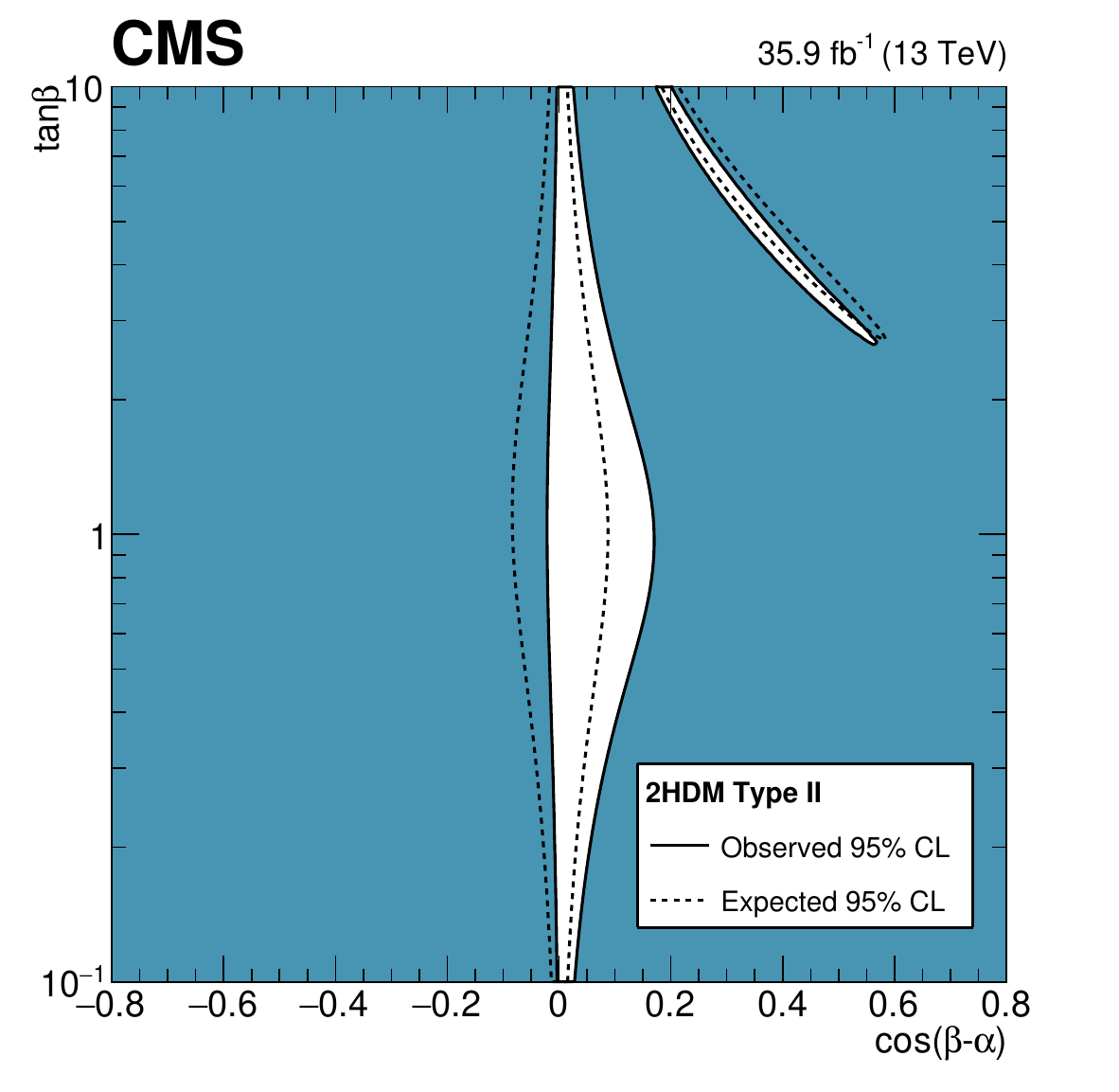}
\caption{\label{fig:2HDM:cba_tb}
   Allowed regions of (\cba, \tb) parameters in 2HDM Type-II, obtained from the compatibility
   with the observed couplings of the $125$ GeV boson, when identified as
   the light Higgs boson, $h$ of the model.
   The plot show the most up-to-date available results from
   ATLAS~\cite{Aad:2688596} and
   CMS~\cite{Sirunyan:2018koj}, seen on the left and right plot respectively.
}
\end{center}
\end{figure}

\noindent
The most up-to-date 125 GeV Higgs combined signal strength analyses from
ATLAS~\cite{Aad:2688596} and CMS~\cite{Sirunyan:2018koj}, interpreted in the 2HDM Type-II can be seen in Fig.~\ref{fig:2HDM:cba_tb}, where it is found that the hypotheses of $\kappa_{hdd}=1$ and $\kappa_{hdd}=-1$ are still both allowed. On the theory side, an interesting study \cite{Basler:2017nzu} based on Renormalisation Group Equations (RGEs) has shown that, if one requires the model to be valid  up to higher energies (beyond 1 TeV), the allowed parameter space shrinks to the positive sign of $\kappa_{huu}/\kappa_{hdd}$, otherwise called the alignment region. Below the TeV energy scale, both the alignment and the wrong sign scenario are valid. From a more phenomenological point of view, many analyses have been performed to constrain these two domains. In particular, the importance of the decay
channels of the two extra neutral Higgses, $A$ and $H$, in the wrong-sign limit of the model has been clearly illustrated in Ref.~\cite{Ferreira:2017bnx}. 
\noindent
Here, we intend to revisit in detail how the constraints onto the 2HDM Type-II
parameter space are normally drawn and how they can be improved upon using our new framework \Magellan. In the next section, we describe the methodology and the tools employed to perform our scans.

\section{Magellan: global scan for bounds extraction and data interpretation}

In this section, we describe a new methodology that can be employed to explore the parameter space of any BSM theory. Our novel framework, \Magellan, is indeed designed for a twofold scope. Firstly, it allows one to easily import any new experimental results so as to interpret these within any given model and derive bounds on the corresponding parameter space. Secondly, \Magellan can quickly predict the regions of the latter that can be accessible in a given search with the actual luminosity at hand, and show therein the characteristics of the new particles to be searched for (e.g. mass, width, branching ratios, etc.) thus allowing to improve the data analysis.

\noindent
In order to illustrate the model exploration approach adopted by \Magellan ~\cite{Magellan-Web}, we take as an example the 2HDM Type-II. Within this new interactive framework, the latest limits on this model are then derived to show the effectiveness of the new method. The key starting point is implementing all existing constraints, from theory and experiment. In order to scan over the 2HDM parameter space, we use a 
MCMC  based on \TTPS~\cite{T3PS} for parallel processing of parameter 
scans. This tool makes use of the standard 
Metropolis-Hastings~\cite{Metropolis,Hastings} algorithm that is briefly 
summarised below.
\begin{itemize}
	\item \textbf{Step 0)}
			Draw a point from the prior distribution $\pi(\theta$), which
			will serve as the starting point of the chain.
			The likelihood corresponding to this point is 
			$\mathcal{L}(\theta|d)$.
	\item \textbf{Step 1)}
			Propose a new candidate point $\theta'$, taken from the proposal
			distribution $q(\theta', \theta)$. In our case $q(\theta', \theta)$ 
			is a Gaussian distribution, centered around the previous point $\theta$ with a 
			standard deviation of $a$, commonly referred to as the \textit{step-size}.
			The likelihood corresponding to the new point is $\mathcal{L}(\theta'|d)$.
	\item \textbf{Step 2)}
			Calculate the ratio of the posterior probabilities corresponding to 
			the two points:
			$r = \frac{ \mathcal{L}(d|\theta') \pi(\theta') q(\theta',\theta) }{ \mathcal{L}(d|\theta) \pi(\theta) q(\theta',\theta) }$.
			In the Metropolis-Hastings algorithm, $q(\theta',\theta)$ is 
			symmetric, therefore it drops out in the ratio.
	\item \textbf{Step 3)}
			If $r \geq 1$, then accept the new proposal, otherwise accept the
			candidate with a probability of $r$. If the point is rejected 
			repeat the process from \textbf{Step 1)}.
	\item \textbf{Step 4)}
			Once a new candidate is found, add it to the chain and repeat the 
			process from \textbf{Step 1)}.
\end{itemize}

\begin{table}[t!]
\begin{center}
\begin{tabular}{|c|c|c|c|}
   \hline
	{Parameter}     & {min} & {max} & {step-size}
	\tabularnewline
   \hline
	$Z_{7}$                & $-10.0$ &   10.0 &  0.2
	\tabularnewline
   \hline
	$m_{H}$ [GeV]                &  150  & 1000.0 & 20.0
	\tabularnewline
   \hline
	$m_{H^{\pm}}$ [GeV]         &  500  & 1000.0 & 20.0
	\tabularnewline
   \hline
	$m_{A}$      [GeV]          &  100  & 1000.0 & 20.0
	\tabularnewline
   \hline
	$\cos(\beta - \alpha)$ & $-1.0$  &    1.0 &  0.03
	\tabularnewline
   \hline
	$\tan(\beta)$          &  0.5  &   30.0 &  0.5
	\tabularnewline
   \hline
\end{tabular}
\caption{Range and step-size of the 6-dimensional 2HDM parameters used in the MCMC scan. }
\label{tab:mcmc_ranges}
\end{center}
\end{table}

\begin{table}[t]
\begin{center}
\begin{tabular}{|c|c|c|c|c|}
   \hline
	$\alpha$     & $\alpha_s$ & $\alpha_{\rm EM}\equiv \alpha(Q^2=0)$ & $m_{t}$ [GeV] & $m_{h}$ [GeV]
	\tabularnewline
   \hline
	1/\alphan      & \alphaS &  1/\alphaE &  \mtop & \mhiggs
	\tabularnewline
   \hline
\end{tabular}
\caption{Physical parameters kept fixed in our scans.}
\label{tab:phys_params}
\end{center}
\end{table}
\noindent
The likelihood function, $\mathcal{L}(d|\theta)$, is constructed using the experimental 
$\chi^{2}$ values coming from the Higgs coupling measurements and the fit to the $S$, $T$ and $U$ parameters of the EW Precision Observables (EWPOs). The likelihood is defined as:
\begin{equation}
	\mathcal{L} = \exp \left( - \frac{\chi^{2}_{\rm tot}}{2} \right),
\end{equation}
\noindent
where $\chi^{2}_{\rm tot} = \chi^{2}_{HS} + \chi^{2}_{ST}$, with $\chi^{2}_{HS}$ being the
$\chi^{2}$ value extracted from measurements of the $h$ couplings entering the production
and decays modes of the SM-like Higgs state discovered at CERN. We have used HiggsSignals-2 \cite{Bechtle:2013xfa} beta version, which includes experimental data tables collected until September 2018. HiggsSignals compares the 2HDM predictions for the scalar sector with the SM-like Higgs signal rate and mass measurements at the LHC, giving rise to a likelihood estimate. Recently, they have released a validation (see Ref. \cite{Bechtle_2021}) against the ATLAS and CMS combined analysis of the LHC Run1 data at 7 and 8 TeV \cite{PhysRevLett.114.191803, Aad_2016}. An update for the Run2 data is still unpublished. We have however checked the consistency of our results, obtained via a link to HiggsSignals, against the experimental fits at Run2. The $S$ and $T$ parameter compatibility measure ($U$ is irrelevant for our purposes) $\chi^{2}_{ST}$ is:
\begin{equation}
	\chi^{2}_{ST} =
	\frac{(S-S^{\textrm{exp}}_{\textrm{best fit}})^{2}}{\sigma^{2}_{S} (1 - \rho^{2}_{ST}) } +
	\frac{(T-T^{\textrm{exp}}_{\textrm{best fit}})^{2}}{\sigma^{2}_{T} (1 - \rho^{2}_{ST}) }
	- 2 \rho_{ST} \frac{(S-S^{\textrm{exp}}_{\textrm{best fit}})(T-T^{\textrm{exp}}_{\textrm{best
	fit}})}{\sigma_{T} \sigma_{S} (1 - \rho^{2}_{ST})},
\end{equation}
\noindent
where the best fit values $S^{\textrm{exp}}_{\textrm{best fit}}$ and
$T^{\textrm{exp}}_{\textrm{best fit}}$, their uncertainties $\sigma_{S/T}$ and
the correlation parameter, $\rho^{2}_{ST}$, are taken from the fit result of the
Gfitter group~\cite{Gfitter}. 

\noindent
One naturally concentrates on the experimental observables where the discovered $h$ state enters. However, searches for additional Higgs states, both neutral and charged (at present yielding null results in either case), once interpreted in a specific theoretical model, can force constraints onto its parameter space. Hence, these ought to be included as well. We have done so here using the program \HiggsBounds 4 \cite{Bechtle:2013wla},  which tests the model against the exclusion limits extracted from the Higgs searches at LEP, Tevatron and LHC. Another constraint, which must be accounted for, comes from the inclusive weak radiative $B$-meson Branching Ratio (BR) that proceeds through the quark-level transition of $b\rightarrow s \gamma$. A recent study \cite{Misiak2017}, using results from the  Belle Collaboration, places a ${95\% ~\text{CL}}$ lower bound on the charged Higgs mass: $m_{H^{\pm}} > 580$ GeV. Therefore, we only select points above this mass value. 

\noindent
The algorithm specified above determines how a Markov chain evolves in the parameter space. Since each chain is independent, the different chains can be run in parallel, reducing the wall-clock time of the scan. The MCMC scan is performed over the 6-dimensional parameter space ($Z_{7}$, $m_{H}$, $m_{H^{\pm}}$, $m_{A}$, $\cos(\beta - \alpha)$, $\tan \beta$). The ranges and step-size of each parameter can be found in Tab.~\ref{tab:mcmc_ranges}. Other physical quantities are kept constant and their chosen values are listed in Tab.~\ref{tab:phys_params}. As the scan is computationally expensive, it is worth specifying what options were chosen for the scan: 400 independent chains were submitted, each for 20 hours on Dual 2.6 GHz Intel Xeon 8 core processor machines. With the given time limit, the setup yields an average chain length of O(10000) steps for each chain. Since the Markov chain first needs to find the minimum of the likelihood, then
converge to thermal equilibrium, we account for this ``warm-up'' period, hence, the first 200 steps are discarded within every chain. The result of the MCMC scan is a data sample consisting of \MCMCnPoints\ points, before applying the theoretical constraints (vacuum stability, unitarity and perturbativity). A key feature of the MCMC scanning method is that the results can be interpreted in the Bayesian statistical framework, that is, the density of the points in the parameter space is proportional to the posterior probability of the model describing the data.

\noindent
A post processing step is then performed where we calculate the production cross-sections and BRs of the (pseudo)scalars using \SusHi \cite{Harlander:2012pb} and \THDMC \cite{2HDMC}, respectively. This allows a direct link between experimental measurements and data interpretation within a given BSM theory, like  (but not only) the 2HDM Type-II, which is the model we are focussing on in this paper.

\section{Bounds on the 2HDM Type-II}

In this section, we discuss the bounds that can be extracted on the 6 independent free parameters of the 2HDM Type-II simultaneously taking into account Higgs coupling strengths,  EWPO and the theoretical constraints.

\subsection{Experimental constraints}
\label{sec:2HDM-EWPO}

\begin{figure}[t]
\begin{center}
\includegraphics[width=0.60\linewidth]{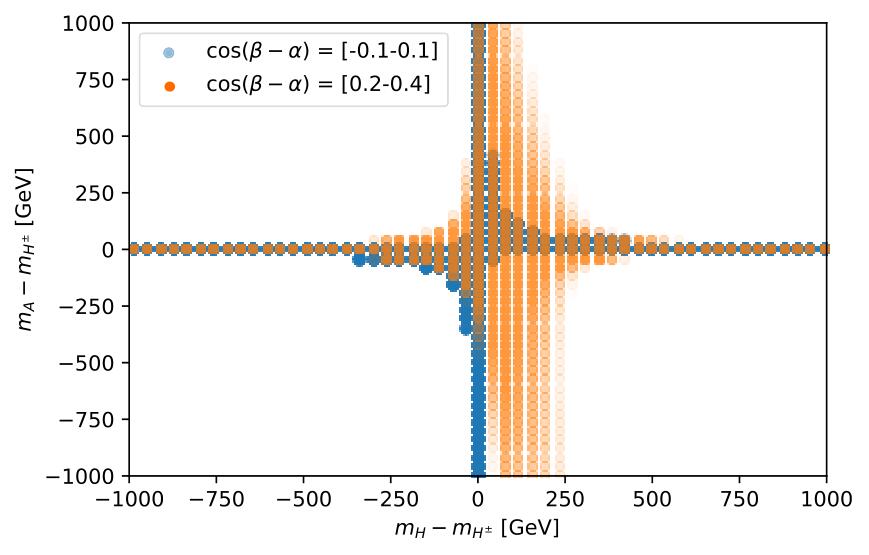}
\caption{\label{fig:2HDm_mA_mH_mHc} 
Parameter points with $-0.04\le T\le 0.24$ in the ($m_{H}-m_{H^\pm}$, $m_{A}-m_{H^\pm}$) plane, for $|\cos(\beta-\alpha)| < 0.1$ and $0.2 < \cos(\beta-\alpha) < 0.4$.
}
\end{center}
\end{figure}
\noindent
The values of the EWPOs, $S$, $T$ and $U$ within the 2HDM are derived in~\cite{Grimus:2008nb,Grimus:2007if} and implemented in \THDMC. The latter depends on the squared masses of the neutral Higgses through the $F$ function~\cite{Veltman:1977kh}, which commonly appears in loop calculations:
\begin{equation}
	F(x,y) = \frac{x+y}{2} - \frac{xy}{x-y} \ln \frac{x}{y}.
\end{equation}
\noindent
$F(x,y)$ is a non-negative function, which is zero for $x=y$, and monotonically
increasing with the difference between $x$ and $y$. To simplify the notation, we use $F(A,B)$ denoting $F(m^{2}_{A}, m^{2}_{B})$. The $T$ parameter in the 2HDM can be
expressed as:
\begin{equation}
\begin{aligned}
	T 
	&=
	c
	\Biggl\{ \Biggr.
	\cos^{2}( \beta - \alpha )
	\Bigl[
	  	  F( H^{\pm} , h )
		- F( A       , h )
		- F( H^{\pm} , H )
      + F( H       , A ) \Bigr. \\
      &+ 3 \left[
			 F(Z,H) - F(W,H)
          \right]
      - 3 \left[
          F(Z,h) - F(W,h)
          \right]
	\Bigl. \Bigr] \\
	&+
	F( H^{\pm}, H ) - F( H , A ) + F( H^{\pm} , A )
	\Biggl. \Biggr\}, 
\end{aligned}
\end{equation}
\noindent
where $c$ is
\begin{equation}
	c = \frac{1}{\alpha_{\rm EM}} \frac{ g^{2} }{ 64 \pi^{2} m_{W}^{2}}.
\end{equation}
\noindent
In the alignment limit, where $\cos(\beta - \alpha) \approx 0$, the $T$ parameter simplifies to:
\begin{equation}
	T
	=
	c
	\left[
	F( H^{\pm}, H ) - F( H , A ) + F( H^{\pm} , A )
	\right].
\end{equation}
\noindent
From this we see that a mass degeneracy between $A$ or $H$ and $H^\pm$ induces a 
vanishing  $T$ parameter: i.e., 
\begin{itemize}
	\item $m_{H^{\pm}} \approx m_{H}$ implies $T \approx 0$;
	\item $m_{H^{\pm}} \approx m_{A}$ implies $T \approx 0$.
\end{itemize}
\noindent
Fixing either $m_H$ or $m_A$ to be equal to the charged Higgs mass is the rule-of-thumb generally taken in the literature to satisfy the EWPO constraints within the 2HDM. However, in the wrong-sign region where $\cos(\beta - \alpha) > 0$, by taking only the leading bound on the $T$ parameter into account, this mass degeneracy can be relaxed to  a large extent. This is shown Fig.~\ref{fig:2HDm_mA_mH_mHc}, where we choose the $T$ value to be in the interval: $-0.04\le T\le 0.24$. This choice is based on the GFitter analysis of Ref.~\cite{Gfitter}, where $U=0$ is imposed for extracting the 95\% CL bounds. As displayed by the orange points, for large $\cos (\alpha - \beta )$ values, that is in the wrong-sign domain, the $m_H$ and $m_A$ masses could simultaneously differ from the charged Higgs mass by roughly 250 GeV (or even more). Very large differences between scalar masses lead to large non-perturbative contributions, though, therefore extreme cases are disfavoured (see later).

\begin{figure}[t!]
	\begin{center}
			\includegraphics[width=0.4\textwidth]{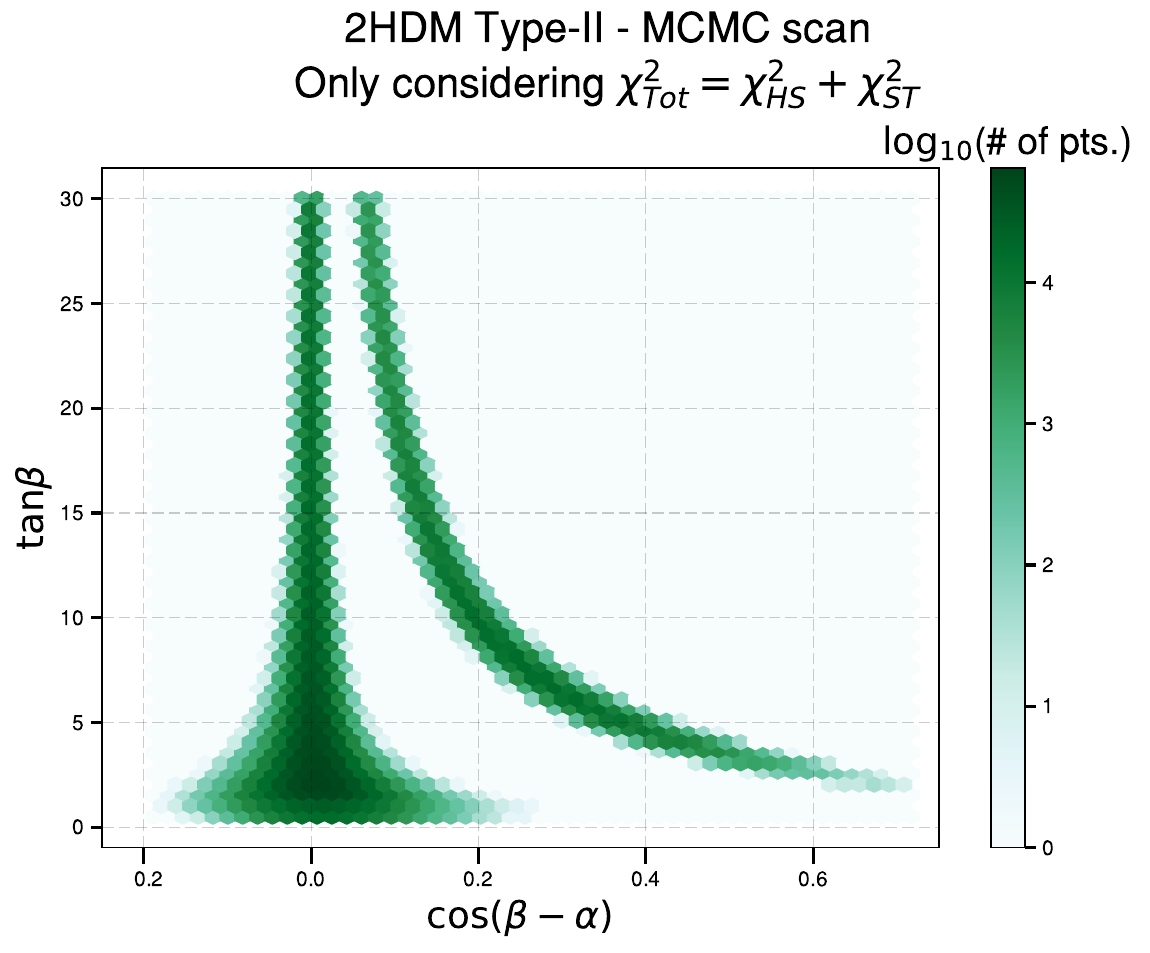}
			\includegraphics[width=0.4\textwidth]{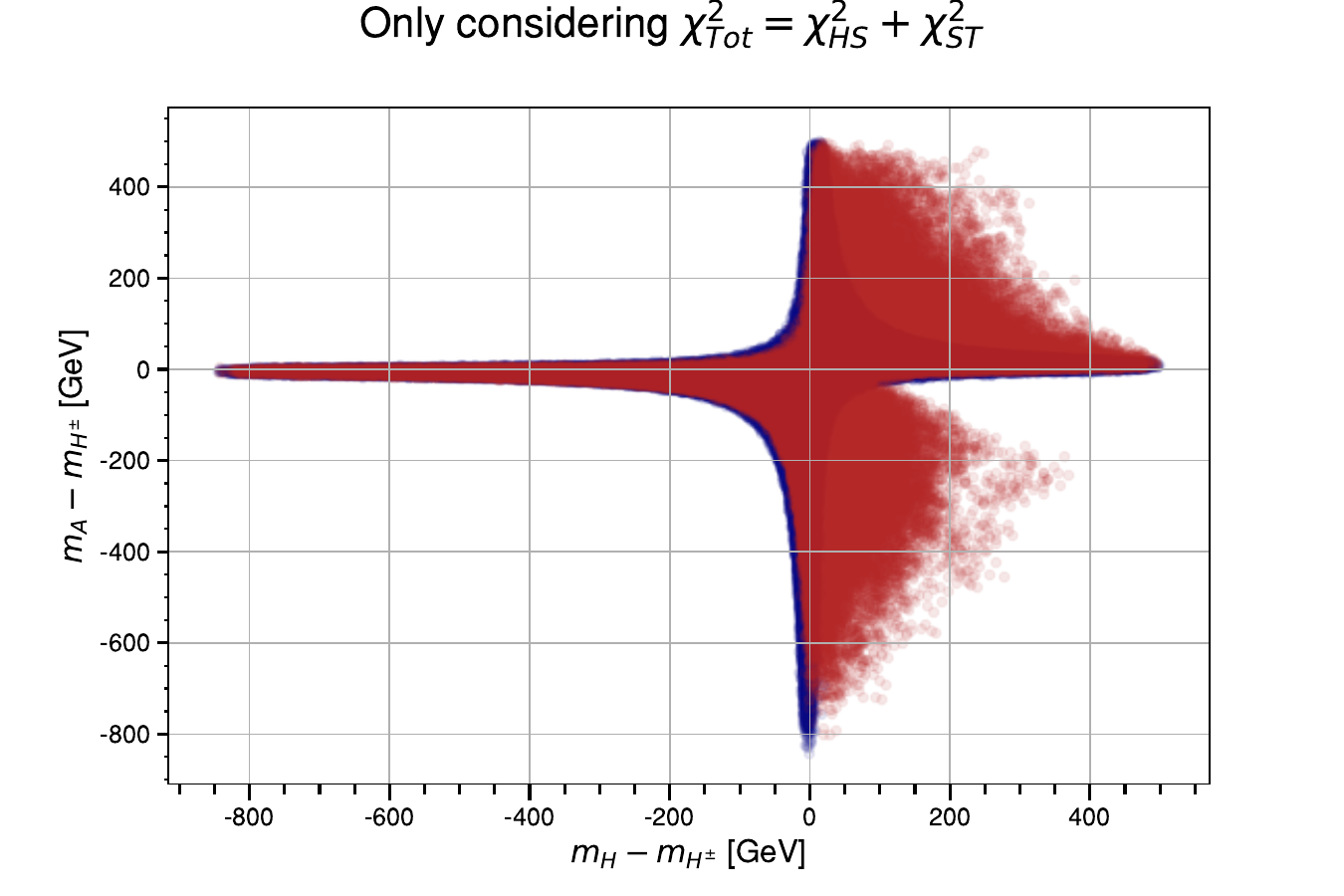}
			\includegraphics[width=0.4\textwidth]{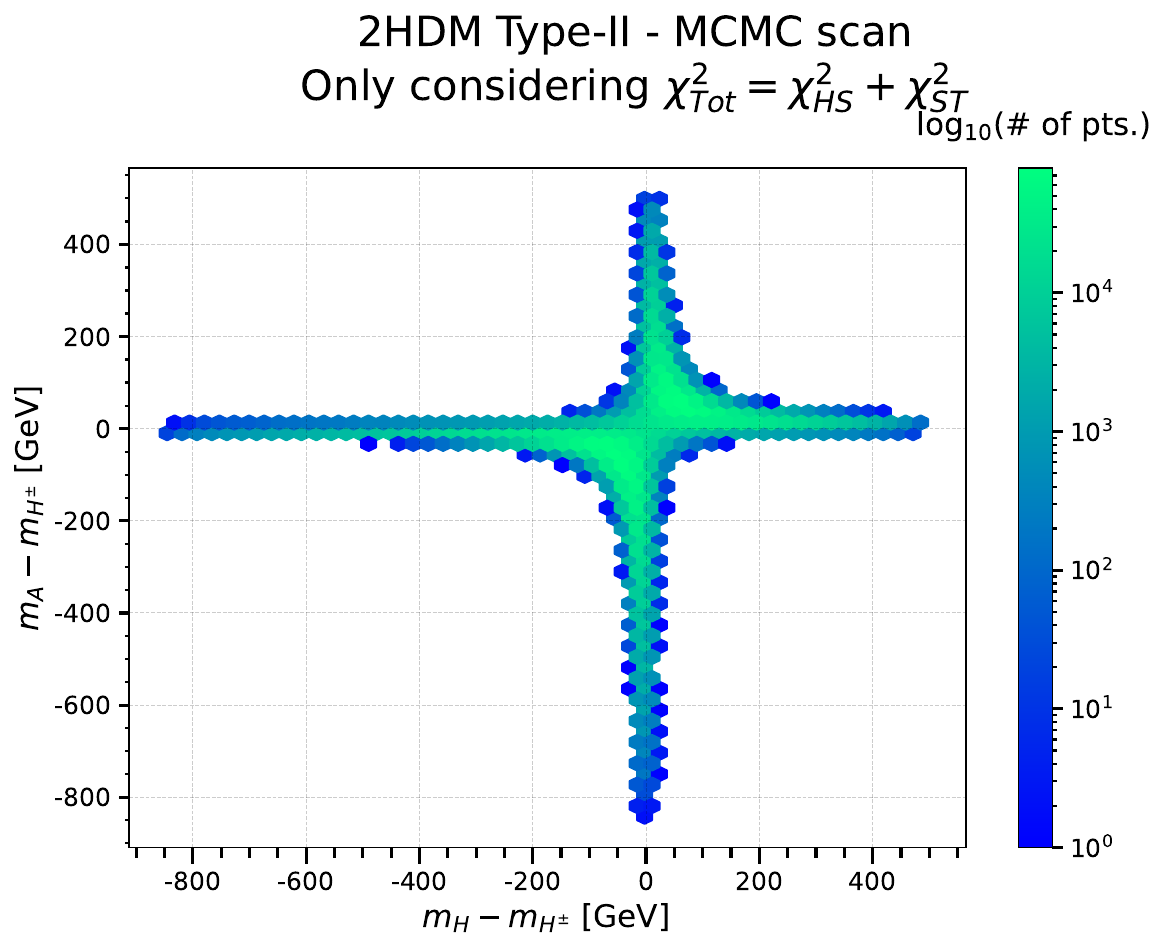}
			\includegraphics[width=0.4\linewidth]{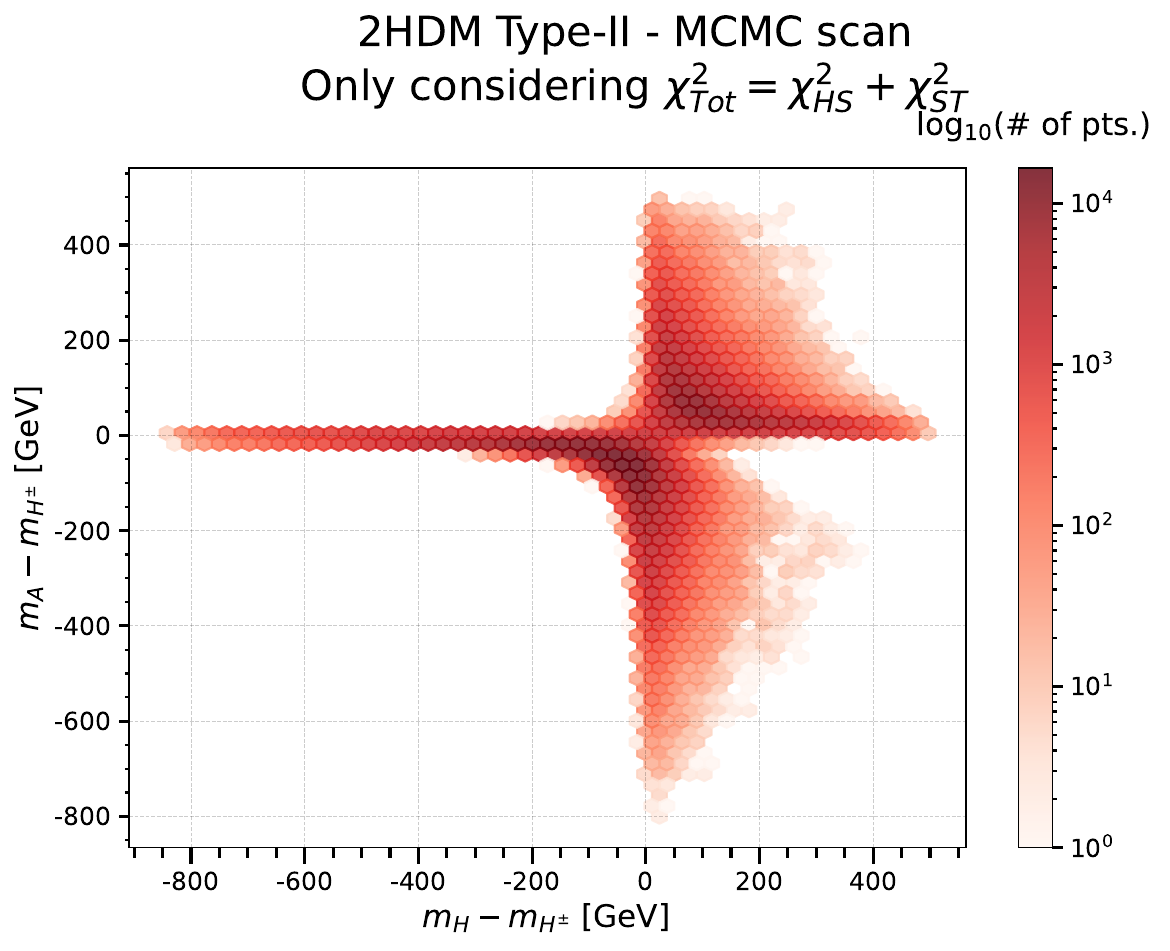}
	\caption{
	Distribution of the parameter space points on the ($\cos (\beta - \alpha), \tan \beta$) (top left) and ($m_H-m_{H^\pm},  m_A-m_{H^\pm}$) (top right) planes from the MCMC scan in the 2HDM Type-II. In the latter plot, the alignment region is represented in blue  while the wrong-sign scenario is superimposed in red. For a clearer picture of the ($m_H-m_{H^\pm},  m_A-m_{H^\pm}$) planes the wrong-sign and alignment points are seperated into a further two plots on the bottom row.The blue-green (bottom left) is the alignment plot while the red-orange (bottom right) is the wrong-sign plot.}
	\label{fig:2HDm_HS_ST} 
	\end{center}
\end{figure}

\noindent
The net result, upon including in the MCMC scan the constraints coming from both the SM-like Higgs boson measurements and the EWPOs, is visualised in Fig.~\ref{fig:2HDm_HS_ST}. There, we plot the allowed points  in two parameter planes: ($\cos (\beta - \alpha), \tan \beta$)  (left) as well as (${m_H-m_{H^\pm},m_A-m_{H^\pm}}$) (right). In the first case, we also display the density of points while, in the second case, we split the point between the alignment and wrong-sign scenarios. The left plot in Fig.~\ref{fig:2HDm_HS_ST} is in fairly good agreement with the experimental fits displayed in Fig.~\ref{fig:2HDM:cba_tb}, thus passing the goodness-of-fit test of the adopted HiggsSignals link.

\subsection{Theoretical constraints}

After discussing the limits on the 2HDM Type-II parameter space coming from direct and indirect experimental searches,  in this section, we analyse the effect of  theoretical constraints. The three major conditions can be concisely summarised as follows.
\begin{itemize}
	\item Unitarity of the $S$ matrix: the upper bound on the eigenvalues $L_i$ of the scattering matrix of all Goldstone and Higgs 2-to-2 channels \cite{Ginzburg2005,Kanemura1993} is fixed to be
\begin{equation}
	|L_{i}| \leq 16 \pi. 
\end{equation}
	\item Perturbativity: the quartic Higgs couplings should be small to justify the perturbative nature of the calculations
	\begin{equation}
		|\lambda_{H_{i} H_{j} H_{k} H_{l}}| \leq 8\pi.
	\end{equation}
	\item Stability of the potential: the quartic Higgs potential terms are bounded from below, in turn implying that \cite{Deshpande1978}
	\begin{equation}
		\lambda_{1} > 0,
		\quad
		\lambda_{2} > 0,
		\quad
		\lambda_{3} + \sqrt{\lambda_{1} \lambda_{2}} > 0, 
		\quad
		\lambda_{3} + \lambda_{4} - |\lambda_{5}| + \sqrt{\lambda_{1} \lambda_{2}} > 0.
        \label{eq:stability}
	\end{equation}
\end{itemize}

\begin{figure}[t]
\begin{center}
\includegraphics[width=0.4\linewidth]{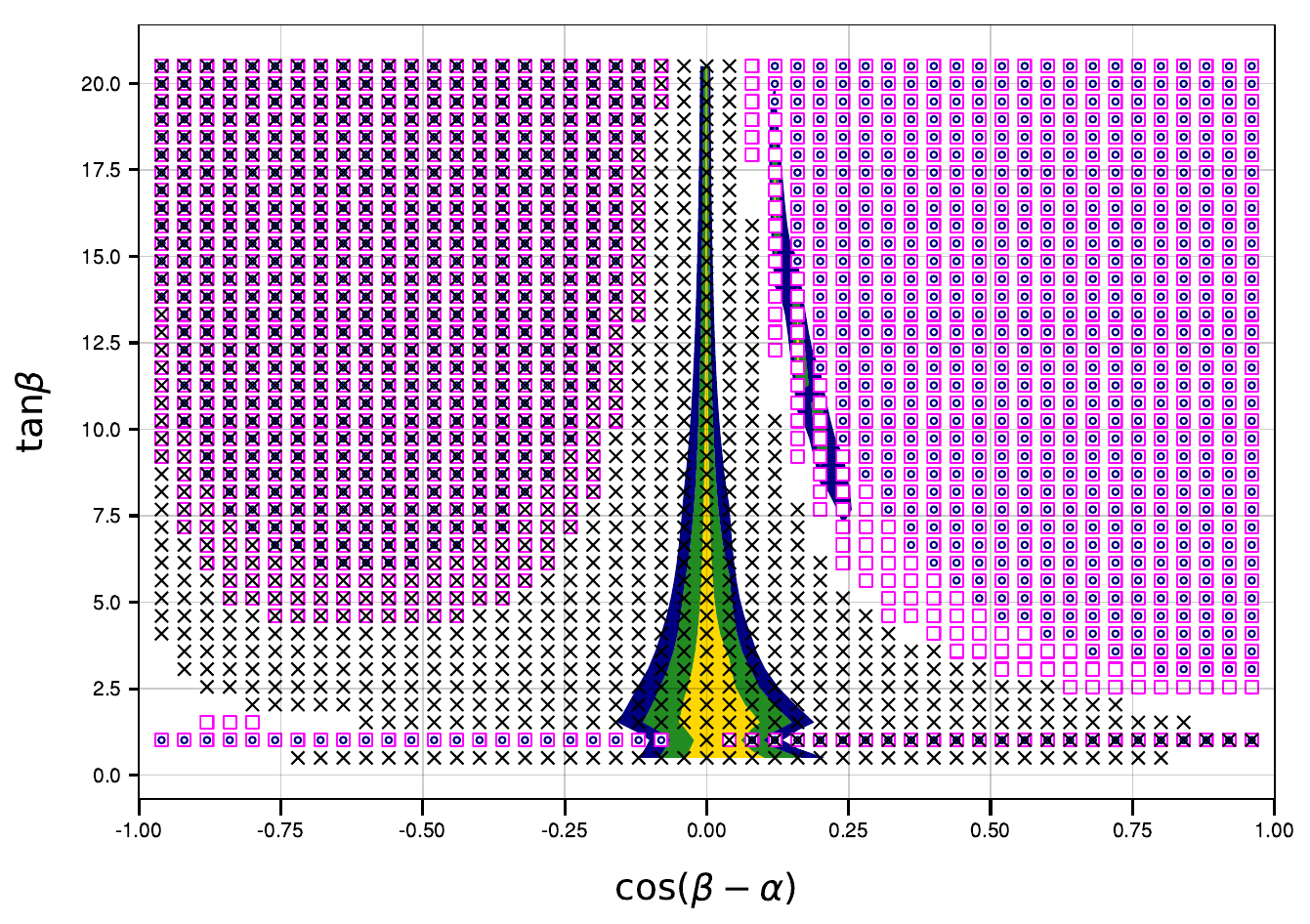}
\includegraphics[width=0.4\linewidth]{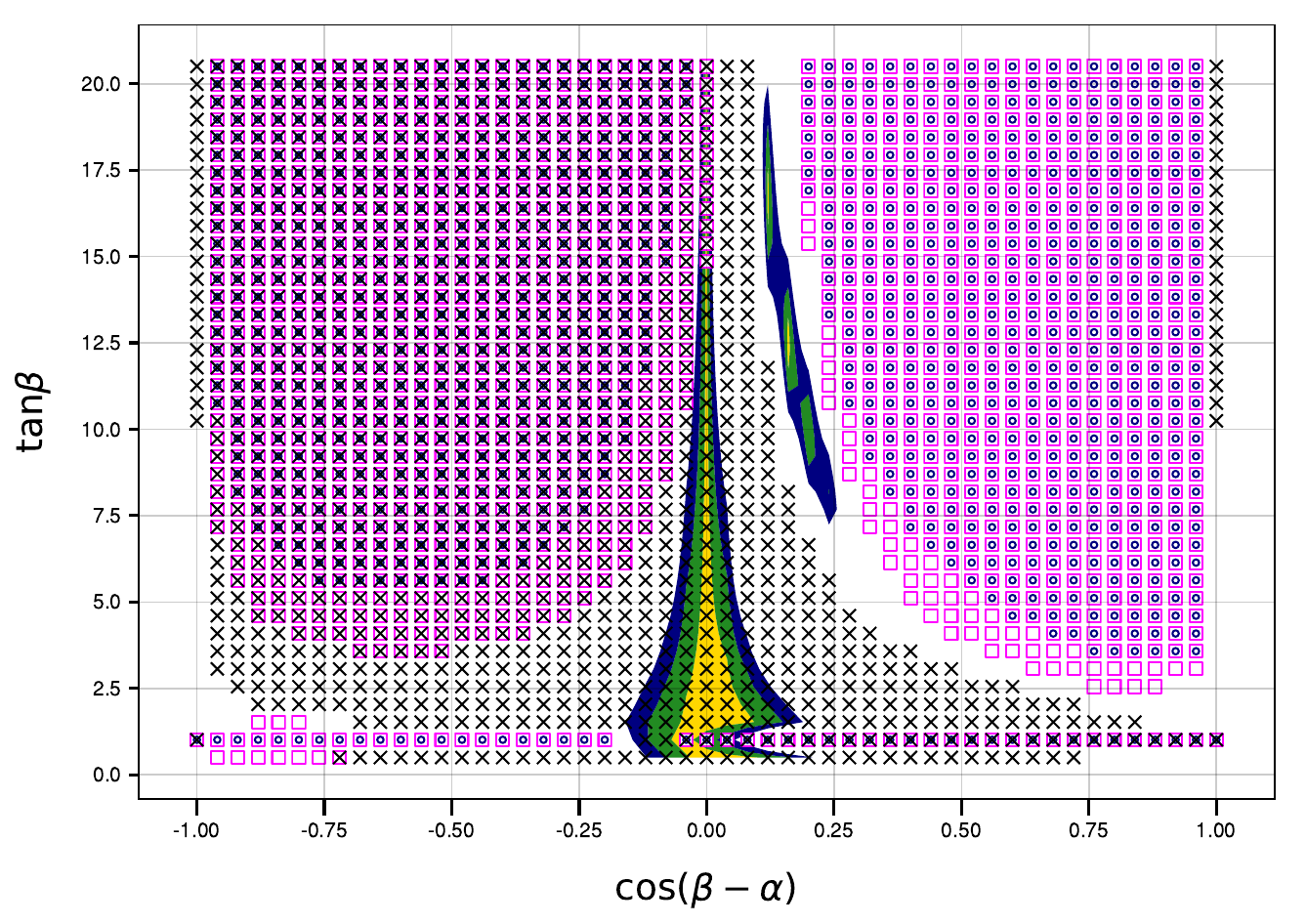}\\
\includegraphics[width=0.4\linewidth]{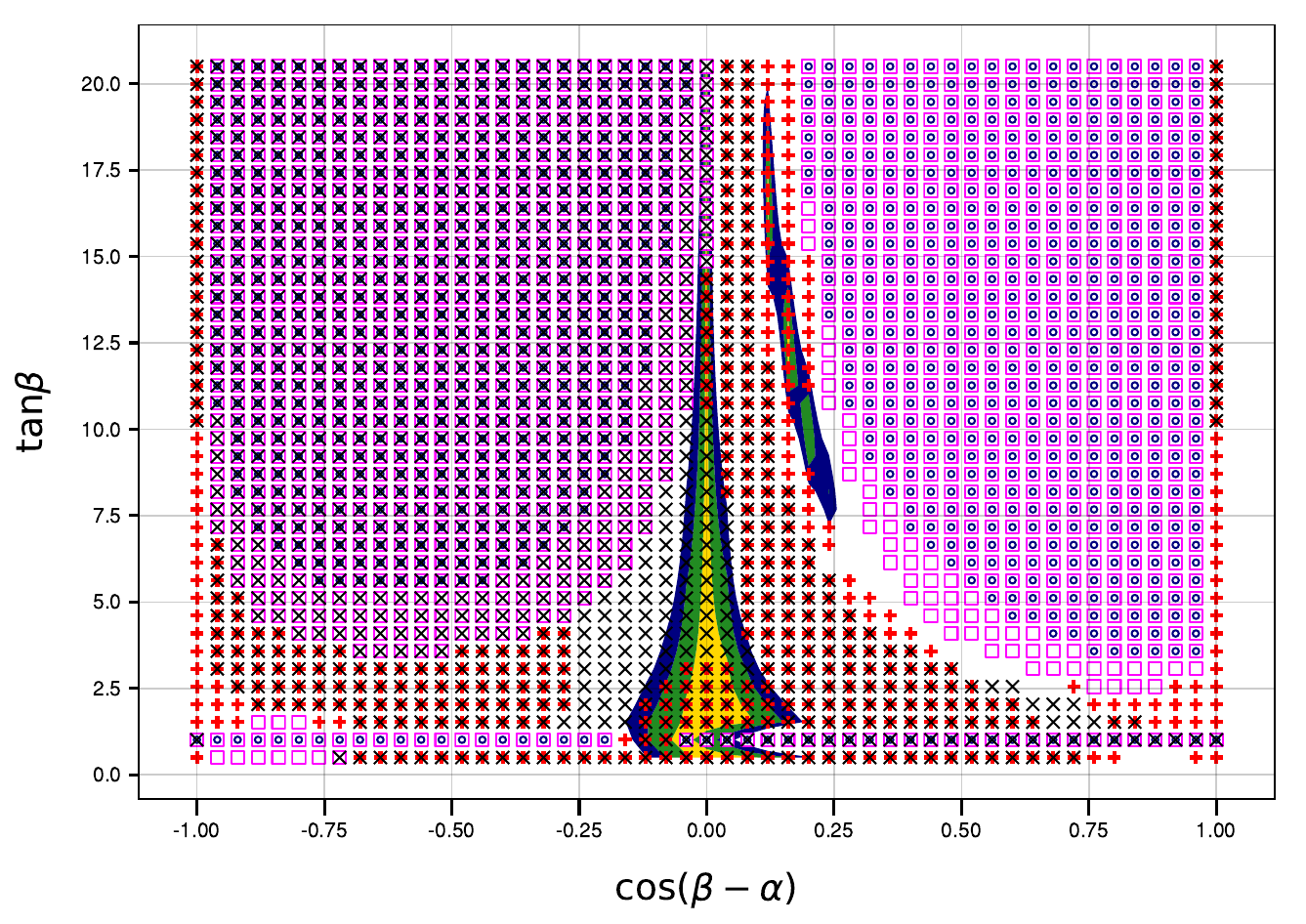}
\includegraphics[width=0.4\linewidth]{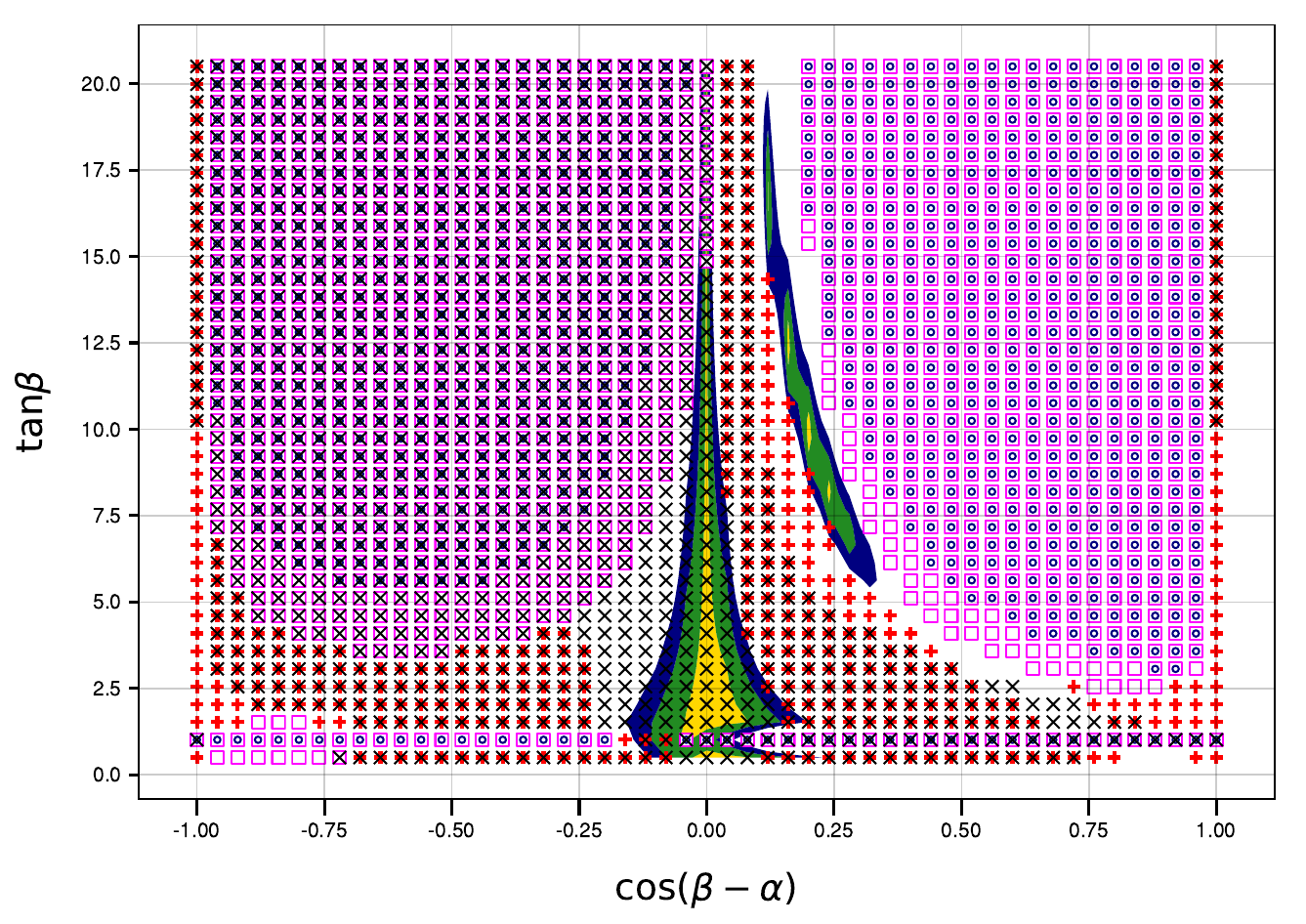}
\caption{
\label{fig:2HDm_Theory}
Distribution of the parameter space points on the ($\cos (\beta - \alpha),  \tan \beta)$ plane excluded by the theoretical constraints of unitarity (blue hollow dots), perturbativity (magenta hollow squares) and stability (black crosses) in the 2HDM Type-II. The masses of the heavy Higgs, $H$ and the charged Higgs $H^{\pm}$ are fixed at $m_{H^\pm}=m_H$ = 600 GeV. The \HiggsSignals and EWPO allowed regions correspond to the yellow, green, blue regions, with 1, 2 and 3$\sigma$ CL compatibility respectively.
Top row: in the left-handed plot, $m_A = 300$ GeV and $Z_7 = 0$; in the right-handed plot, $m_A = 300$ GeV and $Z_7 = 0.6$. Bottom row: in the left-handed plot, $m_A = 300$ GeV and $Z_7 = 0.6$; in the right-handed plot, $m_A = 400$ GeV and $Z_7 = 0.6$. The points excluded by \HiggsBounds are also shown as red crosses.}
\end{center}
\end{figure}

\noindent
The general potential coefficients can be expressed with the help of the masses and the angles $\alpha, \beta$ as follows (we use the notation $c_X, s_X$ and $t_X$ ($X=\alpha,\beta$) to signify $\cos X, \sin X$ and $\tan X$, respectively): 
\begin{equation}
	\lambda_{1} =
	\frac{m^{2}_{H} c^{2}_{\alpha} + m^{2}_{h} s^{2}_{\alpha} - m^{2}_{A} s^{2}_{\beta}} { v^{2} c^{2}_{\beta} }
	- \lambda_{5} t^{2}_{\beta} - 2 \lambda_{6} t_{\beta},
\end{equation}
\begin{equation}
	\lambda_{2} =
	\frac{m^{2}_{H} s^{2}_{\alpha} + m^{2}_{h} c^{2}_{\alpha} - m^{2}_{A} c^{2}_{\beta}} { v^{2} s^{2}_{\beta} }
	- \lambda_{5} t^{-2}_{\beta} - 2 \lambda_{7} t^{-1}_{\beta},
\end{equation}
\begin{equation}
\lambda_{3} =
\frac{ (m^{2}_{H} - m^{2}_{h}) s_{\alpha} c_{\alpha} + (2 m^{2}_{H^{\pm}} - m^{2}_{A}) s_{\beta} c_{\beta} }{ v^{2} s_{\beta} c_{\beta}}
- \lambda_{5} - \lambda_{6} t^{-1}_{\beta} - \lambda_{7} t_{\beta},
\end{equation}
\begin{equation}
\lambda_{4} =
\frac{2 (m^{2}_{A} - m^{2}_{H^{\pm}})} { v^{2} } + \lambda_{5},
\end{equation}

\begin{equation}
\lambda_{4} - |\lambda_{5}| =
\frac{2 (m^{2}_{A} - m^{2}_{H^{\pm}})} { v^{2} } + \lambda_{5} - |\lambda_{5}| =
\begin{cases}
\frac{2 (m^{2}_{A} - m^{2}_{H^{\pm}})} { v^{2} },                   & \text{if } \lambda_{5} > 0,\\
\frac{2 (m^{2}_{A} - m^{2}_{H^{\pm}})} { v^{2} } - 2 |\lambda_{5}|, & \text{if } \lambda_{5} > 0.
\end{cases}
\end{equation}

\noindent
Out of these, the stability and the perturbativity of the potential pose the most severe constraints on the parameter space. In order to give an overview of the bounds coming from the theoretical constraints, in Fig.~\ref{fig:2HDm_Theory}, we display the 2HDM Type-II parameter space regions excluded by the different sources. For illustrative purposes, we have fixed the mass of the (pseudo)scalars to be $m_{H^\pm} = m_H$ = 600 GeV and $m_A$ = 300 and 400 GeV. The blue dots reflect the bounds arising from the requirement of unitarity. The effects are concentrated in the medium-high $\tan \beta$ range and for $|\cos (\beta - \alpha )|\ge$ 0.1. Positive(negative) values of $Z_7$ disfavour the negative(positive) values of $\cos (\beta - \alpha )$, shifting the excluded region on the right-(left-)hand side. The unitarity bounds do not affect the alignment and the wrong-sign domains, allowed by the \HiggsSignals and EWPO constraints and represented by the
blue(green and yellow) region at the 95\% (90\% and 68\%) CL. The perturbativity constraint, represented by the magenta squares, extends the excluded region towards lower values of $\tan (\beta )$ and $|\cos (\beta - \alpha )|$ for the chosen value of the quartic Higgs coupling. The effect of $Z_7$ is the same as for unitarity. As displayed in the two upper plots of Fig.~\ref{fig:2HDm_Theory}, for $Z_7=0$ the wrong-sign contour region is completely excluded by the perturbativity constraint (see left plot). By increasing the $Z_7$ value to $Z_7=0.6$, the parameter space opens up again (see right plot). According to this trend, even if not shown explicitly in the Figure, one can desume that negative $Z_7$ values tend to exclude the alignment region as well.
Finally, the stability of the potential, represented by the black crosses, excludes all the negative values of $\cos (\beta - \alpha )$ and part of the positive values so to suppress almost completely the alignment domain. Summarising the effect of the three constraints coming from unitarity, perturbativity and stability, it is clear that negative values of $Z_7$ are disfavoured. For the chosen setup, no points lie in the alignment region; they are indeed concentrated in the wrong-sign domain (see lower right plot). More generally, by increasing the $m_A$ value, the alignment and the wrong-sign scenarios get both populated again, the alignment contour at extremely low $\tan\beta$ values in particular.

\noindent
The conclusion to be drawn from this exercise is that the stability of the scalar potential enforces a lower bound on the pseudo-scalar mass, $m_A$, in the alignment portion of the parameter space. We analyse this effect in more details in the next section.

\subsection{The role of $m_A$}

\begin{figure}[t]
\begin{center}
\includegraphics[width=0.45\linewidth]{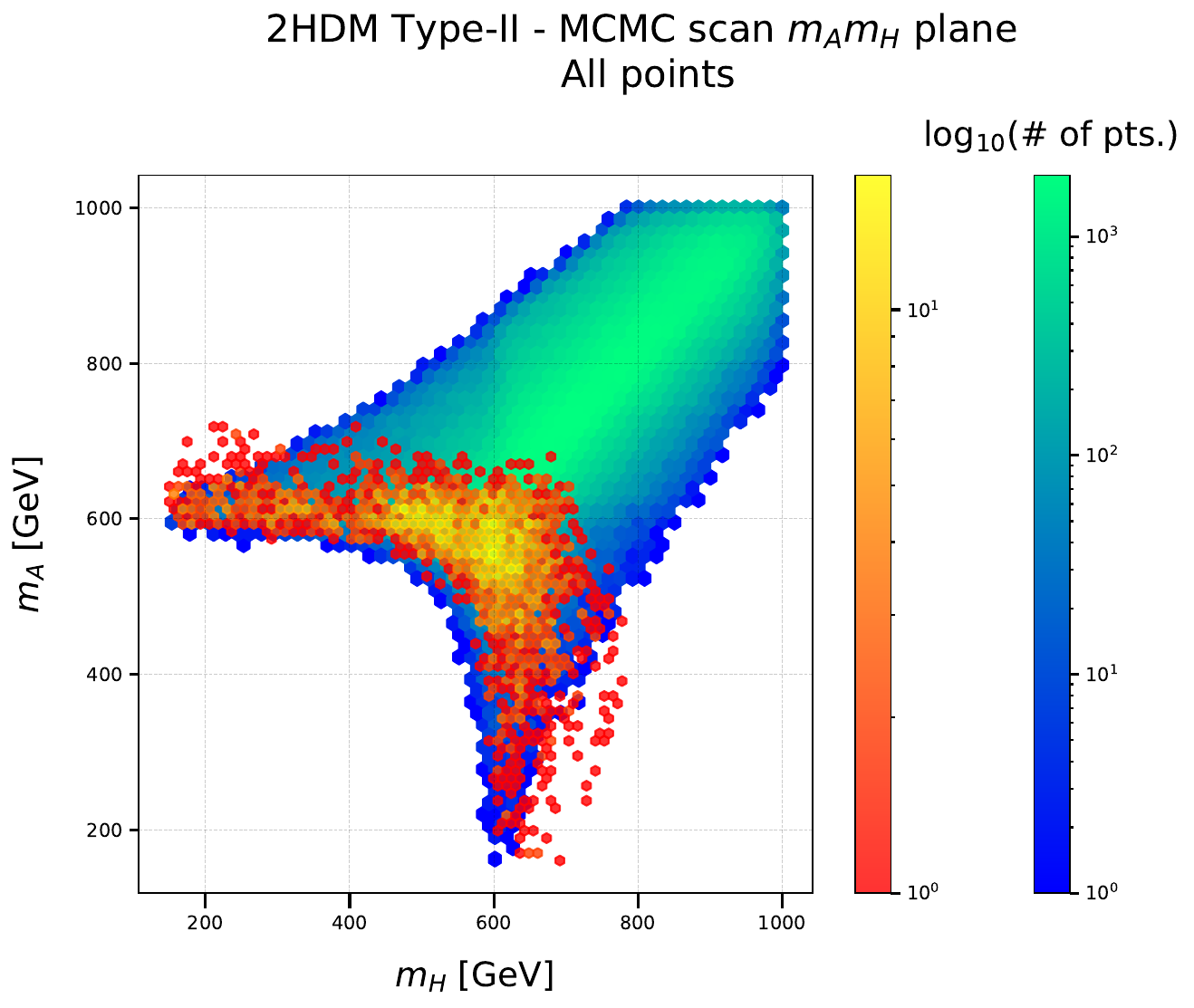}
\includegraphics[width=0.45\linewidth]{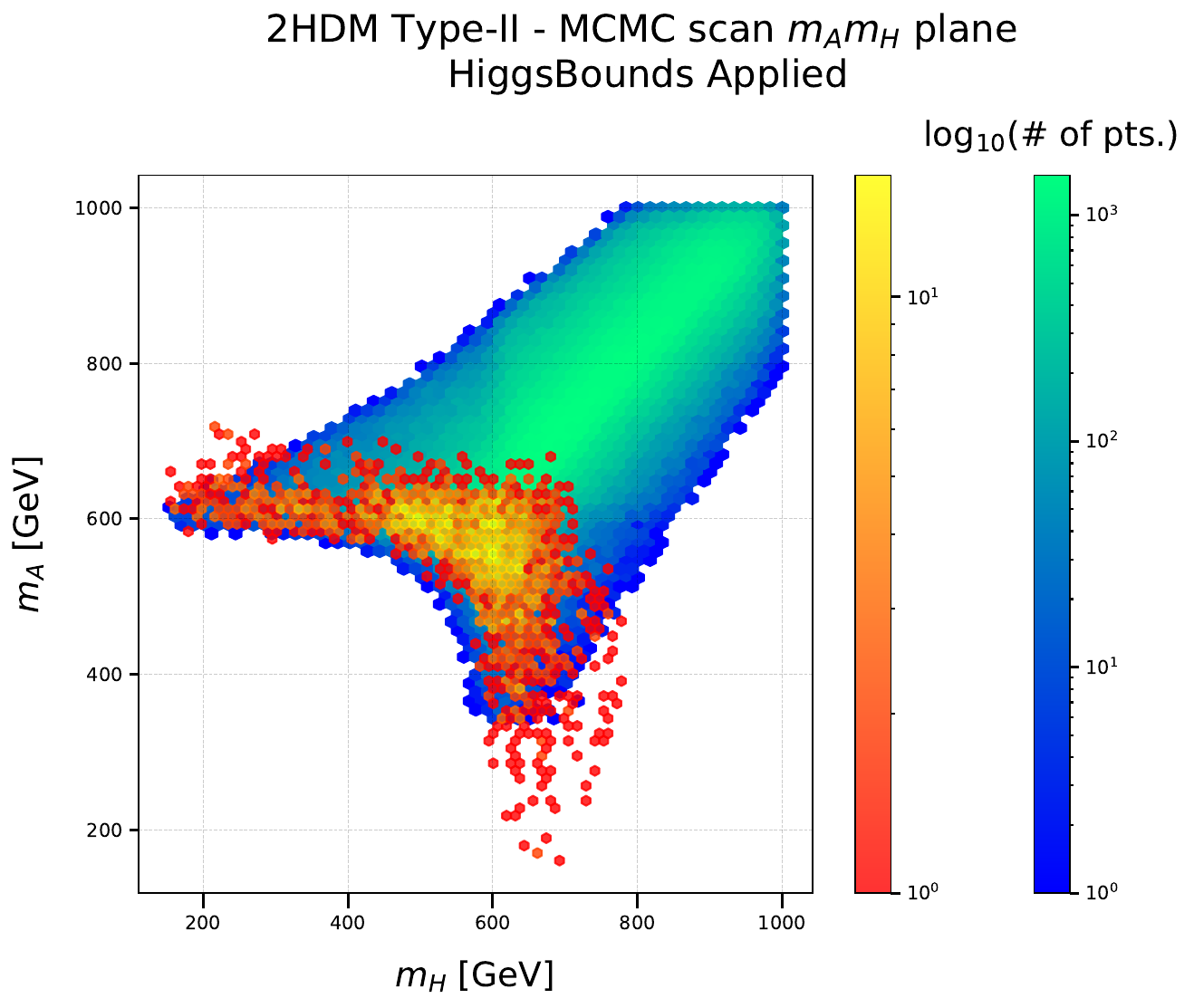}
\caption{
Distribution of the parameter space points on the $(m_A, m_H)$ plane allowed by the theoretical
constraints in the 2HDM Type-II. The bound on the charged Higgs mass is implemented as
$m_{H^\pm}\ge$ 600 GeV. In the  left plot, the \HiggsSignals, EWPOs and theoretical constraints are
enforced. In the right plot, \HiggsBounds limits are also added. The blue/green dots represent the alignment 
region while the red/yellow ones refer to the wrong-sign scenario. Two colour bars are shown next to each plot, both giving a logarithmic scale count to the colours used for the plot.
}\label{fig:2HDm_mABound} 
\end{center}
\end{figure}

In this subsection we investigate the conditions imposed by a stable scalar potential and their effect on the two limits of the model under consideration (2HDM Type-II): the alignment and wrong-sign domains. We use a collection of points from the MCMC scan, which passes the condition $\Delta\chi^{2}_{\rm tot} < (3 \sigma$ CL upper limit) without imposing any other constraints. The stability inequalities in Eq.~(\ref{eq:stability}) are implemented step-by-step to be able to uniquely identify their effect on the parameter space. The following observations can be made.
\begin{itemize}
	\item At the beginning (without imposing any of the stability conditions) there are points present
	in both the alignment and wrong-sign limit regions.
	\item The constraints $\lambda_{1} > 0$ and $\lambda_{2} > 0$ are targeting points from both
	regions irrespectively of the $m_{A}$ value.
	\item There are surviving points in both regions after imposing $\lambda_{1} > 0$ and
	$\lambda_{2} > 0$.
	\item The condition $\lambda_{3} + \sqrt{\lambda_{1} \lambda_{2}} > 0$ does not exclude any
	additional points for low $m_{A}$ values but discards a large number of points exclusively
	from the alignment limit in the high $m_{A}$ domain.
	\item The final constraint of $\lambda_{3} + \sqrt{\lambda_{1} \lambda_{2}} + \lambda_{4} -
	|\lambda_{5}] > 0$ again disfavours points from the alignment region independently on the $m_{A}$
	value. More importantly, this proves to exclude all of the points from the alignment limit region in
	the low-intermediate $m_{A}$ range, with the exception of a handful of points at low $\tan \beta$.
	Contrary to this, the high $m_{A}$ range contains surviving  points in both regions after imposing
	all the conditions.
\end{itemize}

\noindent
This result is visualised in the scatter plots of Fig.~\ref{fig:2HDm_mABound}, where we display the $(m_H, m_A)$ parameter space. The blue dots represent the alignment region while the red ones refer to the wrong-sign scenario. In these plots, we enforce the experimental bounds coming from \HiggsSignals and  EWPOs plus the theoretical constraints discussed above. We moreover set the lower bound on the charged Higgs mass at 600 GeV. In the left  plot, one can see that very few points are left in the alignment region at low $m_A$. Those few are characterised by very small values of $\tan\beta$, as discussed previously. If we superimpose the \HiggsBounds limits, even these remaining points disappear. 

\noindent
The global picture is shown in the right plot of Fig.~\ref{fig:2HDm_mABound}. There one can see that, in the alignment limit of the 2HDM, the pseudoscalar state is required to be rather heavy: $m_A \ge 350$ GeV. Only in the wrong-sign scenario, it can in principle have a mass as light as $m_A \simeq$ 150 GeV (see red dots), when $Z_7$ is rather large and positive definite as shown in Fig.~\ref{fig:2HDm_Theory}. This latter feature is the result of the effects coming from the perturbativity enforcement. This picture depends however on the limit that could be in future set on the charged Higgs mass. Raising the $m_{H^\pm}$ limit pushes the lower bound on $m_A$ further up, in the alignment scenario.
In the wrong-sign domain, one can still have light CP-odd Higgs masses at the price of stretching $Z_7$ towards large and positive values, $Z_7\ge 1$,  typically. This is in agreement with the findings given in Ref.~\cite{Bernon:2015wef}. Here, we have added a more detailed analysis of the effects coming from the individual constraints, highlighting in particular the role of the stability requirement on the scalar potential in setting a lower bound on the CP-odd Higgs mass in the alignment scenario.

\noindent
In this section, we have described the framework and tools to extract the portion of the 2HDM Type-II parameter space that is allowed by the present experimental constraints (summarised by \HiggsSignals, EWPOs and \HiggsBounds) and the theoretical requirements. We are now ready to discuss the possibilities that \Magellan, the global scan tool we are presenting in this paper, offers to interpret the LHC data coming from a variety of up-to-date analyses within this specific model we are focussing on, the 2HDM Type-II. 

\section{Data interpretation}

\begin{figure}[t]
\begin{center}
\includegraphics[width=0.40\linewidth]{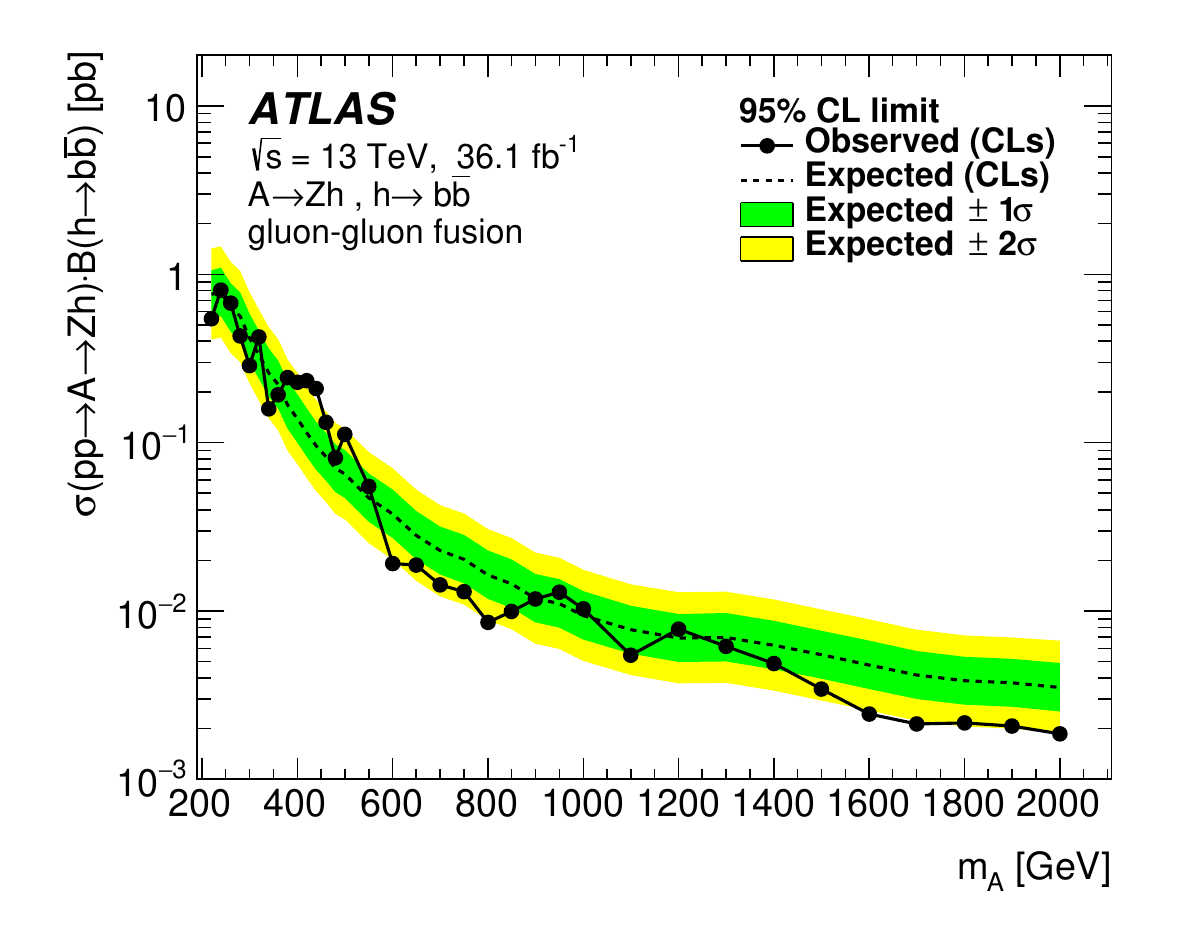}
\includegraphics[width=0.45\linewidth]{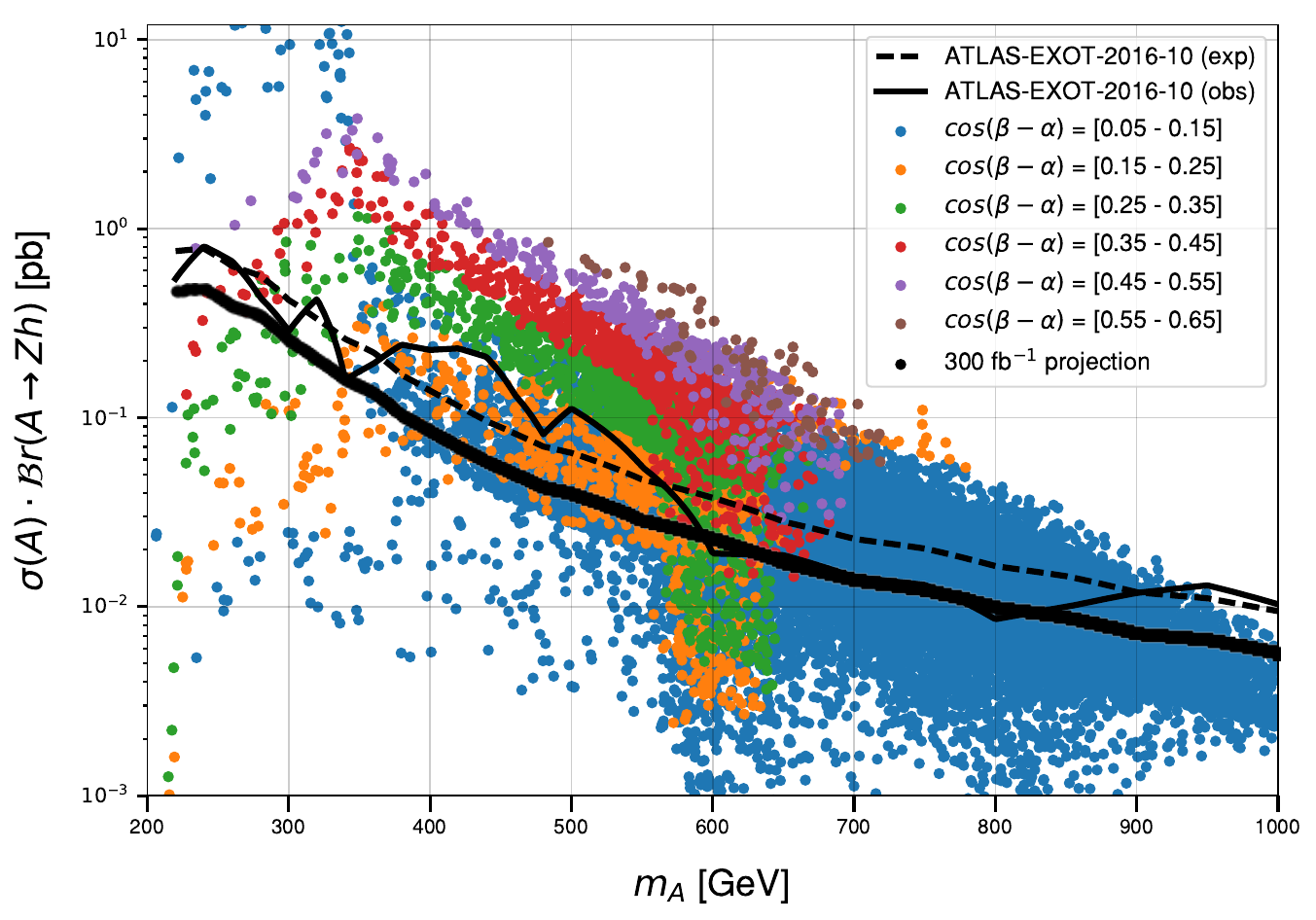}
\caption{Left plot: 95\% CL upper bound on the cross-section times BRs, $\sigma (pp\rightarrow A\to Zh\rightarrow
Zb\bar b)$, as a function of the CP-odd Higgs mass, extracted by ATLAS at the 13 TeV
LHC~\cite{ATLAS_AZh}.
Right-plot: Theoretical predictions for the same process $pp\rightarrow A\to Zh\rightarrow
Zb\bar b$ within the 2HDM Type-II (here, $\sigma_A\equiv \sigma (pp\rightarrow A$)). The
different colours of the points in the scatter plot represent different values of $\cos
(\beta - \alpha)$. Superimposed, there is the ATLAS observed (expected) cross-section
times BR given by the black solid (dashed) line. Finally, the series of black dots show
the projection of the expect limit curve of the ATLAS analysis to a luminosity of $L = 300$ fb$^{-1}$.
}\label{fig:ATLAS-Zh} 
\end{center}
\end{figure}

In this chapter, we apply the methodology of the global scan tool, \Magellan, to interpret the LHC data within the 2HDM Type-II. During the course of the MCMC scan, various experimental and theoretical properties linked to the individual parameter space points are computed and saved. This retained information allows to examine different aspects of the model from the same dataset. Any new unfolded experimental results can be then 
translated into direct bounds on the parameter space of the BSM scenario at hand, the 2HDM Type-II. The experimental results corresponding to a given observable, typically the 95\% CL exclusion bound on the cross-section times BR, can be projected onto any two-dimensional sections of the full parameter space, thus allowing the extraction of limits on different parameters of the theory. The observables, i.e. cross-sections and BRs used for comparison, are computed by making use of \SusHi and \THDMC.

\noindent
As a working example, in the following, we consider the most recent ATLAS analysis of the process $pp\rightarrow A \rightarrow Zh\rightarrow Zb\bar b$~\cite{ATLAS_AZh}. The search for the heavy CP-odd Higgs, $A$, decaying into a $Z$ boson and the 125 GeV Higgs state,  is performed by  looking at final states with either two opposite-sign charge leptons ($l^+l^-$ with $l = e, \mu$) or a neutrino pair ($\nu\bar\nu$) plus two $b$-jets at the 13 TeV LHC with a total integrated  luminosity of $L = 36.1 $ fb$^{-1}$. The 95\% CL upper bound on the cross-section times BR as a function of the CP-odd Higgs mass $m_A$ is shown in the left plot of Fig.~\ref{fig:ATLAS-Zh}. There, it is assumed that the possible signal comes from the pure gluon-gluon fusion production. In the right plot of the same figure, the theoretical cross-section times  BR is computed within the 2HDM Type-II for the same $m_A$ range. The different colours of the scatter points correspond to the values of $\cos (\alpha - \beta)$ shown in the top-right legend. The cross-section times BR depends on this parameter, sensibly. The couplings of the CP-odd Higgs, $A$, with the heavy quarks in the production subprocess and with the $Z$ and $h$ bosons in its subsequent decay all depend on $\cos (\alpha - \beta)$. Superimposed on this scatter-plot, there are the observed and expected curves taken from the ATLAS analysis (see left plot). From direct comparison, one can immediately see the excluded range of the CP-odd Higgs mass as a function of the $\cos (\alpha - \beta)$ value. This comparison can be further extended by taking into account the limit on the cross-section expected in a near future with a
luminosity $L = 300$ fb$^{-1}$. The projected exclusion bounds on the $\cos (\alpha - \beta)$ show indeed a sensible improvement.
\begin{figure}[t!]
\begin{center}
\includegraphics[width=1\linewidth]{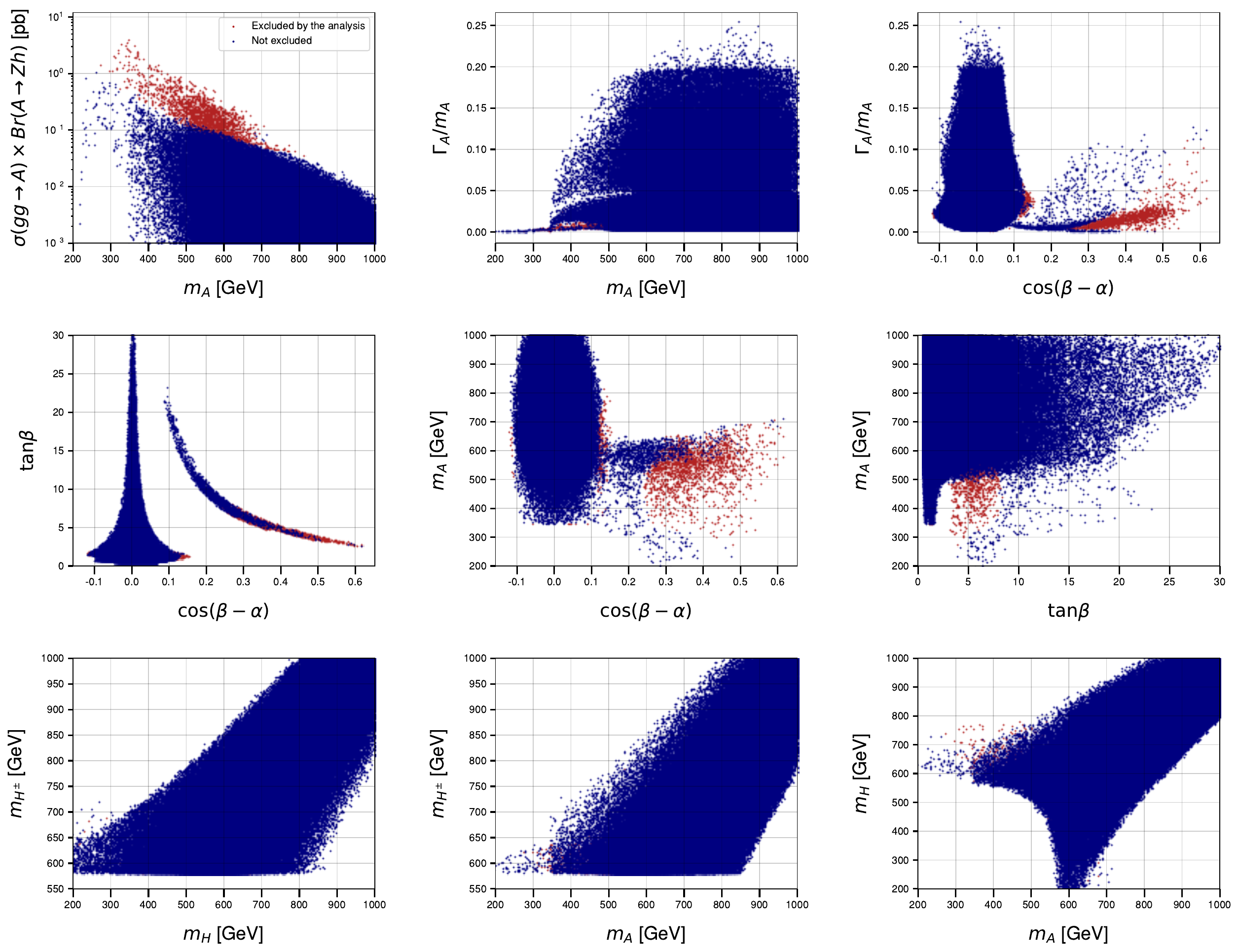}
\caption{\label{fig:AZh_exclusion_current}
Projections of the 2HDM Type-II parameter and observables.
The blue points are those allowed by \HiggsSignals, EWPOs and theoretical constraints. The red ones are those excluded by the ATLAS analysis of the process $pp\rightarrow A\rightarrow Zh\rightarrow Zb\bar b$ with a luminosity of $L = ~36.1$ fb$^{-1}$.
} 
\end{center}
\end{figure}

\begin{figure}[t!]
\begin{center}
\includegraphics[width=1\linewidth]{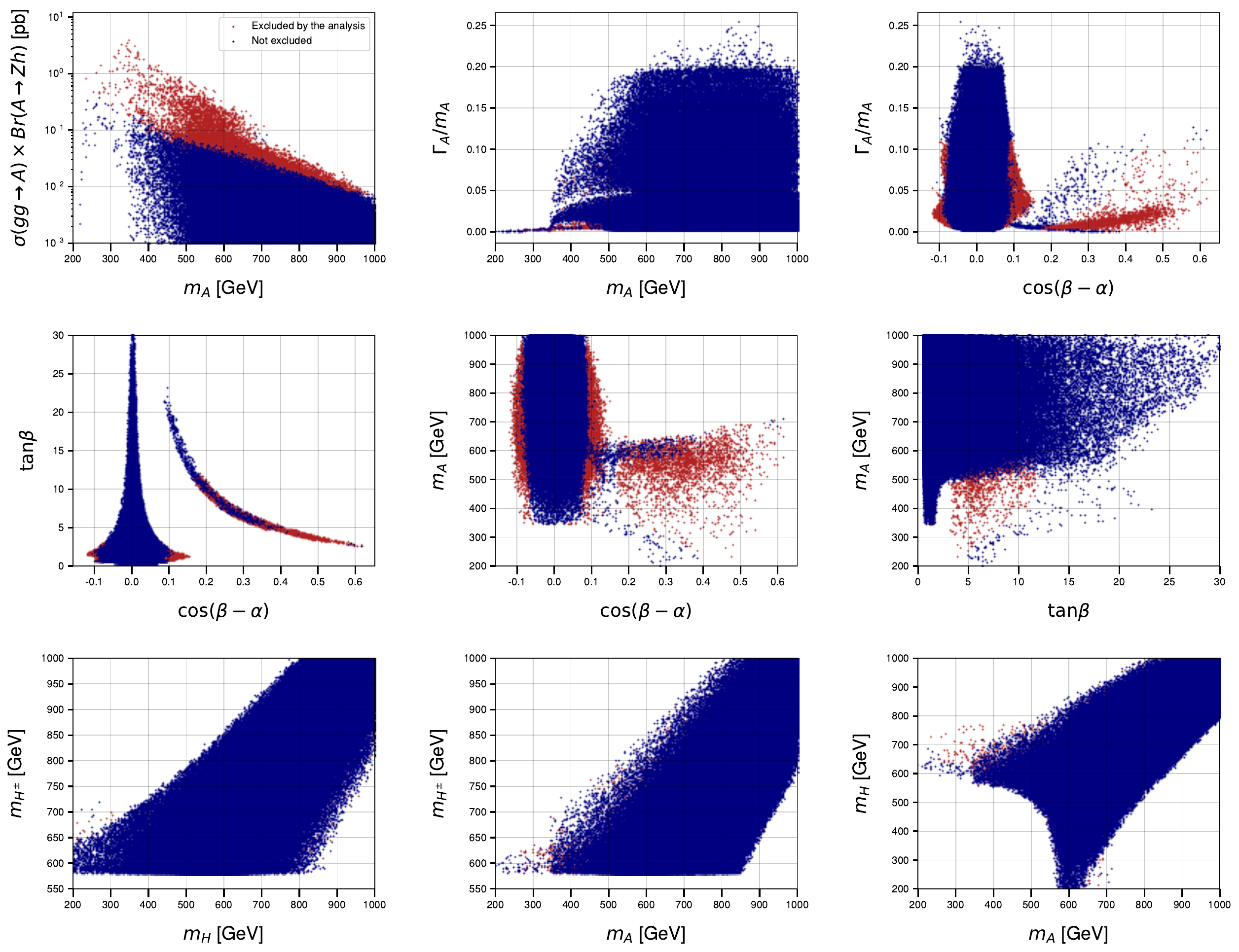}
\caption{\label{fig:ATLAS-projections-HL}
Projections of the 2HDM Type-II parameter and observables. The blue points are those allowed by \HiggsSignals, EWPOs and theoretical constraints. The red ones are those excluded by the ATLAS analysis of the process $pp\rightarrow A\rightarrow Zh
\rightarrow Zb \bar{b}$ projected to a luminosity of $L = ~300$ fb$^{-1}$.} 
\end{center}
\end{figure}

\noindent
Beyond this, \Magellan allows the extraction of a rich variety of information. The toolbox leverages the use of the \DataFrame class of \pandas, making a custom selection on the set
of points relatively easy.  Excluded (or allowed) points by a given theoretical constraint or experimental bound can then be projected onto any other plane, defined by the desired choice of model parameters or observables. In the specific case mentioned above, one can select points above the 95\% CL upper bound on the observed cross-section times BR, given by the black solid line on the right plot of Fig.~\ref{fig:ATLAS-Zh}, and project those points in order to see the effect of that particular model-independent measurement on all the free parameters of the 2HDM Type-II. Note that, as the limits coming from the experimental analyses reported on HEPData ({\tt https://www.hepdata.net/}) depend on the assumption made on the width of the new hypothetical Higgs bosons, when involved, the width of the (pseudo)scalar states is equally taken into account when extracting the bounds on the parameter space \footnote{Experimental limits are available up to $\Gamma_{A}/m_{A} < 11\%$.}.

\noindent
This feature is sketched in Fig.~\ref{fig:AZh_exclusion_current}. Nine different 2D projections of model parameters and observables are shown, where first the points excluded by the analysis (red) are drawn, and then non-excluded points (blue), indicating the region of the parameter space on 2D planes which are likely to be excluded, irrespective to the other hidden parameters. One could also choose to visualise the results in the opposite order, that is first the non-excluded points and then the excluded ones. In this way, the region of the parameter space tested by the specific experimental measurement at hand would stand out. The double option is implemented and shown on the \Magellan interactive webpage \cite{Magellan-Web}.

\noindent
From this subgroup of possible parameter spaces, one can already conclude that the low $\tan\beta$ region is the one being tested by the ATLAS analysis at the 13 TeV
LHC~\cite{ATLAS_AZh}, i.e., $\tan\beta\le$ 5 (\cf the (\cba, \tb) plane). There the range $\cos (\beta - \alpha)\ge$ 0.5 is almost excluded for all $\mA$
masses (c.f. the (\cba, \mA)  projection). Thus, even if initially one built a colourless scatter plot of the ${pp\rightarrow A\rightarrow Zh\rightarrow Zb\bar b}$ rate as a function of $\mA$, with no information on the $\cos (\beta - \alpha)$ value of the individual points, the projection feature could shed light on the range of $\cba$ and 
$\tanb$ that one is testing. Of course, higher luminosities could be sensitive to larger values of $\tan\beta$ and smaller values of $\cos (\beta - \alpha)$, thus extending the search of new physics in particular in the wrong-sign region. 

\noindent
Also, by looking at the top-right plot showing the value of the width of the CP-odd Higgs over its mass as a function of $\cba$, one can see that the present analysis covers a parameter space up to where $\Gamma_A/M_A\le 11\%$. But the possible values of this ratio extend up to $\Gamma_A/M_A\simeq 25\%$. This would imply that future experimental analyses 
should stop relying on the pure narrow width approximation and diversify their approach to include the search for wider resonances.

\noindent
Projecting the points excluded by the expected limit on the production cross-section times BR of the progess $pp\to A\rightarrow Zh$, at an integrated luminosity of $L = 300$ fb$^{-1}$, on the same projection planes as Fig.~\ref{fig:AZh_exclusion_current}, one can see that a very significant  portion of the parameter space could be under scrutiny. This is shown in Fig.~\ref{fig:ATLAS-projections-HL} by the red scatter points. The region $\cos (\beta - \alpha)\ge 0.4$ could already be excluded, as shown by the top-right, middle-left and middle-central plots. Also the alignment region would start to disappear. This already gives a rather good idea of what will happen in the next data taking stages at the LHC.  

\begin{figure}[t]
\centering 
\includegraphics[width=0.80\textwidth]{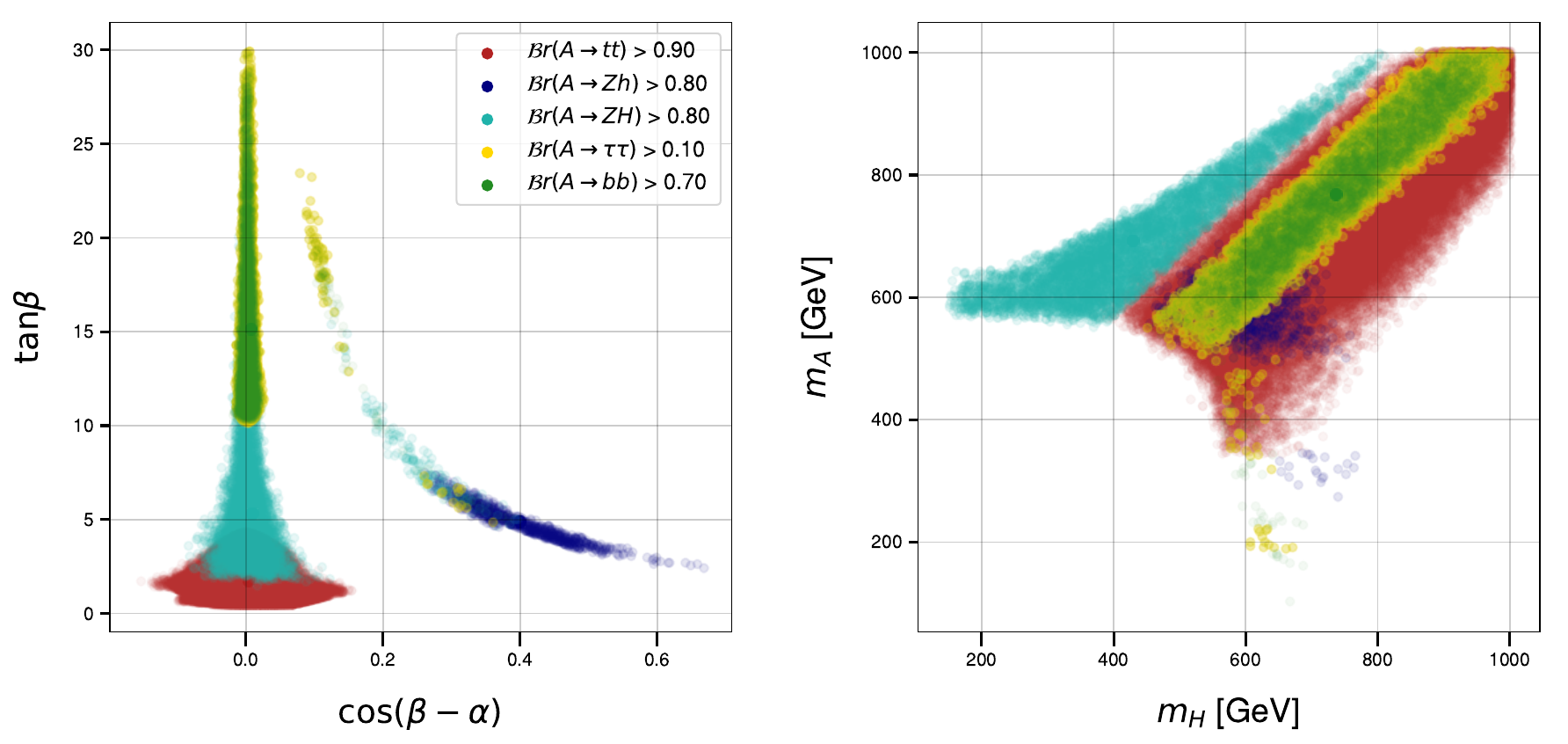}
\includegraphics[width=0.80\textwidth]{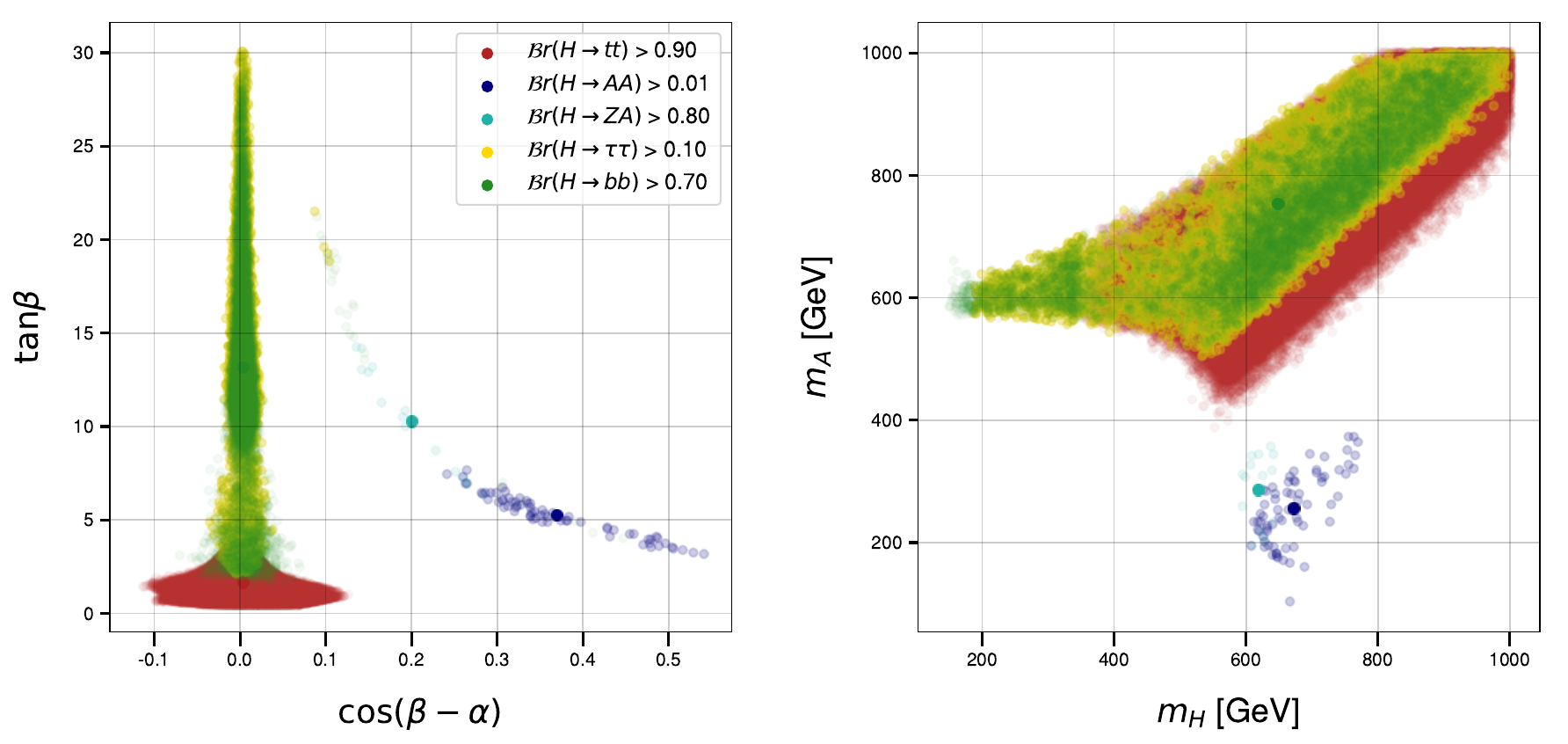}
\caption{Top plots: regions of the two-dimensional parameter spaces with high BRs of the CP-odd Higgs, $A$, in the channels given in the legend of the top-left plot. Bottom plots: same for the heavy CP-even Higgs, $H$, decaying into the channels listed in the legend of the bottom-left plot.}
\label{fig:BrAandH}
\end{figure}

\noindent
This way of interpreting the model-independent experimental data within a given model is much more flexible and complete than the the procedures adopted in the literature.
Referring in particular to the most recent $pp\to A\rightarrow Zh$ search  performed by
ATLAS~\cite{ATLAS_AZh}, one can notice that, for the interpretation of the cross-section times BR limits in the context of the 2HDM, the $H^\pm$, $H$ and $A$ bosons are assumed
to be degenerate. In our analysis, the three masses can differ by 250 GeV and more, as detailed in Section \ref{sec:2HDM-EWPO}. Moreover, the visualisation of the limits at {95\% CL} on the 2HDM parameters as given in Ref.~\cite{ATLAS_AZh} is constrained and therefore partial. Bounds are in fact displayed on the ($\tan\beta$, $\cos (\beta - \alpha)$) plane, at a fixed value of the resonance mass $m_A$, and on the ($\tan\beta$,
$m_A$) plane, at a fixed value of $\cos (\alpha - \beta)$. The global scan presented in this paper can go beyond these limitations and display the full limits on any 2D plane, offering access to a rich variety of information.

\subsection{2HDM sensitivity of different measurements at the LHC}
\begin{figure}[t]
\centering 
\includegraphics[width=0.45\textwidth]{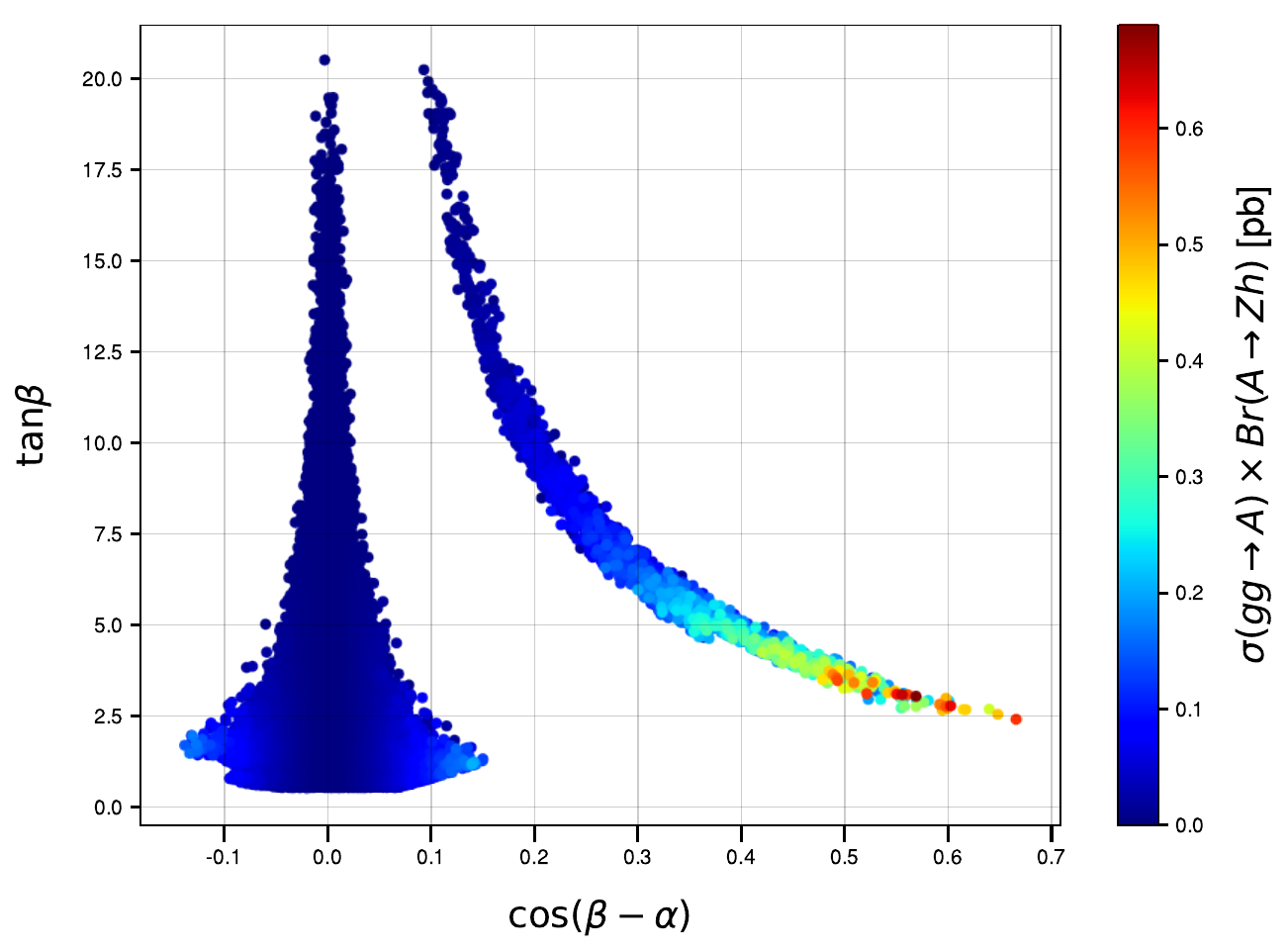}
\includegraphics[width=0.45\textwidth]{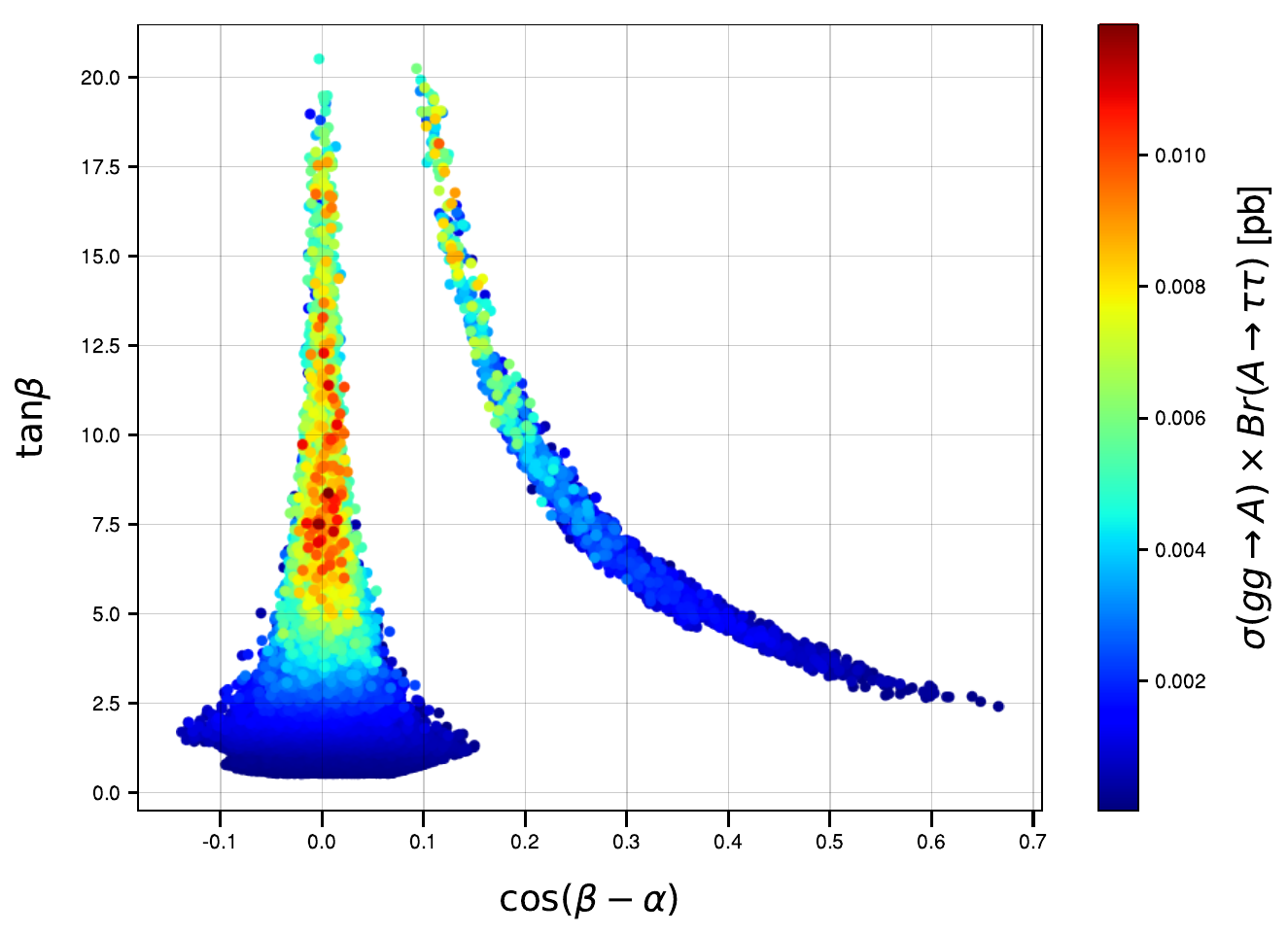}
\includegraphics[width=0.45\textwidth]{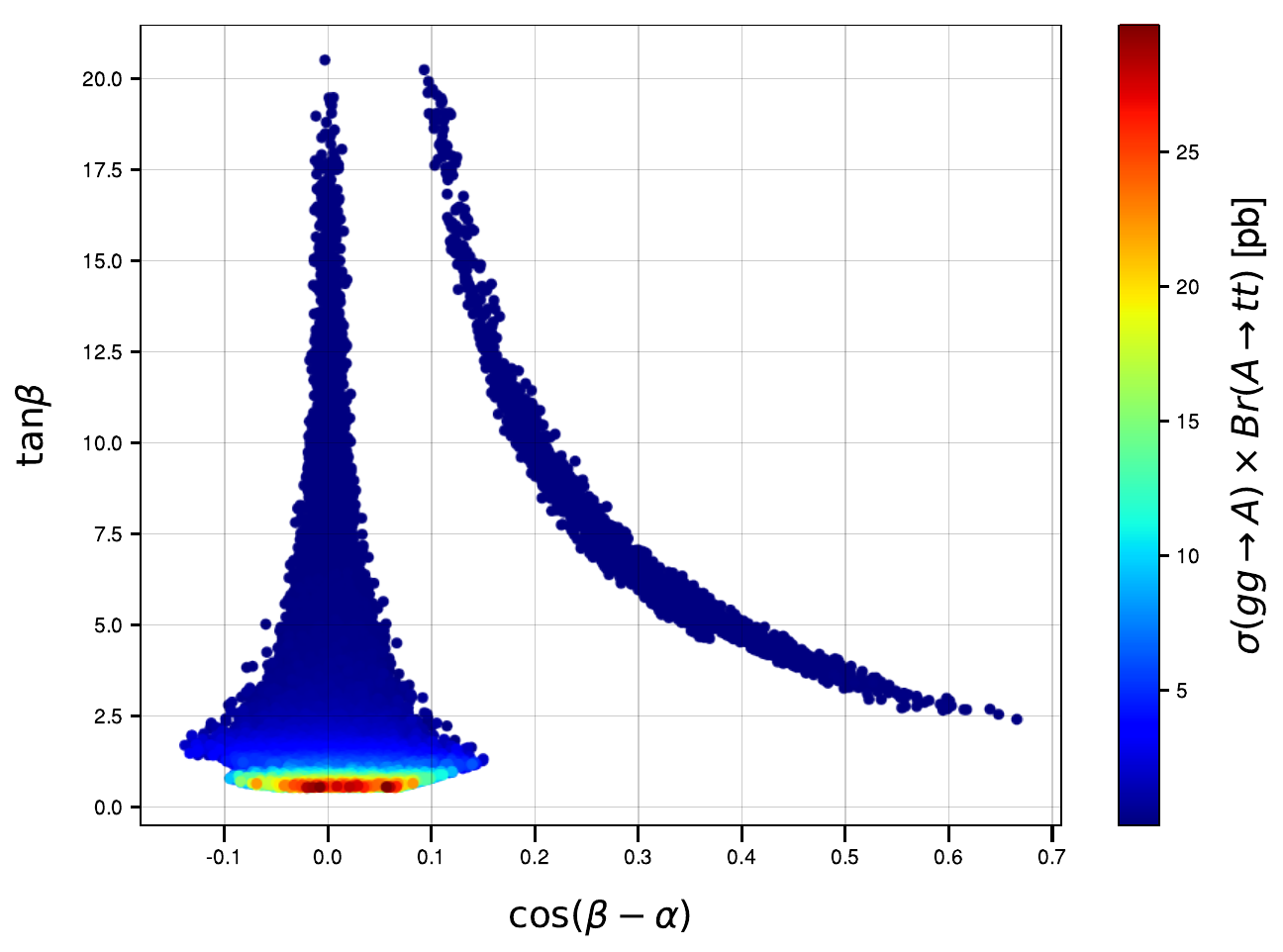}
\caption{Magnitude of the total cross-section times  BRs in the ($\cos (\alpha - \beta),
\tan\beta$) plane for four different processes mediated by the CP-odd Higgs, $A$. From top-left to
bottom (clock-wise): $pp\rightarrow A\rightarrow Zh$, $pp\rightarrow A\rightarrow\tau^-\tau^+$ and
$pp\rightarrow A\rightarrow t\bar t$.}
\label{fig:sigmaA}
\end{figure}

\begin{figure}[t]
\centering 
\includegraphics[width=0.45\textwidth]{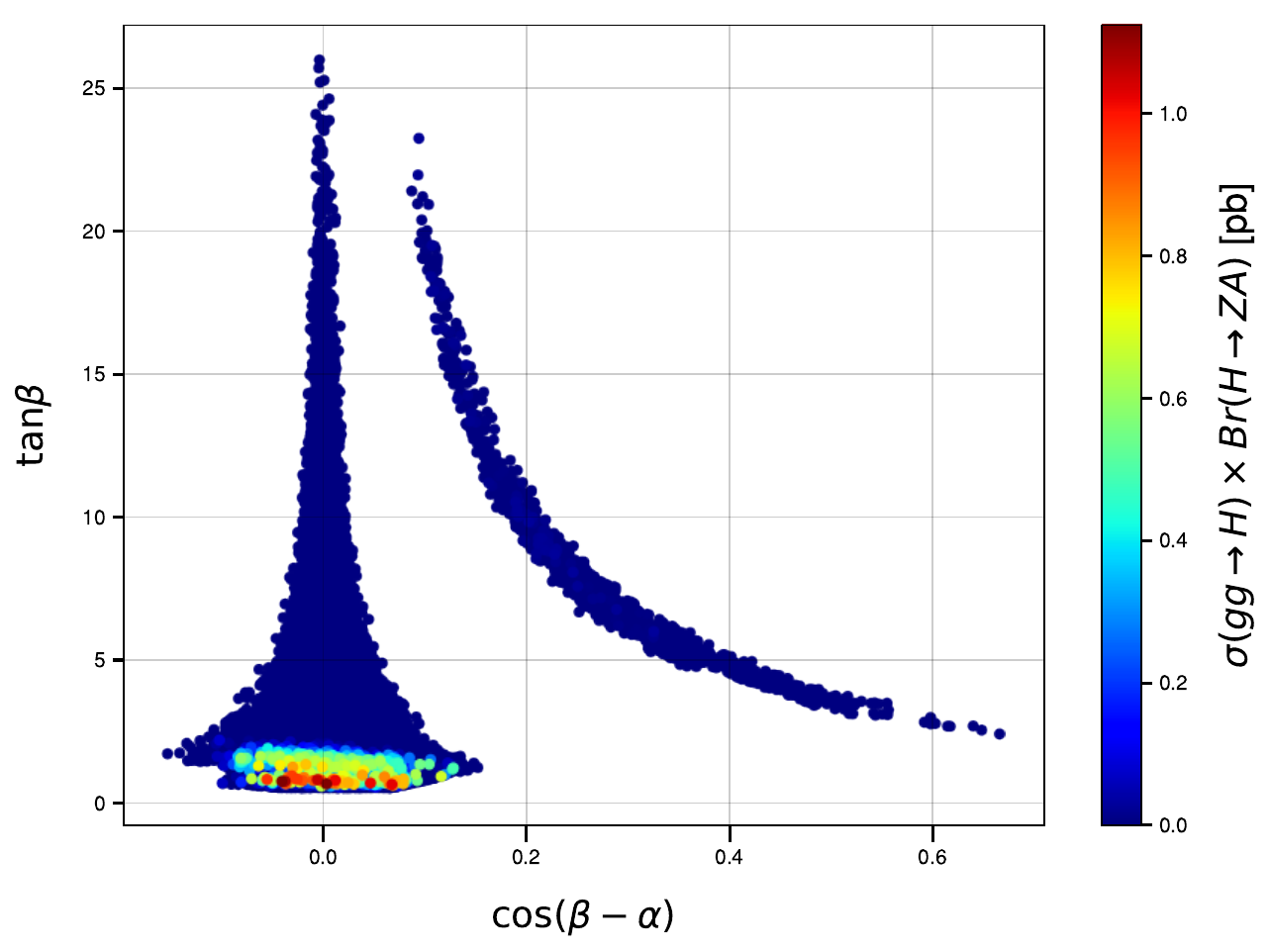}
\includegraphics[width=0.45\textwidth]{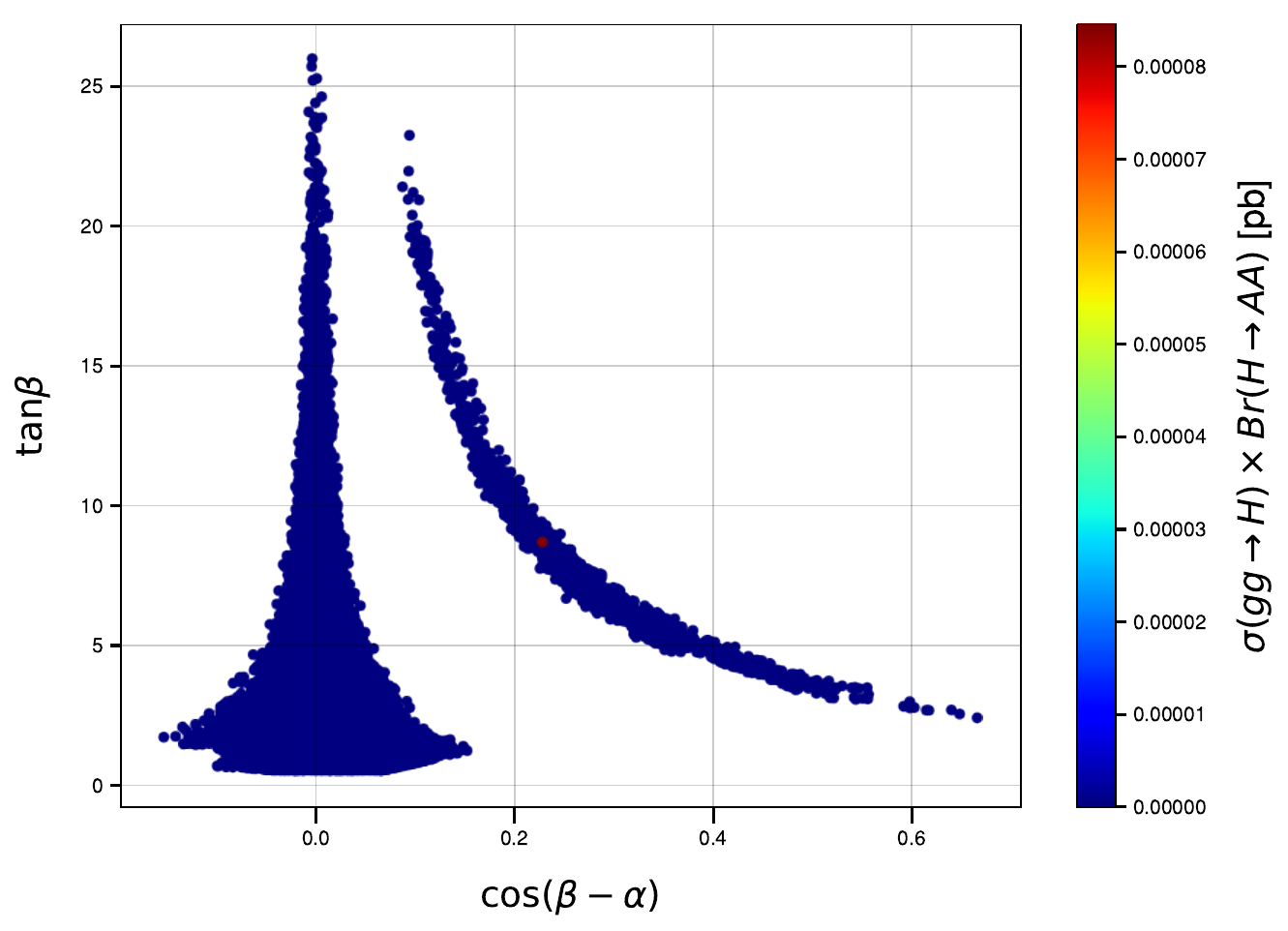}
\includegraphics[width=0.45\textwidth]{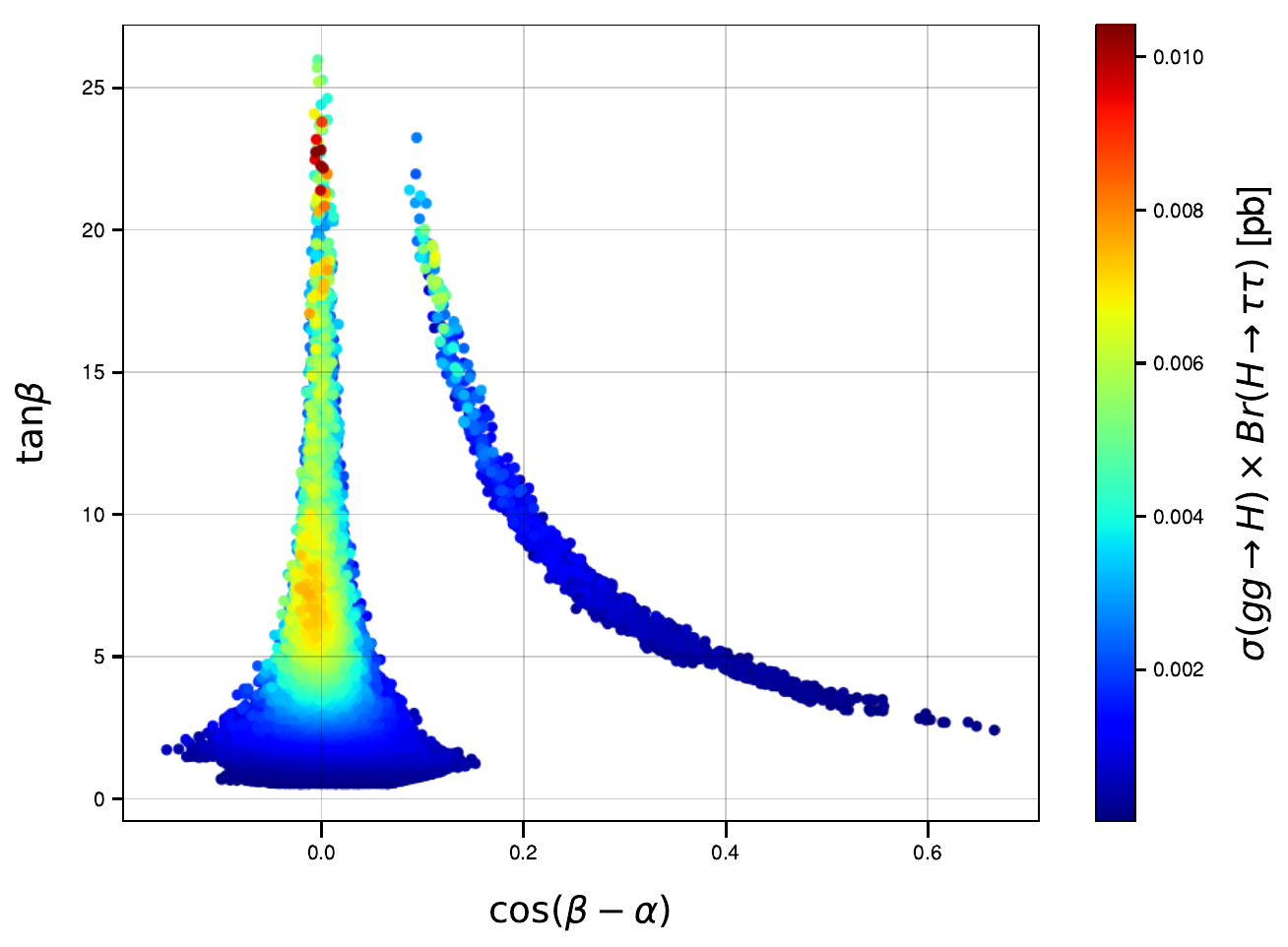}
\includegraphics[width=0.45\textwidth]{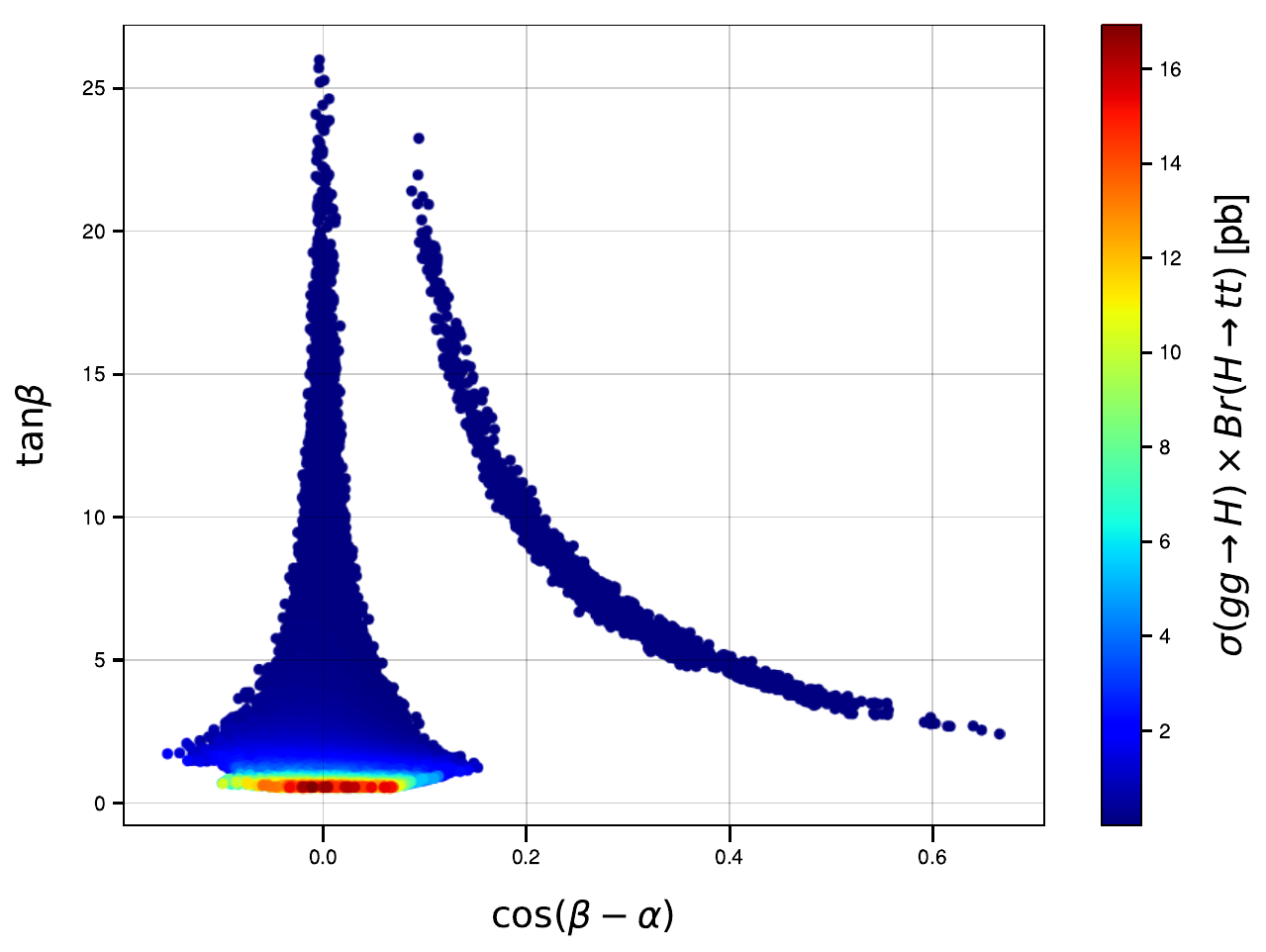}
\caption{Magnitude of the total cross-section times BR in the ($\cos (\alpha - \beta), \tan\beta$) plane for four different processes mediated by the heavy CP-even Higgs, $H$. From top-left to bottom-left (clock-wise): $pp\rightarrow H\rightarrow ZA$, $pp\rightarrow H\rightarrow AA$, $pp\rightarrow H\rightarrow t\bar t$ and $pp\rightarrow H\rightarrow\tau^-\tau^+$. }
\label{fig:sigmaH} 
\end{figure}

In this section, we analyse different possible measurements that can be performed at the LHC with the aim to show their sensitivity to a given set of model parameters within the 2HDM Type-II. We discuss first the relevance of the various channels, which might contain  one or more Higgses as intermediate states, in covering portions of the parameter space via the study of the BRs of the CP-odd $A$ and the CP-even (heavy) $H$. Some of these portions show a partial overlap, some others are disjoint, as displayed in Fig.~\ref{fig:BrAandH}. There, by looking at the top row, one can clearly see that the $A\rightarrow t\bar t$ (red) and $A\rightarrow ZH$ (light blue) channels are quite complementary. The first one is sensitive to low $\tan\beta$ values (see top-left plot) and can cover a broad range of the mass spectrum where the $A$ and $H$ masses do not differ more than 200 GeV from each other and no hierarchy between them is made explicit (see top-right plot). On the contrary, the latter becomes relevant for low to medium
$\tan\beta$ values and when an explicit hierarchy is in place. The $A$ decay into down type particles, $b$-quarks or $\tau$-leptons, is enhanced at medium-to-high values of $\tan\beta$, as displayed by the green and yellow points in the top-left plot. Finally, the $A\rightarrow Zh$ mode is particularly sensitive to the large $\cos (\beta -\alpha)$ region and low to medium $\tan\beta$ values. If we instead look at the $H$ decay modes (see bottom row), we see that they are dominated by the decays into $b\bar b$, $\tau^+\tau^-$ at high $\tan\beta$ and $t\bar t$, $ZA$ at  low $\tan\beta$. These decays are concentrated in the alignment region. This means that the processes mediated by the heavy $H$ scalar are not sensitive to the region of  large $\cos (\beta - \alpha)$. For probing or excluding this portion of the parameter space, that is, the wrong-sign scenario, one needs to rely on  processes mediated by the $A$ state, in particular $A\rightarrow Zh$. 

\noindent 
The decay modes give of course only a partial picture of the sensitivity of the experimental searches to the free parameters of the theory. One should consider the total rate, that is, production cross-section times  BR(s), in order to have a complete view. This is displayed in Fig.~\ref{fig:sigmaA}, where we plot the gluon-gluon induced cross-section for the CP-odd Higgs in the bi-dimensional ($\cos (\beta - \alpha),  \tan\beta$) plane, and in Fig.~\ref{fig:sigmaH}, where we display the same observable for the heavy CP-even Higgs mediated processes. The magnitude of the total cross-section is given following the colour code on the right columns. For the $A$ mediated processes, the cross-section can range from the order of 30 pb, corresponding to $gg\rightarrow A\rightarrow t\bar t$, to the order of a few fb, corresponding to the $\tau^+\tau^-$ channel. Analogous results hold for the $H$ mediated processes.

\section{Summary}
In this paper, we have tensioned the 2HDM Type-II against data stemming from a variety of experimental contexts. We have included a wide range of results spanning from the old high precision LEP and SLC data, encoded into the so-called EWPOs, to the latest measurements performed at the LHC. 

\noindent
Compared to the existing literature, in this paper we have applied for the first time a new method that can improve the commonly used procedure for extracting bounds on the 2HDM parameter space. Our new numerical framework, called \Magellan, and statistical techniques can be applied to any BSM scenarios. Here, we have taken as testing ground the 2HDM for two main reasons. Firstly, the 2HDM group representation of the scalar sector is representative of a large variety of BSM theories, where it is found to emerge in a natural way. Secondly, the 2HDM is characterised by a far from trivial multi-dimensional parameter space where the effectiveness of the new methods can be robustly proved. 

\noindent
\Magellan is based on a Markov Chain Monte Carlo technique exploiting the Metropolis-Hastings algorithm (via \TTPS), which features the following key elements: use of parallel processing when doing parameter scans, efficient data storage with fast I/O and interactive visualisation. This allows the user to explore any model in a complete and efficient way. The novelty of the proposed method is that the parameter space of any BSM theory can still be projected onto any bi-dimensional plane but one can map any portion of this sub-space into any other bi-dimensional plane thus having a full control of the whole parameter space at once. The toolbox \Magellan leverages the use of the \DataFrame class of \pandas, making a custom selection on the set of points relatively easy. The outcome is that excluded (or allowed) points by a given theoretical constraints or experimental results corresponding to a given observable, typically the 95\% C.L. exclusion bound on the cross-section times BR, can be therefore projected onto any 2D sections of the full parameter space, thus allowing the simultaneous extraction of limits on all the different parameters of the theory. A further scope of \Magellan is that it can quickly predict the regions of the parameter space that can be accessible in a given search with the actual luminosity at hand, and show therein the characteristics of the new particles to be searched for (e.g. mass, width, branching ratios, etc.) thus allowing to improve the data analysis. This way of interpreting the model-independent experimental data within a given BSM theory is much more flexible and complete than the procedures adopted in the literature until now. \Magellan is not published. However, its website interactive dashboards can be accessed via a public link. Through this website, the user can explore the full parameter space and exploit the phenomenological features of the model with ease. 

\noindent
In this paper, we have demonstrated some of its capabilities in relation to the mapping of the present and future LHC sensitivity to the aforementioned dynamics of the 2HDM Type-II. 
In this case, \Magellan has been linked to the external packages enabling one to test the 2HDM against experimental data, i.e. \HiggsBounds and \HiggsSignals, as well as to those enabling the prediction of the Higgs production and decay observables such as \SusHi and
\THDMC. This was done to assess whether the enlarged Higgs sector embedded in the 2HDM construct has survived experimental scrutiny to date and can thus be taken as a solid theoretical framework in which searches for new Higgs signals can be pursued at the LHC in the near future. In particular,  we have shown that two distinct configurations of the parameter space of the 2HDM Type-II are currently compliant with all such data and also satisfy internal consistency requirements of the model, namely, the so-called `wrong-sign' scenario (up to 1 TeV scale) and the `alignment' limit. Both of these can be probed during the upcoming runs of the LHC. The dynamics enabling one doing so are the production channels $pp\to A$ and $pp\to H$, i.e., those yielding, respectively, the heavy CP-even and CP-odd Higgs states belonging to the 2HDM Type-II spectrum. These extra Higgs bosons can in turn decay into a variety of modes, including chain decays of one Higgs boson into another, e.g. $A\to Zh$ and $H\to ZA$. These processes contain all the neutral Higgs bosons of such a BSM scenario ($h$ represents the discovered SM-like Higgs state). The sensitivity of future LHC stages to all such production and decay modes was studied and it was argued that a combination of these could potentially pave the way to the detection of all such neutral states of the 2HDM Type-II. In particular, the discovery of a low-mid mass CP-odd Higgs boson, $m_A\le$ 400 GeV, could exclude the alignment limit of the 2HDM Type-II.

We have therefore equipped ourselves and readers with a new powerful and flexible framework, capable to test the hypothesis of an enlarged Higgs sector existing in Nature, as the \Magellan voyage undertaken here can easily be repeated within any other BSM theory.

\section*{Acknowledgements}
\noindent
The authors are immensely grateful to Alex Owen at QMUL for having restored the lost website and dashboard of the Magellan toolkit. We would like to thank also Ciara Byers for contributing to maintain Magellan and to revalidate the plots in the paper with the most recent data entered in HiggsBounds and HiggsSignals.
The authors acknowledge the use of the IRIDIS High Performance Computing Facility and associated support services at the University of Southampton. EA, DE and SM are supported in part through the NExT Institute. EA and SM also acknowledge support from the STFC Consolidated grant ST/L000296/1. 
\bibliographystyle{apsrev4-1}
\bibliography{./2HDM,./flavour,./tools,./higgs,./general,./stat,./EWPO}

\begin{thebibliography}{72}%
\makeatletter
\providecommand \@ifxundefined [1]{%
 \@ifx{#1\undefined}
}%
\providecommand \@ifnum [1]{%
 \ifnum #1\expandafter \@firstoftwo
 \else \expandafter \@secondoftwo
 \fi
}%
\providecommand \@ifx [1]{%
 \ifx #1\expandafter \@firstoftwo
 \else \expandafter \@secondoftwo
 \fi
}%
\providecommand \natexlab [1]{#1}%
\providecommand \enquote  [1]{``#1''}%
\providecommand \bibnamefont  [1]{#1}%
\providecommand \bibfnamefont [1]{#1}%
\providecommand \citenamefont [1]{#1}%
\providecommand \href@noop [0]{\@secondoftwo}%
\providecommand \href [0]{\begingroup \@sanitize@url \@href}%
\providecommand \@href[1]{\@@startlink{#1}\@@href}%
\providecommand \@@href[1]{\endgroup#1\@@endlink}%
\providecommand \@sanitize@url [0]{\catcode `\\12\catcode `\$12\catcode
  `\&12\catcode `\#12\catcode `\^12\catcode `\_12\catcode `\%12\relax}%
\providecommand \@@startlink[1]{}%
\providecommand \@@endlink[0]{}%
\providecommand \url  [0]{\begingroup\@sanitize@url \@url }%
\providecommand \@url [1]{\endgroup\@href {#1}{\urlprefix }}%
\providecommand \urlprefix  [0]{URL }%
\providecommand \Eprint [0]{\href }%
\providecommand \doibase [0]{http://dx.doi.org/}%
\providecommand \selectlanguage [0]{\@gobble}%
\providecommand \bibinfo  [0]{\@secondoftwo}%
\providecommand \bibfield  [0]{\@secondoftwo}%
\providecommand \translation [1]{[#1]}%
\providecommand \BibitemOpen [0]{}%
\providecommand \bibitemStop [0]{}%
\providecommand \bibitemNoStop [0]{.\EOS\space}%
\providecommand \EOS [0]{\spacefactor3000\relax}%
\providecommand \BibitemShut  [1]{\csname bibitem#1\endcsname}%
\let\auto@bib@innerbib\@empty
\bibitem [{\citenamefont {{ATLAS Collaboration}}(2012)}]{ATLAS-Higgs}%
  \BibitemOpen
  \bibfield  {author} {\bibinfo {author} {\bibnamefont {{ATLAS
  Collaboration}}},\ }\bibfield  {title} {\emph {\enquote {\bibinfo {title}
  {{Observation of a new particle in the search for the Standard Model Higgs
  boson with the ATLAS detector at the LHC}},}\ }}\href {\doibase
  http://dx.doi.org/10.1016/j.physletb.2012.08.020} {\bibfield  {journal}
  {\bibinfo  {journal} {Physics Letters B}\ }\textbf {\bibinfo {volume}
  {716}},\ \bibinfo {pages} {1 } (\bibinfo {year} {2012})},\ \bibinfo {note}
  {{Observation of a new particle in the search for the Standard Model Higgs
  boson with the ATLAS detector at the LHC}}\BibitemShut {NoStop}%
\bibitem [{\citenamefont {Chatrchyan}\ \emph {et~al.}(2012)\citenamefont
  {Chatrchyan} \emph {et~al.}}]{CMS-Higgs}%
  \BibitemOpen
  \bibfield  {author} {\bibinfo {author} {\bibfnamefont {S.}~\bibnamefont
  {Chatrchyan}} \emph {et~al.},\ }\bibfield  {title} {\emph {\enquote {\bibinfo
  {title} {Observation of a new boson at a mass of 125 gev with the cms
  experiment at the lhc},}\ }}\href {\doibase
  http://dx.doi.org/10.1016/j.physletb.2012.08.021} {\bibfield  {journal}
  {\bibinfo  {journal} {Physics Letters B}\ }\textbf {\bibinfo {volume}
  {716}},\ \bibinfo {pages} {30 } (\bibinfo {year} {2012})},\ \bibinfo {note}
  {{Observation of a new boson at a mass of 125 GeV with the CMS experiment at
  the LHC}}\BibitemShut {NoStop}%
\bibitem [{\citenamefont {Celis}\ \emph {et~al.}(2014)\citenamefont {Celis},
  \citenamefont {Fuentes-Martín},\ and\ \citenamefont
  {Serôdio}}]{Celis2014185}%
  \BibitemOpen
  \bibfield  {author} {\bibinfo {author} {\bibfnamefont {A.}~\bibnamefont
  {Celis}}, \bibinfo {author} {\bibfnamefont {J.}~\bibnamefont
  {Fuentes-Martín}}, \ and\ \bibinfo {author} {\bibfnamefont {H.}~\bibnamefont
  {Serôdio}},\ }\bibfield  {title} {\emph {\enquote {\bibinfo {title}
  {{Effective aligned 2HDM with a DFSZ-like invisible axion}},}\ }}\href
  {\doibase http://dx.doi.org/10.1016/j.physletb.2014.08.032} {\bibfield
  {journal} {\bibinfo  {journal} {{Physics Letters B}}\ }\textbf {\bibinfo
  {volume} {{737}}},\ \bibinfo {pages} {185 } (\bibinfo {year} {2014})},\
  \bibinfo {note} {effective aligned 2HDM with a DFSZ-like invisible
  axion}\BibitemShut {NoStop}%
\bibitem [{\citenamefont {Kim}(1987)}]{Axion1987}%
  \BibitemOpen
  \bibfield  {author} {\bibinfo {author} {\bibfnamefont {J.~E.}\ \bibnamefont
  {Kim}},\ }\bibfield  {title} {\emph {\enquote {\bibinfo {title} {{Light
  pseudoscalars, particle physics and cosmology}},}\ }}\href {\doibase
  https://doi.org/10.1016/0370-1573(87)90017-2} {\bibfield  {journal} {\bibinfo
   {journal} {{Physics Reports}}\ }\textbf {\bibinfo {volume} {{150}}},\
  \bibinfo {pages} {1} (\bibinfo {year} {1987})}\BibitemShut {NoStop}%
\bibitem [{\citenamefont {Mrazek}\ \emph {et~al.}(2011)\citenamefont {Mrazek},
  \citenamefont {Pomarol}, \citenamefont {Rattazzi}, \citenamefont {Redi},
  \citenamefont {Serra},\ and\ \citenamefont {Wulzer}}]{Mrazek20111}%
  \BibitemOpen
  \bibfield  {author} {\bibinfo {author} {\bibfnamefont {J.}~\bibnamefont
  {Mrazek}}, \bibinfo {author} {\bibfnamefont {A.}~\bibnamefont {Pomarol}},
  \bibinfo {author} {\bibfnamefont {R.}~\bibnamefont {Rattazzi}}, \bibinfo
  {author} {\bibfnamefont {M.}~\bibnamefont {Redi}}, \bibinfo {author}
  {\bibfnamefont {J.}~\bibnamefont {Serra}}, \ and\ \bibinfo {author}
  {\bibfnamefont {A.}~\bibnamefont {Wulzer}},\ }\bibfield  {title} {\emph
  {\enquote {\bibinfo {title} {{The other natural two Higgs doublet model}},}\
  }}\href {\doibase http://dx.doi.org/10.1016/j.nuclphysb.2011.07.008}
  {\bibfield  {journal} {\bibinfo  {journal} {Nuclear Physics B}\ }\textbf
  {\bibinfo {volume} {853}},\ \bibinfo {pages} {1 } (\bibinfo {year}
  {2011})}\BibitemShut {NoStop}%
\bibitem [{\citenamefont {Bertuzzo}\ \emph {et~al.}(2013)\citenamefont
  {Bertuzzo}, \citenamefont {Ray}, \citenamefont {de~Sandes},\ and\
  \citenamefont {Savoy}}]{Bertuzzo2013}%
  \BibitemOpen
  \bibfield  {author} {\bibinfo {author} {\bibfnamefont {E.}~\bibnamefont
  {Bertuzzo}}, \bibinfo {author} {\bibfnamefont {T.~S.}\ \bibnamefont {Ray}},
  \bibinfo {author} {\bibfnamefont {H.}~\bibnamefont {de~Sandes}}, \ and\
  \bibinfo {author} {\bibfnamefont {C.~A.}\ \bibnamefont {Savoy}},\ }\bibfield
  {title} {\emph {\enquote {\bibinfo {title} {{On composite two Higgs doublet
  models}},}\ }}\href {\doibase 10.1007/JHEP05(2013)153} {\bibfield  {journal}
  {\bibinfo  {journal} {Journal of High Energy Physics}\ }\textbf {\bibinfo
  {volume} {2013}},\ \bibinfo {pages} {1} (\bibinfo {year} {2013})}\BibitemShut
  {NoStop}%
\bibitem [{\citenamefont {Agashe}\ \emph {et~al.}(2005)\citenamefont {Agashe},
  \citenamefont {Contino},\ and\ \citenamefont {Pomarol}}]{Agashe2005165}%
  \BibitemOpen
  \bibfield  {author} {\bibinfo {author} {\bibfnamefont {K.}~\bibnamefont
  {Agashe}}, \bibinfo {author} {\bibfnamefont {R.}~\bibnamefont {Contino}}, \
  and\ \bibinfo {author} {\bibfnamefont {A.}~\bibnamefont {Pomarol}},\
  }\bibfield  {title} {\emph {\enquote {\bibinfo {title} {{The minimal
  composite Higgs model}},}\ }}\href {\doibase
  http://dx.doi.org/10.1016/j.nuclphysb.2005.04.035} {\bibfield  {journal}
  {\bibinfo  {journal} {Nuclear Physics B}\ }\textbf {\bibinfo {volume}
  {719}},\ \bibinfo {pages} {165 } (\bibinfo {year} {2005})}\BibitemShut
  {NoStop}%
\bibitem [{\citenamefont {Curtis}\ \emph
  {et~al.}(2018{\natexlab{a}})\citenamefont {Curtis}, \citenamefont {Rose},
  \citenamefont {Moretti},\ and\ \citenamefont {Yagyu}}]{DeCurtis:2018iqd}%
  \BibitemOpen
  \bibfield  {author} {\bibinfo {author} {\bibfnamefont {S.~D.}\ \bibnamefont
  {Curtis}}, \bibinfo {author} {\bibfnamefont {L.~D.}\ \bibnamefont {Rose}},
  \bibinfo {author} {\bibfnamefont {S.}~\bibnamefont {Moretti}}, \ and\
  \bibinfo {author} {\bibfnamefont {K.}~\bibnamefont {Yagyu}},\ }\bibfield
  {title} {\emph {\enquote {\bibinfo {title} {{Supersymmetry versus
  Compositeness: 2HDMs tell the story}},}\ }}\href {\doibase
  10.1016/j.physletb.2018.09.042} {\bibfield  {journal} {\bibinfo  {journal}
  {Physics Letters B}\ }\textbf {\bibinfo {volume} {786}},\ \bibinfo {pages}
  {189} (\bibinfo {year} {2018}{\natexlab{a}})}\BibitemShut {NoStop}%
\bibitem [{\citenamefont {Curtis}\ \emph
  {et~al.}(2018{\natexlab{b}})\citenamefont {Curtis}, \citenamefont {Rose},
  \citenamefont {Moretti},\ and\ \citenamefont {Yagyu}}]{DeCurtis:2018zvh}%
  \BibitemOpen
  \bibfield  {author} {\bibinfo {author} {\bibfnamefont {S.~D.}\ \bibnamefont
  {Curtis}}, \bibinfo {author} {\bibfnamefont {L.~D.}\ \bibnamefont {Rose}},
  \bibinfo {author} {\bibfnamefont {S.}~\bibnamefont {Moretti}}, \ and\
  \bibinfo {author} {\bibfnamefont {K.}~\bibnamefont {Yagyu}},\ }\bibfield
  {title} {\emph {\enquote {\bibinfo {title} {{A concrete composite 2-Higgs
  doublet model}},}\ }}\href {\doibase 10.1007/jhep12(2018)051} {\bibfield
  {journal} {\bibinfo  {journal} {Journal of High Energy Physics}\ }\textbf
  {\bibinfo {volume} {2018}} (\bibinfo {year} {2018}{\natexlab{b}}),\
  10.1007/jhep12(2018)051}\BibitemShut {NoStop}%
\bibitem [{\citenamefont {Aoki}\ \emph {et~al.}(2009)\citenamefont {Aoki},
  \citenamefont {Kanemura},\ and\ \citenamefont {Seto}}]{Neutrino-mass}%
  \BibitemOpen
  \bibfield  {author} {\bibinfo {author} {\bibfnamefont {M.}~\bibnamefont
  {Aoki}}, \bibinfo {author} {\bibfnamefont {S.}~\bibnamefont {Kanemura}}, \
  and\ \bibinfo {author} {\bibfnamefont {O.}~\bibnamefont {Seto}},\ }\bibfield
  {title} {\emph {\enquote {\bibinfo {title} {{Neutrino Mass, Dark Matter, and
  Baryon Asymmetry via TeV-Scale Physics without Fine-Tuning}},}\ }}\href
  {\doibase 10.1103/PhysRevLett.102.051805} {\bibfield  {journal} {\bibinfo
  {journal} {Phys. Rev. Lett.}\ }\textbf {\bibinfo {volume} {102}},\ \bibinfo
  {pages} {051805} (\bibinfo {year} {2009})}\BibitemShut {NoStop}%
\bibitem [{\citenamefont {Ko}\ \emph {et~al.}(2014)\citenamefont {Ko},
  \citenamefont {Omura},\ and\ \citenamefont {Yu}}]{Ko2014}%
  \BibitemOpen
  \bibfield  {author} {\bibinfo {author} {\bibfnamefont {P.}~\bibnamefont
  {Ko}}, \bibinfo {author} {\bibfnamefont {Y.}~\bibnamefont {Omura}}, \ and\
  \bibinfo {author} {\bibfnamefont {C.}~\bibnamefont {Yu}},\ }\bibfield
  {title} {\emph {\enquote {\bibinfo {title} {{Dark matter and dark force in
  the type-I inert 2HDM with local U(1)$_{H}$ gauge symmetry}},}\ }}\href
  {\doibase 10.1007/JHEP11(2014)054} {\bibfield  {journal} {\bibinfo  {journal}
  {Journal of High Energy Physics}\ }\textbf {\bibinfo {volume} {2014}},\
  \bibinfo {pages} {1} (\bibinfo {year} {2014})}\BibitemShut {NoStop}%
\bibitem [{\citenamefont {Cao}\ \emph {et~al.}(2009)\citenamefont {Cao},
  \citenamefont {Wan}, \citenamefont {Wu},\ and\ \citenamefont
  {Yang}}]{muong2_lepton}%
  \BibitemOpen
  \bibfield  {author} {\bibinfo {author} {\bibfnamefont {J.}~\bibnamefont
  {Cao}}, \bibinfo {author} {\bibfnamefont {P.}~\bibnamefont {Wan}}, \bibinfo
  {author} {\bibfnamefont {L.}~\bibnamefont {Wu}}, \ and\ \bibinfo {author}
  {\bibfnamefont {J.~M.}\ \bibnamefont {Yang}},\ }\bibfield  {title} {\emph
  {\enquote {\bibinfo {title} {{Lepton-specific two-Higgs-doublet model:
  Experimental constraints and implication on Higgs phenomenology}},}\ }}\href
  {\doibase 10.1103/PhysRevD.80.071701} {\bibfield  {journal} {\bibinfo
  {journal} {Phys. Rev. D}\ }\textbf {\bibinfo {volume} {80}},\ \bibinfo
  {pages} {071701} (\bibinfo {year} {2009})}\BibitemShut {NoStop}%
\bibitem [{\citenamefont {Broggio}\ \emph {et~al.}(2014)\citenamefont
  {Broggio}, \citenamefont {Chun}, \citenamefont {Passera}, \citenamefont
  {Patel},\ and\ \citenamefont {Vempati}}]{muong2_limiting}%
  \BibitemOpen
  \bibfield  {author} {\bibinfo {author} {\bibfnamefont {A.}~\bibnamefont
  {Broggio}}, \bibinfo {author} {\bibfnamefont {E.~J.}\ \bibnamefont {Chun}},
  \bibinfo {author} {\bibfnamefont {M.}~\bibnamefont {Passera}}, \bibinfo
  {author} {\bibfnamefont {K.~M.}\ \bibnamefont {Patel}}, \ and\ \bibinfo
  {author} {\bibfnamefont {S.~K.}\ \bibnamefont {Vempati}},\ }\bibfield
  {title} {\emph {\enquote {\bibinfo {title} {Limiting two-higgs-doublet
  models},}\ }}\href {\doibase 10.1007/JHEP11(2014)058} {\bibfield  {journal}
  {\bibinfo  {journal} {Journal of High Energy Physics}\ }\textbf {\bibinfo
  {volume} {2014}},\ \bibinfo {pages} {58} (\bibinfo {year}
  {2014})}\BibitemShut {NoStop}%
\bibitem [{\citenamefont {Wang}\ and\ \citenamefont {Han}(2015)}]{Wang2015}%
  \BibitemOpen
  \bibfield  {author} {\bibinfo {author} {\bibfnamefont {L.}~\bibnamefont
  {Wang}}\ and\ \bibinfo {author} {\bibfnamefont {X.-F.}\ \bibnamefont {Han}},\
  }\bibfield  {title} {\emph {\enquote {\bibinfo {title} {A light pseudoscalar
  of 2hdm confronted with muon g-2 and experimental constraints},}\ }}\href
  {\doibase 10.1007/JHEP05(2015)039} {\bibfield  {journal} {\bibinfo  {journal}
  {Journal of High Energy Physics}\ }\textbf {\bibinfo {volume} {2015}},\
  \bibinfo {pages} {39} (\bibinfo {year} {2015})}\BibitemShut {NoStop}%
\bibitem [{Hig(1990)}]{HiggsHunters}%
  \BibitemOpen
  \bibfield  {title} {\emph {\enquote {\bibinfo {title} {{The Higgs Hunter's
  Guide}},}\ }}\href@noop {} {\bibfield  {journal} {\bibinfo  {journal}
  {{Front. Phys.}}\ }\textbf {\bibinfo {volume} {{80}}},\ \bibinfo {pages} {1}
  (\bibinfo {year} {1990})},\ \bibinfo {note} {the Higgs Hunter's
  Guide}\BibitemShut {NoStop}%
\bibitem [{\citenamefont {Djouadi}(2008)}]{DjouadiMSSM}%
  \BibitemOpen
  \bibfield  {author} {\bibinfo {author} {\bibfnamefont {A.}~\bibnamefont
  {Djouadi}},\ }\bibfield  {title} {\emph {\enquote {\bibinfo {title} {{The
  anatomy of electroweak symmetry breaking Tome II: The Higgs bosons in the
  Minimal Supersymmetric Model}},}\ }}\href {\doibase
  http://dx.doi.org/10.1016/j.physrep.2007.10.005} {\bibfield  {journal}
  {\bibinfo  {journal} {{Physics Reports}}\ }\textbf {\bibinfo {volume}
  {{459}}},\ \bibinfo {pages} {1 } (\bibinfo {year} {2008})},\ \bibinfo {note}
  {the anatomy of electroweak symmetry breaking Tome II: The Higgs bosons in
  the Minimal Supersymmetric Model}\BibitemShut {NoStop}%
\bibitem [{\citenamefont {Branco}\ \emph {et~al.}(2012)\citenamefont {Branco},
  \citenamefont {Ferreira}, \citenamefont {Lavoura}, \citenamefont {Rebelo},
  \citenamefont {Sher},\ and\ \citenamefont {Silva}}]{Branco2012}%
  \BibitemOpen
  \bibfield  {author} {\bibinfo {author} {\bibfnamefont {G.}~\bibnamefont
  {Branco}}, \bibinfo {author} {\bibfnamefont {P.}~\bibnamefont {Ferreira}},
  \bibinfo {author} {\bibfnamefont {L.}~\bibnamefont {Lavoura}}, \bibinfo
  {author} {\bibfnamefont {M.}~\bibnamefont {Rebelo}}, \bibinfo {author}
  {\bibfnamefont {M.}~\bibnamefont {Sher}}, \ and\ \bibinfo {author}
  {\bibfnamefont {J.~P.}\ \bibnamefont {Silva}},\ }\bibfield  {title} {\emph
  {\enquote {\bibinfo {title} {{Theory and phenomenology of two-Higgs-doublet
  models}},}\ }}\href {\doibase
  http://dx.doi.org/10.1016/j.physrep.2012.02.002} {\bibfield  {journal}
  {\bibinfo  {journal} {{Physics Reports}}\ }\textbf {\bibinfo {volume}
  {{516}}},\ \bibinfo {pages} {1 } (\bibinfo {year} {2012})},\ \bibinfo {note}
  {theory and phenomenology of two-Higgs-doublet models}\BibitemShut {NoStop}%
\bibitem [{\citenamefont {Baak}\ \emph {et~al.}(2012)\citenamefont {Baak},
  \citenamefont {Goebel}, \citenamefont {Haller}, \citenamefont {Hoecker},
  \citenamefont {Kennedy}, \citenamefont {Mönig}, \citenamefont {Schott},\
  and\ \citenamefont {Stelzer}}]{Baak_2012}%
  \BibitemOpen
  \bibfield  {author} {\bibinfo {author} {\bibfnamefont {M.}~\bibnamefont
  {Baak}}, \bibinfo {author} {\bibfnamefont {M.}~\bibnamefont {Goebel}},
  \bibinfo {author} {\bibfnamefont {J.}~\bibnamefont {Haller}}, \bibinfo
  {author} {\bibfnamefont {A.}~\bibnamefont {Hoecker}}, \bibinfo {author}
  {\bibfnamefont {D.}~\bibnamefont {Kennedy}}, \bibinfo {author} {\bibfnamefont
  {K.}~\bibnamefont {Mönig}}, \bibinfo {author} {\bibfnamefont
  {M.}~\bibnamefont {Schott}}, \ and\ \bibinfo {author} {\bibfnamefont
  {J.}~\bibnamefont {Stelzer}},\ }\bibfield  {title} {\emph {\enquote {\bibinfo
  {title} {Updated status of the global electroweak fit and constraints on new
  physics},}\ }}\href {\doibase 10.1140/epjc/s10052-012-2003-4} {\bibfield
  {journal} {\bibinfo  {journal} {The European Physical Journal C}\ }\textbf
  {\bibinfo {volume} {72}} (\bibinfo {year} {2012}),\
  10.1140/epjc/s10052-012-2003-4}\BibitemShut {NoStop}%
\bibitem [{\citenamefont {Lafaye}\ \emph {et~al.}(2004)\citenamefont {Lafaye},
  \citenamefont {Plehn},\ and\ \citenamefont {Zerwas}}]{Lafaye:2004cn}%
  \BibitemOpen
  \bibfield  {author} {\bibinfo {author} {\bibfnamefont {R.}~\bibnamefont
  {Lafaye}}, \bibinfo {author} {\bibfnamefont {T.}~\bibnamefont {Plehn}}, \
  and\ \bibinfo {author} {\bibfnamefont {D.}~\bibnamefont {Zerwas}},\
  }\bibfield  {title} {\emph {\enquote {\bibinfo {title} {{SFITTER: SUSY
  parameter analysis at LHC and LC}},}\ }}\href@noop {} {\  (\bibinfo {year}
  {2004})},\ \Eprint {http://arxiv.org/abs/hep-ph/0404282}
  {arXiv:hep-ph/0404282} \BibitemShut {NoStop}%
\bibitem [{\citenamefont {Austri}\ \emph {et~al.}(2006)\citenamefont {Austri},
  \citenamefont {Trotta},\ and\ \citenamefont {Roszkowski}}]{Austri_2006}%
  \BibitemOpen
  \bibfield  {author} {\bibinfo {author} {\bibfnamefont {R.~R.~d.}\
  \bibnamefont {Austri}}, \bibinfo {author} {\bibfnamefont {R.}~\bibnamefont
  {Trotta}}, \ and\ \bibinfo {author} {\bibfnamefont {L.}~\bibnamefont
  {Roszkowski}},\ }\bibfield  {title} {\emph {\enquote {\bibinfo {title} {A
  markov chain monte carlo analysis of the cmssm},}\ }}\href {\doibase
  10.1088/1126-6708/2006/05/002} {\bibfield  {journal} {\bibinfo  {journal}
  {Journal of High Energy Physics}\ }\textbf {\bibinfo {volume} {2006}},\
  \bibinfo {pages} {002–002} (\bibinfo {year} {2006})}\BibitemShut {NoStop}%
\bibitem [{\citenamefont {Strege}\ \emph {et~al.}(2012)\citenamefont {Strege},
  \citenamefont {Bertone}, \citenamefont {Cerdeño}, \citenamefont {Fornasa},
  \citenamefont {Austri},\ and\ \citenamefont {Trotta}}]{Strege_2012}%
  \BibitemOpen
  \bibfield  {author} {\bibinfo {author} {\bibfnamefont {C.}~\bibnamefont
  {Strege}}, \bibinfo {author} {\bibfnamefont {G.}~\bibnamefont {Bertone}},
  \bibinfo {author} {\bibfnamefont {D.}~\bibnamefont {Cerdeño}}, \bibinfo
  {author} {\bibfnamefont {M.}~\bibnamefont {Fornasa}}, \bibinfo {author}
  {\bibfnamefont {R.~R.~d.}\ \bibnamefont {Austri}}, \ and\ \bibinfo {author}
  {\bibfnamefont {R.}~\bibnamefont {Trotta}},\ }\bibfield  {title} {\emph
  {\enquote {\bibinfo {title} {Updated global fits of the cmssm including the
  latest lhc susy and higgs searches and xenon100 data},}\ }}\href {\doibase
  10.1088/1475-7516/2012/03/030} {\bibfield  {journal} {\bibinfo  {journal}
  {Journal of Cosmology and Astroparticle Physics}\ }\textbf {\bibinfo {volume}
  {2012}},\ \bibinfo {pages} {030–030} (\bibinfo {year} {2012})}\BibitemShut
  {NoStop}%
\bibitem [{\citenamefont {Strege}\ \emph {et~al.}(2013)\citenamefont {Strege},
  \citenamefont {Bertone}, \citenamefont {Feroz}, \citenamefont {Fornasa},
  \citenamefont {Austri},\ and\ \citenamefont {Trotta}}]{Strege_2013}%
  \BibitemOpen
  \bibfield  {author} {\bibinfo {author} {\bibfnamefont {C.}~\bibnamefont
  {Strege}}, \bibinfo {author} {\bibfnamefont {G.}~\bibnamefont {Bertone}},
  \bibinfo {author} {\bibfnamefont {F.}~\bibnamefont {Feroz}}, \bibinfo
  {author} {\bibfnamefont {M.}~\bibnamefont {Fornasa}}, \bibinfo {author}
  {\bibfnamefont {R.~R.~d.}\ \bibnamefont {Austri}}, \ and\ \bibinfo {author}
  {\bibfnamefont {R.}~\bibnamefont {Trotta}},\ }\bibfield  {title} {\emph
  {\enquote {\bibinfo {title} {Global fits of the cmssm and nuhm including the
  lhc higgs discovery and new xenon100 constraints},}\ }}\href {\doibase
  10.1088/1475-7516/2013/04/013} {\bibfield  {journal} {\bibinfo  {journal}
  {Journal of Cosmology and Astroparticle Physics}\ }\textbf {\bibinfo {volume}
  {2013}},\ \bibinfo {pages} {013–013} (\bibinfo {year} {2013})}\BibitemShut
  {NoStop}%
\bibitem [{\citenamefont {Bechtle}\ \emph {et~al.}(2006)\citenamefont
  {Bechtle}, \citenamefont {Desch},\ and\ \citenamefont
  {Wienemann}}]{Bechtle_2006}%
  \BibitemOpen
  \bibfield  {author} {\bibinfo {author} {\bibfnamefont {P.}~\bibnamefont
  {Bechtle}}, \bibinfo {author} {\bibfnamefont {K.}~\bibnamefont {Desch}}, \
  and\ \bibinfo {author} {\bibfnamefont {P.}~\bibnamefont {Wienemann}},\
  }\bibfield  {title} {\emph {\enquote {\bibinfo {title} {Fittino, a program
  for determining mssm parameters from collider observables using an iterative
  method},}\ }}\href {\doibase 10.1016/j.cpc.2005.09.002} {\bibfield  {journal}
  {\bibinfo  {journal} {Computer Physics Communications}\ }\textbf {\bibinfo
  {volume} {174}},\ \bibinfo {pages} {47–70} (\bibinfo {year}
  {2006})}\BibitemShut {NoStop}%
\bibitem [{\citenamefont {Bechtle}\ \emph {et~al.}(2010)\citenamefont
  {Bechtle}, \citenamefont {Desch}, \citenamefont {Uhlenbrock},\ and\
  \citenamefont {Wienemann}}]{Bechtle_2010}%
  \BibitemOpen
  \bibfield  {author} {\bibinfo {author} {\bibfnamefont {P.}~\bibnamefont
  {Bechtle}}, \bibinfo {author} {\bibfnamefont {K.}~\bibnamefont {Desch}},
  \bibinfo {author} {\bibfnamefont {M.}~\bibnamefont {Uhlenbrock}}, \ and\
  \bibinfo {author} {\bibfnamefont {P.}~\bibnamefont {Wienemann}},\ }\bibfield
  {title} {\emph {\enquote {\bibinfo {title} {Constraining susy models with
  fittino using measurements before, with and beyond the lhc},}\ }}\href
  {\doibase 10.1140/epjc/s10052-009-1228-3} {\bibfield  {journal} {\bibinfo
  {journal} {The European Physical Journal C}\ }\textbf {\bibinfo {volume}
  {66}},\ \bibinfo {pages} {215–259} (\bibinfo {year} {2010})}\BibitemShut
  {NoStop}%
\bibitem [{\citenamefont {Bechtle}\ \emph {et~al.}(2012)\citenamefont
  {Bechtle}, \citenamefont {Bringmann}, \citenamefont {Desch}, \citenamefont
  {Dreiner}, \citenamefont {Hamer}, \citenamefont {Hensel}, \citenamefont
  {Krämer}, \citenamefont {Nguyen}, \citenamefont {Porod}, \citenamefont
  {Prudent},\ and\ \citenamefont {et~al.}}]{Bechtle_2012}%
  \BibitemOpen
  \bibfield  {author} {\bibinfo {author} {\bibfnamefont {P.}~\bibnamefont
  {Bechtle}}, \bibinfo {author} {\bibfnamefont {T.}~\bibnamefont {Bringmann}},
  \bibinfo {author} {\bibfnamefont {K.}~\bibnamefont {Desch}}, \bibinfo
  {author} {\bibfnamefont {H.}~\bibnamefont {Dreiner}}, \bibinfo {author}
  {\bibfnamefont {M.}~\bibnamefont {Hamer}}, \bibinfo {author} {\bibfnamefont
  {C.}~\bibnamefont {Hensel}}, \bibinfo {author} {\bibfnamefont
  {M.}~\bibnamefont {Krämer}}, \bibinfo {author} {\bibfnamefont
  {N.}~\bibnamefont {Nguyen}}, \bibinfo {author} {\bibfnamefont
  {W.}~\bibnamefont {Porod}}, \bibinfo {author} {\bibfnamefont
  {X.}~\bibnamefont {Prudent}}, \ and\ \bibinfo {author} {\bibnamefont
  {et~al.}},\ }\bibfield  {title} {\emph {\enquote {\bibinfo {title}
  {Constrained supersymmetry after two years of lhc data: a global view with
  fittino},}\ }}\href {\doibase 10.1007/jhep06(2012)098} {\bibfield  {journal}
  {\bibinfo  {journal} {Journal of High Energy Physics}\ }\textbf {\bibinfo
  {volume} {2012}} (\bibinfo {year} {2012}),\
  10.1007/jhep06(2012)098}\BibitemShut {NoStop}%
\bibitem [{\citenamefont {Kraml}\ \emph {et~al.}(2019)\citenamefont {Kraml},
  \citenamefont {Quang~Loc}, \citenamefont {Nhung},\ and\ \citenamefont
  {Ninh}}]{Kraml_2019}%
  \BibitemOpen
  \bibfield  {author} {\bibinfo {author} {\bibfnamefont {S.}~\bibnamefont
  {Kraml}}, \bibinfo {author} {\bibfnamefont {T.}~\bibnamefont {Quang~Loc}},
  \bibinfo {author} {\bibfnamefont {D.~T.}\ \bibnamefont {Nhung}}, \ and\
  \bibinfo {author} {\bibfnamefont {L.~D.}\ \bibnamefont {Ninh}},\ }\bibfield
  {title} {\emph {\enquote {\bibinfo {title} {Constraining new physics from
  higgs measurements with lilith: update to lhc run 2 results},}\ }}\href
  {\doibase 10.21468/scipostphys.7.4.052} {\bibfield  {journal} {\bibinfo
  {journal} {SciPost Physics}\ }\textbf {\bibinfo {volume} {7}} (\bibinfo
  {year} {2019}),\ 10.21468/scipostphys.7.4.052}\BibitemShut {NoStop}%
\bibitem [{\citenamefont {Bernon}\ and\ \citenamefont
  {Dumont}(2015)}]{Bernon_2015}%
  \BibitemOpen
  \bibfield  {author} {\bibinfo {author} {\bibfnamefont {J.}~\bibnamefont
  {Bernon}}\ and\ \bibinfo {author} {\bibfnamefont {B.}~\bibnamefont
  {Dumont}},\ }\bibfield  {title} {\emph {\enquote {\bibinfo {title} {Lilith: a
  tool for constraining new physics from higgs measurements},}\ }}\href
  {\doibase 10.1140/epjc/s10052-015-3645-9} {\bibfield  {journal} {\bibinfo
  {journal} {The European Physical Journal C}\ }\textbf {\bibinfo {volume}
  {75}} (\bibinfo {year} {2015}),\ 10.1140/epjc/s10052-015-3645-9}\BibitemShut
  {NoStop}%
\bibitem [{\citenamefont {Buchmueller}\ \emph {et~al.}(2009)\citenamefont
  {Buchmueller}, \citenamefont {Cavanaugh}, \citenamefont {De~Roeck},
  \citenamefont {Ellis}, \citenamefont {Flaecher}, \citenamefont {Heinemeyer},
  \citenamefont {Isidori}, \citenamefont {Olive}, \citenamefont {Ronga},\ and\
  \citenamefont {Weiglein}}]{Buchmueller_2009}%
  \BibitemOpen
  \bibfield  {author} {\bibinfo {author} {\bibfnamefont {O.}~\bibnamefont
  {Buchmueller}}, \bibinfo {author} {\bibfnamefont {R.}~\bibnamefont
  {Cavanaugh}}, \bibinfo {author} {\bibfnamefont {A.}~\bibnamefont {De~Roeck}},
  \bibinfo {author} {\bibfnamefont {J.~R.}\ \bibnamefont {Ellis}}, \bibinfo
  {author} {\bibfnamefont {H.}~\bibnamefont {Flaecher}}, \bibinfo {author}
  {\bibfnamefont {S.}~\bibnamefont {Heinemeyer}}, \bibinfo {author}
  {\bibfnamefont {G.}~\bibnamefont {Isidori}}, \bibinfo {author} {\bibfnamefont
  {K.~A.}\ \bibnamefont {Olive}}, \bibinfo {author} {\bibfnamefont {F.~J.}\
  \bibnamefont {Ronga}}, \ and\ \bibinfo {author} {\bibfnamefont
  {G.}~\bibnamefont {Weiglein}},\ }\bibfield  {title} {\emph {\enquote
  {\bibinfo {title} {Likelihood functions for supersymmetric observables in
  frequentist analyses of the cmssm and nuhm1},}\ }}\href {\doibase
  10.1140/epjc/s10052-009-1159-z} {\bibfield  {journal} {\bibinfo  {journal}
  {The European Physical Journal C}\ }\textbf {\bibinfo {volume} {64}},\
  \bibinfo {pages} {391–415} (\bibinfo {year} {2009})}\BibitemShut {NoStop}%
\bibitem [{\citenamefont {Buchmueller}\ \emph {et~al.}(2011)\citenamefont
  {Buchmueller}, \citenamefont {Cavanaugh}, \citenamefont {Colling},
  \citenamefont {De~Roeck}, \citenamefont {Dolan}, \citenamefont {Ellis},
  \citenamefont {Flächer}, \citenamefont {Heinemeyer}, \citenamefont
  {Isidori}, \citenamefont {Olive},\ and\ \citenamefont
  {et~al.}}]{Buchmueller_2011}%
  \BibitemOpen
  \bibfield  {author} {\bibinfo {author} {\bibfnamefont {O.}~\bibnamefont
  {Buchmueller}}, \bibinfo {author} {\bibfnamefont {R.}~\bibnamefont
  {Cavanaugh}}, \bibinfo {author} {\bibfnamefont {D.}~\bibnamefont {Colling}},
  \bibinfo {author} {\bibfnamefont {A.}~\bibnamefont {De~Roeck}}, \bibinfo
  {author} {\bibfnamefont {M.~J.}\ \bibnamefont {Dolan}}, \bibinfo {author}
  {\bibfnamefont {J.~R.}\ \bibnamefont {Ellis}}, \bibinfo {author}
  {\bibfnamefont {H.}~\bibnamefont {Flächer}}, \bibinfo {author}
  {\bibfnamefont {S.}~\bibnamefont {Heinemeyer}}, \bibinfo {author}
  {\bibfnamefont {G.}~\bibnamefont {Isidori}}, \bibinfo {author} {\bibfnamefont
  {K.}~\bibnamefont {Olive}}, \ and\ \bibinfo {author} {\bibnamefont
  {et~al.}},\ }\bibfield  {title} {\emph {\enquote {\bibinfo {title}
  {Implications of initial lhc searches for supersymmetry},}\ }}\href {\doibase
  10.1140/epjc/s10052-011-1634-1} {\bibfield  {journal} {\bibinfo  {journal}
  {The European Physical Journal C}\ }\textbf {\bibinfo {volume} {71}}
  (\bibinfo {year} {2011}),\ 10.1140/epjc/s10052-011-1634-1}\BibitemShut
  {NoStop}%
\bibitem [{\citenamefont {Buchmueller}\ \emph
  {et~al.}(2012{\natexlab{a}})\citenamefont {Buchmueller}, \citenamefont
  {Cavanaugh}, \citenamefont {De~Roeck}, \citenamefont {Dolan}, \citenamefont
  {Ellis}, \citenamefont {Flächer}, \citenamefont {Heinemeyer}, \citenamefont
  {Isidori}, \citenamefont {Martínez~Santos}, \citenamefont {Olive},\ and\
  \citenamefont {et~al.}}]{Buchmueller_2012_2}%
  \BibitemOpen
  \bibfield  {author} {\bibinfo {author} {\bibfnamefont {O.}~\bibnamefont
  {Buchmueller}}, \bibinfo {author} {\bibfnamefont {R.}~\bibnamefont
  {Cavanaugh}}, \bibinfo {author} {\bibfnamefont {A.}~\bibnamefont {De~Roeck}},
  \bibinfo {author} {\bibfnamefont {M.~J.}\ \bibnamefont {Dolan}}, \bibinfo
  {author} {\bibfnamefont {J.~R.}\ \bibnamefont {Ellis}}, \bibinfo {author}
  {\bibfnamefont {H.}~\bibnamefont {Flächer}}, \bibinfo {author}
  {\bibfnamefont {S.}~\bibnamefont {Heinemeyer}}, \bibinfo {author}
  {\bibfnamefont {G.}~\bibnamefont {Isidori}}, \bibinfo {author} {\bibfnamefont
  {D.}~\bibnamefont {Martínez~Santos}}, \bibinfo {author} {\bibfnamefont
  {K.~A.}\ \bibnamefont {Olive}}, \ and\ \bibinfo {author} {\bibnamefont
  {et~al.}},\ }\bibfield  {title} {\emph {\enquote {\bibinfo {title}
  {Supersymmetry in light of 1/fb of lhc data},}\ }}\href {\doibase
  10.1140/epjc/s10052-012-1878-4} {\bibfield  {journal} {\bibinfo  {journal}
  {The European Physical Journal C}\ }\textbf {\bibinfo {volume} {72}}
  (\bibinfo {year} {2012}{\natexlab{a}}),\
  10.1140/epjc/s10052-012-1878-4}\BibitemShut {NoStop}%
\bibitem [{\citenamefont {Buchmueller}\ \emph
  {et~al.}(2012{\natexlab{b}})\citenamefont {Buchmueller}, \citenamefont
  {Cavanaugh}, \citenamefont {De~Roeck}, \citenamefont {Dolan}, \citenamefont
  {Ellis}, \citenamefont {Flächer}, \citenamefont {Heinemeyer}, \citenamefont
  {Isidori}, \citenamefont {Marrouche}, \citenamefont {Martínez~Santos},\ and\
  \citenamefont {et~al.}}]{Buchmueller_2012_6}%
  \BibitemOpen
  \bibfield  {author} {\bibinfo {author} {\bibfnamefont {O.}~\bibnamefont
  {Buchmueller}}, \bibinfo {author} {\bibfnamefont {R.}~\bibnamefont
  {Cavanaugh}}, \bibinfo {author} {\bibfnamefont {A.}~\bibnamefont {De~Roeck}},
  \bibinfo {author} {\bibfnamefont {M.~J.}\ \bibnamefont {Dolan}}, \bibinfo
  {author} {\bibfnamefont {J.~R.}\ \bibnamefont {Ellis}}, \bibinfo {author}
  {\bibfnamefont {H.}~\bibnamefont {Flächer}}, \bibinfo {author}
  {\bibfnamefont {S.}~\bibnamefont {Heinemeyer}}, \bibinfo {author}
  {\bibfnamefont {G.}~\bibnamefont {Isidori}}, \bibinfo {author} {\bibfnamefont
  {J.}~\bibnamefont {Marrouche}}, \bibinfo {author} {\bibfnamefont
  {D.}~\bibnamefont {Martínez~Santos}}, \ and\ \bibinfo {author} {\bibnamefont
  {et~al.}},\ }\bibfield  {title} {\emph {\enquote {\bibinfo {title} {Higgs and
  supersymmetry},}\ }}\href {\doibase 10.1140/epjc/s10052-012-2020-3}
  {\bibfield  {journal} {\bibinfo  {journal} {The European Physical Journal C}\
  }\textbf {\bibinfo {volume} {72}} (\bibinfo {year} {2012}{\natexlab{b}}),\
  10.1140/epjc/s10052-012-2020-3}\BibitemShut {NoStop}%
\bibitem [{\citenamefont {Buchmueller}\ \emph {et~al.}(2014)\citenamefont
  {Buchmueller}, \citenamefont {Cavanaugh}, \citenamefont {Roeck},
  \citenamefont {Dolan}, \citenamefont {Ellis}, \citenamefont {Flächer},
  \citenamefont {Heinemeyer}, \citenamefont {Isidori}, \citenamefont
  {Marrouche}, \citenamefont {Santos},\ and\ \citenamefont
  {et~al.}}]{Buchmueller_2014}%
  \BibitemOpen
  \bibfield  {author} {\bibinfo {author} {\bibfnamefont {O.}~\bibnamefont
  {Buchmueller}}, \bibinfo {author} {\bibfnamefont {R.}~\bibnamefont
  {Cavanaugh}}, \bibinfo {author} {\bibfnamefont {A.~D.}\ \bibnamefont
  {Roeck}}, \bibinfo {author} {\bibfnamefont {M.~J.}\ \bibnamefont {Dolan}},
  \bibinfo {author} {\bibfnamefont {J.~R.}\ \bibnamefont {Ellis}}, \bibinfo
  {author} {\bibfnamefont {H.}~\bibnamefont {Flächer}}, \bibinfo {author}
  {\bibfnamefont {S.}~\bibnamefont {Heinemeyer}}, \bibinfo {author}
  {\bibfnamefont {G.}~\bibnamefont {Isidori}}, \bibinfo {author} {\bibfnamefont
  {J.}~\bibnamefont {Marrouche}}, \bibinfo {author} {\bibfnamefont {D.~M.}\
  \bibnamefont {Santos}}, \ and\ \bibinfo {author} {\bibnamefont {et~al.}},\
  }\bibfield  {title} {\emph {\enquote {\bibinfo {title} {The cmssm and nuhm1
  after lhc run 1},}\ }}\href {\doibase 10.1140/epjc/s10052-014-2922-3}
  {\bibfield  {journal} {\bibinfo  {journal} {The European Physical Journal C}\
  }\textbf {\bibinfo {volume} {74}} (\bibinfo {year} {2014}),\
  10.1140/epjc/s10052-014-2922-3}\BibitemShut {NoStop}%
\bibitem [{\citenamefont {Athron}\ \emph
  {et~al.}(2017{\natexlab{a}})\citenamefont {Athron}, \citenamefont {Balazs},
  \citenamefont {Bringmann}, \citenamefont {Buckley}, \citenamefont
  {Chrząszcz}, \citenamefont {Conrad}, \citenamefont {Cornell}, \citenamefont
  {Dal}, \citenamefont {Dickinson},\ and\ \citenamefont
  {et~al.}}]{Athron_2017_11}%
  \BibitemOpen
  \bibfield  {author} {\bibinfo {author} {\bibfnamefont {P.}~\bibnamefont
  {Athron}}, \bibinfo {author} {\bibfnamefont {C.}~\bibnamefont {Balazs}},
  \bibinfo {author} {\bibfnamefont {T.}~\bibnamefont {Bringmann}}, \bibinfo
  {author} {\bibfnamefont {A.}~\bibnamefont {Buckley}}, \bibinfo {author}
  {\bibfnamefont {M.}~\bibnamefont {Chrząszcz}}, \bibinfo {author}
  {\bibfnamefont {J.}~\bibnamefont {Conrad}}, \bibinfo {author} {\bibfnamefont
  {J.~M.}\ \bibnamefont {Cornell}}, \bibinfo {author} {\bibfnamefont {L.~A.}\
  \bibnamefont {Dal}}, \bibinfo {author} {\bibfnamefont {H.}~\bibnamefont
  {Dickinson}}, \ and\ \bibinfo {author} {\bibnamefont {et~al.}},\ }\bibfield
  {title} {\emph {\enquote {\bibinfo {title} {Gambit: the global and modular
  beyond-the-standard-model inference tool},}\ }}\href {\doibase
  10.1140/epjc/s10052-017-5321-8} {\bibfield  {journal} {\bibinfo  {journal}
  {The European Physical Journal C}\ }\textbf {\bibinfo {volume} {77}}
  (\bibinfo {year} {2017}{\natexlab{a}}),\
  10.1140/epjc/s10052-017-5321-8}\BibitemShut {NoStop}%
\bibitem [{\citenamefont {Athron}\ \emph
  {et~al.}(2017{\natexlab{b}})\citenamefont {Athron}, \citenamefont {Balázs},
  \citenamefont {Bringmann}, \citenamefont {Buckley}, \citenamefont
  {Chrząszcz}, \citenamefont {Conrad}, \citenamefont {Cornell}, \citenamefont
  {Dal}, \citenamefont {Edsjö},\ and\ \citenamefont
  {et~al.}}]{Athron_2017_12}%
  \BibitemOpen
  \bibfield  {author} {\bibinfo {author} {\bibfnamefont {P.}~\bibnamefont
  {Athron}}, \bibinfo {author} {\bibfnamefont {C.}~\bibnamefont {Balázs}},
  \bibinfo {author} {\bibfnamefont {T.}~\bibnamefont {Bringmann}}, \bibinfo
  {author} {\bibfnamefont {A.}~\bibnamefont {Buckley}}, \bibinfo {author}
  {\bibfnamefont {M.}~\bibnamefont {Chrząszcz}}, \bibinfo {author}
  {\bibfnamefont {J.}~\bibnamefont {Conrad}}, \bibinfo {author} {\bibfnamefont
  {J.~M.}\ \bibnamefont {Cornell}}, \bibinfo {author} {\bibfnamefont {L.~A.}\
  \bibnamefont {Dal}}, \bibinfo {author} {\bibfnamefont {J.}~\bibnamefont
  {Edsjö}}, \ and\ \bibinfo {author} {\bibnamefont {et~al.}},\ }\bibfield
  {title} {\emph {\enquote {\bibinfo {title} {A global fit of the mssm with
  gambit},}\ }}\href {\doibase 10.1140/epjc/s10052-017-5196-8} {\bibfield
  {journal} {\bibinfo  {journal} {The European Physical Journal C}\ }\textbf
  {\bibinfo {volume} {77}} (\bibinfo {year} {2017}{\natexlab{b}}),\
  10.1140/epjc/s10052-017-5196-8}\BibitemShut {NoStop}%
\bibitem [{\citenamefont {Rajec}\ \emph {et~al.}(2020)\citenamefont {Rajec},
  \citenamefont {Su}, \citenamefont {White},\ and\ \citenamefont
  {Williams}}]{Rajec_2020}%
  \BibitemOpen
  \bibfield  {author} {\bibinfo {author} {\bibfnamefont {F.}~\bibnamefont
  {Rajec}}, \bibinfo {author} {\bibfnamefont {W.}~\bibnamefont {Su}}, \bibinfo
  {author} {\bibfnamefont {M.}~\bibnamefont {White}}, \ and\ \bibinfo {author}
  {\bibfnamefont {A.~G.}\ \bibnamefont {Williams}},\ }\bibfield  {title} {\emph
  {\enquote {\bibinfo {title} {Exploring the 2hdm with global fits in
  gambit},}\ }}\href {\doibase 10.1051/epjconf/202024506022} {\bibfield
  {journal} {\bibinfo  {journal} {EPJ Web of Conferences}\ }\textbf {\bibinfo
  {volume} {245}},\ \bibinfo {pages} {06022} (\bibinfo {year}
  {2020})}\BibitemShut {NoStop}%
\bibitem [{\citenamefont {Maurer}(2015)}]{T3PS}%
  \BibitemOpen
  \bibfield  {author} {\bibinfo {author} {\bibfnamefont {V.}~\bibnamefont
  {Maurer}},\ }\bibfield  {title} {\emph {\enquote {\bibinfo {title} {{T3PS:}
  tool for parallel processing in parameter scans},}\ }}\href
  {http://arxiv.org/abs/1503.01073} {\  (\bibinfo {year} {2015})},\ \Eprint
  {http://arxiv.org/abs/1503.01073} {arXiv:1503.01073} \BibitemShut {NoStop}%
\bibitem [{\citenamefont {{Wes McKinney}}(2010)}]{pandas}%
  \BibitemOpen
  \bibfield  {author} {\bibinfo {author} {\bibnamefont {{Wes McKinney}}},\
  }\bibfield  {title} {\emph {\enquote {\bibinfo {title} {{Data Structures for
  Statistical Computing in Python}},}\ }}\bibfield  {booktitle} {\emph
  {\bibinfo {booktitle} {Proceedings of the 9th Python in Science
  Conference}},\ }\href@noop {} {\ ,\ \bibinfo {pages} {51 } (\bibinfo {year}
  {2010})}\BibitemShut {NoStop}%
\bibitem [{\citenamefont {Hunter}(2007)}]{matplotlib}%
  \BibitemOpen
  \bibfield  {author} {\bibinfo {author} {\bibfnamefont {J.~D.}\ \bibnamefont
  {Hunter}},\ }\bibfield  {title} {\emph {\enquote {\bibinfo {title}
  {Matplotlib: A 2d graphics environment},}\ }}\href {\doibase
  10.1109/MCSE.2007.55} {\bibfield  {journal} {\bibinfo  {journal} {Computing
  In Science \& Engineering}\ }\textbf {\bibinfo {volume} {9}},\ \bibinfo
  {pages} {90} (\bibinfo {year} {2007})}\BibitemShut {NoStop}%
\bibitem [{\citenamefont {{Bokeh Development Team}}(2014)}]{bokeh}%
  \BibitemOpen
  \bibfield  {author} {\bibinfo {author} {\bibnamefont {{Bokeh Development
  Team}}},\ }\href {http://www.bokeh.pydata.org} {\emph {\bibinfo {title}
  {Bokeh: Python library for interactive visualization}}} (\bibinfo {year}
  {2014})\BibitemShut {NoStop}%
\bibitem [{\citenamefont {Stevens}\ \emph {et~al.}(2015)\citenamefont
  {Stevens}, \citenamefont {Rudiger},\ and\ \citenamefont
  {Bednar}}]{holoviews}%
  \BibitemOpen
  \bibfield  {author} {\bibinfo {author} {\bibfnamefont {J.-L.~R.}\
  \bibnamefont {Stevens}}, \bibinfo {author} {\bibfnamefont {P.}~\bibnamefont
  {Rudiger}}, \ and\ \bibinfo {author} {\bibfnamefont {J.~A.}\ \bibnamefont
  {Bednar}},\ }in\ \href {\doibase 10.25080/Majora-7b98e3ed-00a} {\emph
  {\bibinfo {booktitle} {Proceedings of the 14th {P}ython in Science
  Conference}}},\ \bibinfo {editor} {edited by\ \bibinfo {editor}
  {\bibfnamefont {K.}~\bibnamefont {Huff}}\ and\ \bibinfo {editor}
  {\bibfnamefont {J.}~\bibnamefont {Bergstra}}}\ (\bibinfo {year} {2015})\ pp.\
  \bibinfo {pages} {59--66}\BibitemShut {NoStop}%
\bibitem [{\citenamefont {Byers}\ \emph {et~al.}(in)\citenamefont {Byers},
  \citenamefont {Englert}, \citenamefont {Jain}, \citenamefont {Moretti},\ and\
  \citenamefont {Olaiya}}]{preparation}%
  \BibitemOpen
  \bibfield  {author} {\bibinfo {author} {\bibfnamefont {C.}~\bibnamefont
  {Byers}}, \bibinfo {author} {\bibfnamefont {D.}~\bibnamefont {Englert}},
  \bibinfo {author} {\bibfnamefont {S.}~\bibnamefont {Jain}}, \bibinfo {author}
  {\bibfnamefont {S.}~\bibnamefont {Moretti}}, \ and\ \bibinfo {author}
  {\bibfnamefont {E.}~\bibnamefont {Olaiya}},\ }\href@noop {} {\  (\bibinfo
  {year} {in})}\BibitemShut {NoStop}%
\bibitem [{\citenamefont {{The Magellan Collaboration}}()}]{Magellan-Web}%
  \BibitemOpen
  \bibfield  {author} {\bibinfo {author} {\bibnamefont {{The Magellan
  Collaboration}}},\ }\href@noop {} {\enquote {\bibinfo {title} {{Interactive
  dashboards}},}\ }\bibinfo {howpublished}
  {\url{https://pprc.qmul.ac.uk/projects/magellan/2HDM/}}\BibitemShut {NoStop}%
\bibitem [{\citenamefont {{ATLAS
  Collaboration}}(2020{\natexlab{a}})}]{Aad_2020}%
  \BibitemOpen
  \bibfield  {author} {\bibinfo {author} {\bibnamefont {{ATLAS
  Collaboration}}},\ }\bibfield  {title} {\emph {\enquote {\bibinfo {title}
  {Combination of searches for higgs boson pairs in pp collisions at s=13tev
  with the atlas detector},}\ }}\href {\doibase
  https://doi.org/10.1016/j.physletb.2019.135103} {\bibfield  {journal}
  {\bibinfo  {journal} {Physics Letters B}\ }\textbf {\bibinfo {volume}
  {800}},\ \bibinfo {pages} {135103} (\bibinfo {year}
  {2020}{\natexlab{a}})}\BibitemShut {NoStop}%
\bibitem [{\citenamefont {Accomando}\ \emph {et~al.}(2021)\citenamefont
  {Accomando}, \citenamefont {Chapman}, \citenamefont {Maury},\ and\
  \citenamefont {Moretti}}]{Accomando:2020vbo}%
  \BibitemOpen
  \bibfield  {author} {\bibinfo {author} {\bibfnamefont {E.}~\bibnamefont
  {Accomando}}, \bibinfo {author} {\bibfnamefont {M.}~\bibnamefont {Chapman}},
  \bibinfo {author} {\bibfnamefont {A.}~\bibnamefont {Maury}}, \ and\ \bibinfo
  {author} {\bibfnamefont {S.}~\bibnamefont {Moretti}},\ }\bibfield  {title}
  {\emph {\enquote {\bibinfo {title} {{Below-threshold CP-odd Higgs boson
  search via $A\rightarrow Z^*h$ at the LHC}},}\ }}\href {\doibase
  10.1016/j.physletb.2021.136342} {\bibfield  {journal} {\bibinfo  {journal}
  {Phys. Lett. B}\ }\textbf {\bibinfo {volume} {818}},\ \bibinfo {pages}
  {136342} (\bibinfo {year} {2021})},\ \Eprint
  {http://arxiv.org/abs/2002.07038} {arXiv:2002.07038 [hep-ph]} \BibitemShut
  {NoStop}%
\bibitem [{\citenamefont {Davidson}\ and\ \citenamefont
  {Haber}(2005)}]{Davidson:2005cw}%
  \BibitemOpen
  \bibfield  {author} {\bibinfo {author} {\bibfnamefont {S.}~\bibnamefont
  {Davidson}}\ and\ \bibinfo {author} {\bibfnamefont {H.~E.}\ \bibnamefont
  {Haber}},\ }\bibfield  {title} {\emph {\enquote {\bibinfo {title}
  {Basis-independent methods for the two-higgs-doublet model},}\ }}\href
  {\doibase 10.1103/PhysRevD.72.035004} {\bibfield  {journal} {\bibinfo
  {journal} {Phys. Rev. D}\ }\textbf {\bibinfo {volume} {72}},\ \bibinfo
  {pages} {035004} (\bibinfo {year} {2005})}\BibitemShut {NoStop}%
\bibitem [{\citenamefont {Glashow}\ and\ \citenamefont
  {Weinberg}(1977)}]{GWP1}%
  \BibitemOpen
  \bibfield  {author} {\bibinfo {author} {\bibfnamefont {S.~L.}\ \bibnamefont
  {Glashow}}\ and\ \bibinfo {author} {\bibfnamefont {S.}~\bibnamefont
  {Weinberg}},\ }\bibfield  {title} {\emph {\enquote {\bibinfo {title}
  {{Natural conservation laws for neutral currents}},}\ }}\href {\doibase
  10.1103/PhysRevD.15.1958} {\bibfield  {journal} {\bibinfo  {journal} {{Phys.
  Rev. D}}\ }\textbf {\bibinfo {volume} {{15}}},\ \bibinfo {pages} {1958}
  (\bibinfo {year} {1977})}\BibitemShut {NoStop}%
\bibitem [{\citenamefont {Paschos}(1977)}]{GWP2}%
  \BibitemOpen
  \bibfield  {author} {\bibinfo {author} {\bibfnamefont {E.~A.}\ \bibnamefont
  {Paschos}},\ }\bibfield  {title} {\emph {\enquote {\bibinfo {title}
  {{Diagonal neutral currents}},}\ }}\href {\doibase 10.1103/PhysRevD.15.1966}
  {\bibfield  {journal} {\bibinfo  {journal} {{Phys. Rev. D}}\ }\textbf
  {\bibinfo {volume} {15}},\ \bibinfo {pages} {1966} (\bibinfo {year}
  {1977})}\BibitemShut {NoStop}%
\bibitem [{\citenamefont {Haber}\ and\ \citenamefont {Stal}(2015)}]{Haber2015}%
  \BibitemOpen
  \bibfield  {author} {\bibinfo {author} {\bibfnamefont {H.~E.}\ \bibnamefont
  {Haber}}\ and\ \bibinfo {author} {\bibfnamefont {O.}~\bibnamefont {Stal}},\
  }\bibfield  {title} {\emph {\enquote {\bibinfo {title} {{New LHC benchmarks
  for the CP-conserving two-Higgs-doublet model}},}\ }}\href {\doibase
  10.1140/epjc/s10052-015-3697-x} {\bibfield  {journal} {\bibinfo  {journal}
  {European Physical Journal C}\ }\textbf {\bibinfo {volume} {75}},\ \bibinfo
  {pages} {1} (\bibinfo {year} {2015})},\ \Eprint
  {http://arxiv.org/abs/1507.04281} {arXiv:1507.04281} \BibitemShut {NoStop}%
\bibitem [{\citenamefont {Ferreira}\ \emph {et~al.}(2014)\citenamefont
  {Ferreira}, \citenamefont {Santos}, \citenamefont {Gunion},\ and\
  \citenamefont {Haber}}]{Ferreira2014}%
  \BibitemOpen
  \bibfield  {author} {\bibinfo {author} {\bibfnamefont {P.~M.}\ \bibnamefont
  {Ferreira}}, \bibinfo {author} {\bibfnamefont {R.}~\bibnamefont {Santos}},
  \bibinfo {author} {\bibfnamefont {J.~F.}\ \bibnamefont {Gunion}}, \ and\
  \bibinfo {author} {\bibfnamefont {H.~E.}\ \bibnamefont {Haber}},\ }\bibfield
  {title} {\emph {\enquote {\bibinfo {title} {{Probing wrong-sign Yukawa
  couplings at the LHC and a future linear collider}},}\ }}\href {\doibase
  10.1103/PhysRevD.89.115003} {\bibfield  {journal} {\bibinfo  {journal}
  {Physical Review D - Particles, Fields, Gravitation and Cosmology}\ }\textbf
  {\bibinfo {volume} {89}} (\bibinfo {year} {2014}),\
  10.1103/PhysRevD.89.115003},\ \Eprint {http://arxiv.org/abs/1403.4736}
  {arXiv:1403.4736} \BibitemShut {NoStop}%
\bibitem [{\citenamefont {{ATLAS
  Collaboration}}(2020{\natexlab{b}})}]{Aad:2688596}%
  \BibitemOpen
  \bibfield  {author} {\bibinfo {author} {\bibnamefont {{ATLAS
  Collaboration}}},\ }\bibfield  {title} {\emph {\enquote {\bibinfo {title}
  {Combined measurements of higgs boson production and decay using up to $80$
  fb$^{-1}$ of proton-proton collision data at $\sqrt{s}=$ 13 tev collected
  with the atlas experiment},}\ }}\href {\doibase 10.1103/PhysRevD.101.012002}
  {\bibfield  {journal} {\bibinfo  {journal} {Phys. Rev. D}\ }\textbf {\bibinfo
  {volume} {101}},\ \bibinfo {pages} {012002} (\bibinfo {year}
  {2020}{\natexlab{b}})}\BibitemShut {NoStop}%
\bibitem [{\citenamefont {Sirunyan}\ \emph {et~al.}(2019)\citenamefont
  {Sirunyan} \emph {et~al.}}]{Sirunyan:2018koj}%
  \BibitemOpen
  \bibfield  {author} {\bibinfo {author} {\bibfnamefont {A.~M.}\ \bibnamefont
  {Sirunyan}} \emph {et~al.},\ }\bibfield  {title} {\emph {\enquote {\bibinfo
  {title} {{Combined measurements of Higgs boson couplings in
  proton\textendash{}proton collisions at $\sqrt{s}=13\,\text {Te}\text {V}
  $}},}\ }}\href {\doibase 10.1140/epjc/s10052-019-6909-y} {\bibfield
  {journal} {\bibinfo  {journal} {Eur. Phys. J. C}\ }\textbf {\bibinfo {volume}
  {79}},\ \bibinfo {pages} {421} (\bibinfo {year} {2019})},\ \Eprint
  {http://arxiv.org/abs/1809.10733} {arXiv:1809.10733 [hep-ex]} \BibitemShut
  {NoStop}%
\bibitem [{\citenamefont {Basler}\ \emph {et~al.}(2018)\citenamefont {Basler},
  \citenamefont {Ferreira}, \citenamefont {Mühlleitner},\ and\ \citenamefont
  {Santos}}]{Basler:2017nzu}%
  \BibitemOpen
  \bibfield  {author} {\bibinfo {author} {\bibfnamefont {P.}~\bibnamefont
  {Basler}}, \bibinfo {author} {\bibfnamefont {P.~M.}\ \bibnamefont
  {Ferreira}}, \bibinfo {author} {\bibfnamefont {M.}~\bibnamefont
  {Mühlleitner}}, \ and\ \bibinfo {author} {\bibfnamefont {R.}~\bibnamefont
  {Santos}},\ }\bibfield  {title} {\emph {\enquote {\bibinfo {title} {{High
  scale impact in alignment and decoupling in two-Higgs doublet models}},}\
  }}\href {\doibase 10.1103/PhysRevD.97.095024} {\bibfield  {journal} {\bibinfo
   {journal} {Phys. Rev.}\ }\textbf {\bibinfo {volume} {D97}},\ \bibinfo
  {pages} {095024} (\bibinfo {year} {2018})},\ \Eprint
  {http://arxiv.org/abs/1710.10410} {arXiv:1710.10410 [hep-ph]} \BibitemShut
  {NoStop}%
\bibitem [{\citenamefont {Ferreira}\ \emph {et~al.}(2018)\citenamefont
  {Ferreira}, \citenamefont {Liebler},\ and\ \citenamefont
  {Wittbrodt}}]{Ferreira:2017bnx}%
  \BibitemOpen
  \bibfield  {author} {\bibinfo {author} {\bibfnamefont {P.~M.}\ \bibnamefont
  {Ferreira}}, \bibinfo {author} {\bibfnamefont {S.}~\bibnamefont {Liebler}}, \
  and\ \bibinfo {author} {\bibfnamefont {J.}~\bibnamefont {Wittbrodt}},\
  }\bibfield  {title} {\emph {\enquote {\bibinfo {title} {{$pp\to A\to Zh$ and
  the wrong-sign limit of the two-Higgs-doublet model}},}\ }}\href {\doibase
  10.1103/PhysRevD.97.055008} {\bibfield  {journal} {\bibinfo  {journal} {Phys.
  Rev.}\ }\textbf {\bibinfo {volume} {D97}},\ \bibinfo {pages} {055008}
  (\bibinfo {year} {2018})},\ \Eprint {http://arxiv.org/abs/1711.00024}
  {arXiv:1711.00024 [hep-ph]} \BibitemShut {NoStop}%
\bibitem [{\citenamefont {Metropolis}\ \emph {et~al.}(1953)\citenamefont
  {Metropolis}, \citenamefont {Rosenbluth}, \citenamefont {Rosenbluth},
  \citenamefont {Teller},\ and\ \citenamefont {Teller}}]{Metropolis}%
  \BibitemOpen
  \bibfield  {author} {\bibinfo {author} {\bibfnamefont {N.}~\bibnamefont
  {Metropolis}}, \bibinfo {author} {\bibfnamefont {A.~W.}\ \bibnamefont
  {Rosenbluth}}, \bibinfo {author} {\bibfnamefont {M.~N.}\ \bibnamefont
  {Rosenbluth}}, \bibinfo {author} {\bibfnamefont {A.~H.}\ \bibnamefont
  {Teller}}, \ and\ \bibinfo {author} {\bibfnamefont {E.}~\bibnamefont
  {Teller}},\ }\bibfield  {title} {\emph {\enquote {\bibinfo {title} {Equation
  of state calculations by fast computing machines},}\ }}\href {\doibase
  10.1063/1.1699114} {\bibfield  {journal} {\bibinfo  {journal} {The Journal of
  Chemical Physics}\ }\textbf {\bibinfo {volume} {21}},\ \bibinfo {pages}
  {1087} (\bibinfo {year} {1953})},\ \Eprint
  {http://arxiv.org/abs/http://dx.doi.org/10.1063/1.1699114}
  {http://dx.doi.org/10.1063/1.1699114} \BibitemShut {NoStop}%
\bibitem [{\citenamefont {Hastings}(1970)}]{Hastings}%
  \BibitemOpen
  \bibfield  {author} {\bibinfo {author} {\bibfnamefont {W.~K.}\ \bibnamefont
  {Hastings}},\ }\bibfield  {title} {\emph {\enquote {\bibinfo {title} {Monte
  carlo sampling methods using markov chains and their applications},}\ }}\href
  {http://www.jstor.org/stable/2334940} {\bibfield  {journal} {\bibinfo
  {journal} {Biometrika}\ }\textbf {\bibinfo {volume} {57}},\ \bibinfo {pages}
  {97} (\bibinfo {year} {1970})}\BibitemShut {NoStop}%
\bibitem [{\citenamefont {Bechtle}\ \emph
  {et~al.}(2014{\natexlab{a}})\citenamefont {Bechtle}, \citenamefont
  {Heinemeyer}, \citenamefont {St{\aa}l}, \citenamefont {Stefaniak},\ and\
  \citenamefont {Weiglein}}]{Bechtle:2013xfa}%
  \BibitemOpen
  \bibfield  {author} {\bibinfo {author} {\bibfnamefont {P.}~\bibnamefont
  {Bechtle}}, \bibinfo {author} {\bibfnamefont {S.}~\bibnamefont {Heinemeyer}},
  \bibinfo {author} {\bibfnamefont {O.}~\bibnamefont {St{\aa}l}}, \bibinfo
  {author} {\bibfnamefont {T.}~\bibnamefont {Stefaniak}}, \ and\ \bibinfo
  {author} {\bibfnamefont {G.}~\bibnamefont {Weiglein}},\ }\bibfield  {title}
  {\emph {\enquote {\bibinfo {title} {{$HiggsSignals$: Confronting arbitrary
  Higgs sectors with measurements at the Tevatron and the LHC}},}\ }}\href
  {\doibase 10.1140/epjc/s10052-013-2711-4} {\bibfield  {journal} {\bibinfo
  {journal} {Eur. Phys. J.}\ }\textbf {\bibinfo {volume} {C74}},\ \bibinfo
  {pages} {2711} (\bibinfo {year} {2014}{\natexlab{a}})},\ \Eprint
  {http://arxiv.org/abs/1305.1933} {arXiv:1305.1933 [hep-ph]} \BibitemShut
  {NoStop}%
\bibitem [{\citenamefont {Bechtle}\ \emph {et~al.}(2021)\citenamefont
  {Bechtle}, \citenamefont {Heinemeyer}, \citenamefont {Klingl}, \citenamefont
  {Stefaniak}, \citenamefont {Weiglein},\ and\ \citenamefont
  {Wittbrodt}}]{Bechtle_2021}%
  \BibitemOpen
  \bibfield  {author} {\bibinfo {author} {\bibfnamefont {P.}~\bibnamefont
  {Bechtle}}, \bibinfo {author} {\bibfnamefont {S.}~\bibnamefont {Heinemeyer}},
  \bibinfo {author} {\bibfnamefont {T.}~\bibnamefont {Klingl}}, \bibinfo
  {author} {\bibfnamefont {T.}~\bibnamefont {Stefaniak}}, \bibinfo {author}
  {\bibfnamefont {G.}~\bibnamefont {Weiglein}}, \ and\ \bibinfo {author}
  {\bibfnamefont {J.}~\bibnamefont {Wittbrodt}},\ }\bibfield  {title} {\emph
  {\enquote {\bibinfo {title} {Higgssignals-2: probing new physics with
  precision higgs measurements in the lhc 13 tev era},}\ }}\href {\doibase
  10.1140/epjc/s10052-021-08942-y} {\bibfield  {journal} {\bibinfo  {journal}
  {The European Physical Journal C}\ }\textbf {\bibinfo {volume} {81}}
  (\bibinfo {year} {2021}),\ 10.1140/epjc/s10052-021-08942-y}\BibitemShut
  {NoStop}%
\bibitem [{\citenamefont {ATLAS}\ and\ \citenamefont
  {CMS}(2015)}]{PhysRevLett.114.191803}%
  \BibitemOpen
  \bibfield  {author} {\bibinfo {author} {\bibnamefont {ATLAS}}\ and\ \bibinfo
  {author} {\bibnamefont {CMS}},\ }\bibfield  {title} {\emph {\enquote
  {\bibinfo {title} {Combined measurement of the higgs boson mass in $pp$
  collisions at $\sqrt{s}=7$ and 8 tev with the atlas and cms experiments},}\
  }}\href {\doibase 10.1103/PhysRevLett.114.191803} {\bibfield  {journal}
  {\bibinfo  {journal} {Phys. Rev. Lett.}\ }\textbf {\bibinfo {volume} {114}},\
  \bibinfo {pages} {191803} (\bibinfo {year} {2015})}\BibitemShut {NoStop}%
\bibitem [{\citenamefont {Aad}\ \emph {et~al.}(2016)\citenamefont {Aad},
  \citenamefont {Abbott}, \citenamefont {Abdallah}, \citenamefont {Abdinov},
  \citenamefont {Abeloos}, \citenamefont {Aben}, \citenamefont {AbouZeid},
  \citenamefont {Abraham}, \citenamefont {Abramowicz},\ and\ \citenamefont
  {et~al.}}]{Aad_2016}%
  \BibitemOpen
  \bibfield  {author} {\bibinfo {author} {\bibfnamefont {G.}~\bibnamefont
  {Aad}}, \bibinfo {author} {\bibfnamefont {B.}~\bibnamefont {Abbott}},
  \bibinfo {author} {\bibfnamefont {J.}~\bibnamefont {Abdallah}}, \bibinfo
  {author} {\bibfnamefont {O.}~\bibnamefont {Abdinov}}, \bibinfo {author}
  {\bibfnamefont {B.}~\bibnamefont {Abeloos}}, \bibinfo {author} {\bibfnamefont
  {R.}~\bibnamefont {Aben}}, \bibinfo {author} {\bibfnamefont {O.~S.}\
  \bibnamefont {AbouZeid}}, \bibinfo {author} {\bibfnamefont {N.~L.}\
  \bibnamefont {Abraham}}, \bibinfo {author} {\bibfnamefont {H.}~\bibnamefont
  {Abramowicz}}, \ and\ \bibinfo {author} {\bibnamefont {et~al.}},\ }\bibfield
  {title} {\emph {\enquote {\bibinfo {title} {Measurements of the higgs boson
  production and decay rates and constraints on its couplings from a combined
  atlas and cms analysis of the lhc pp collision data at $\sqrt{s}$=7 and 8
  tev},}\ }}\href {\doibase 10.1007/jhep08(2016)045} {\bibfield  {journal}
  {\bibinfo  {journal} {Journal of High Energy Physics}\ }\textbf {\bibinfo
  {volume} {2016}} (\bibinfo {year} {2016}),\
  10.1007/jhep08(2016)045}\BibitemShut {NoStop}%
\bibitem [{\citenamefont {Baak}\ \emph {et~al.}(2014)\citenamefont {Baak},
  \citenamefont {C{\'u}th}, \citenamefont {Haller}, \citenamefont {Hoecker},
  \citenamefont {Kogler}, \citenamefont {M{\"o}nig}, \citenamefont {Schott},\
  and\ \citenamefont {Stelzer}}]{Gfitter}%
  \BibitemOpen
  \bibfield  {author} {\bibinfo {author} {\bibfnamefont {M.}~\bibnamefont
  {Baak}}, \bibinfo {author} {\bibfnamefont {J.}~\bibnamefont {C{\'u}th}},
  \bibinfo {author} {\bibfnamefont {J.}~\bibnamefont {Haller}}, \bibinfo
  {author} {\bibfnamefont {A.}~\bibnamefont {Hoecker}}, \bibinfo {author}
  {\bibfnamefont {R.}~\bibnamefont {Kogler}}, \bibinfo {author} {\bibfnamefont
  {K.}~\bibnamefont {M{\"o}nig}}, \bibinfo {author} {\bibfnamefont
  {M.}~\bibnamefont {Schott}}, \ and\ \bibinfo {author} {\bibfnamefont
  {J.}~\bibnamefont {Stelzer}},\ }\bibfield  {title} {\emph {\enquote {\bibinfo
  {title} {The global electroweak fit at nnlo and prospects for the lhc and
  ilc},}\ }}\href {\doibase 10.1140/epjc/s10052-014-3046-5} {\bibfield
  {journal} {\bibinfo  {journal} {The European Physical Journal C}\ }\textbf
  {\bibinfo {volume} {74}},\ \bibinfo {pages} {3046} (\bibinfo {year}
  {2014})}\BibitemShut {NoStop}%
\bibitem [{\citenamefont {Bechtle}\ \emph
  {et~al.}(2014{\natexlab{b}})\citenamefont {Bechtle}, \citenamefont {Brein},
  \citenamefont {Heinemeyer}, \citenamefont {St{\aa}l}, \citenamefont
  {Stefaniak}, \citenamefont {Weiglein},\ and\ \citenamefont
  {Williams}}]{Bechtle:2013wla}%
  \BibitemOpen
  \bibfield  {author} {\bibinfo {author} {\bibfnamefont {P.}~\bibnamefont
  {Bechtle}}, \bibinfo {author} {\bibfnamefont {O.}~\bibnamefont {Brein}},
  \bibinfo {author} {\bibfnamefont {S.}~\bibnamefont {Heinemeyer}}, \bibinfo
  {author} {\bibfnamefont {O.}~\bibnamefont {St{\aa}l}}, \bibinfo {author}
  {\bibfnamefont {T.}~\bibnamefont {Stefaniak}}, \bibinfo {author}
  {\bibfnamefont {G.}~\bibnamefont {Weiglein}}, \ and\ \bibinfo {author}
  {\bibfnamefont {K.~E.}\ \bibnamefont {Williams}},\ }\bibfield  {title} {\emph
  {\enquote {\bibinfo {title} {{$\mathsf{HiggsBounds}-4$: Improved Tests of
  Extended Higgs Sectors against Exclusion Bounds from LEP, the Tevatron and
  the LHC}},}\ }}\href {\doibase 10.1140/epjc/s10052-013-2693-2} {\bibfield
  {journal} {\bibinfo  {journal} {Eur. Phys. J.}\ }\textbf {\bibinfo {volume}
  {C74}},\ \bibinfo {pages} {2693} (\bibinfo {year} {2014}{\natexlab{b}})},\
  \Eprint {http://arxiv.org/abs/1311.0055} {arXiv:1311.0055 [hep-ph]}
  \BibitemShut {NoStop}%
\bibitem [{\citenamefont {Misiak}\ and\ \citenamefont
  {Steinhauser}(2017)}]{Misiak2017}%
  \BibitemOpen
  \bibfield  {author} {\bibinfo {author} {\bibfnamefont {M.}~\bibnamefont
  {Misiak}}\ and\ \bibinfo {author} {\bibfnamefont {M.}~\bibnamefont
  {Steinhauser}},\ }\bibfield  {title} {\emph {\enquote {\bibinfo {title} {Weak
  radiative decays of the b meson and bounds on $m_{H^{\pm}}$ in the
  two-higgs-doublet model},}\ }}\href {\doibase 10.1140/epjc/s10052-017-4776-y}
  {\bibfield  {journal} {\bibinfo  {journal} {The European Physical Journal C}\
  }\textbf {\bibinfo {volume} {77}},\ \bibinfo {pages} {201} (\bibinfo {year}
  {2017})}\BibitemShut {NoStop}%
\bibitem [{\citenamefont {Harlander}\ \emph {et~al.}(2013)\citenamefont
  {Harlander}, \citenamefont {Liebler},\ and\ \citenamefont
  {Mantler}}]{Harlander:2012pb}%
  \BibitemOpen
  \bibfield  {author} {\bibinfo {author} {\bibfnamefont {R.~V.}\ \bibnamefont
  {Harlander}}, \bibinfo {author} {\bibfnamefont {S.}~\bibnamefont {Liebler}},
  \ and\ \bibinfo {author} {\bibfnamefont {H.}~\bibnamefont {Mantler}},\
  }\bibfield  {title} {\emph {\enquote {\bibinfo {title} {{SusHi: A program for
  the calculation of Higgs production in gluon fusion and bottom-quark
  annihilation in the Standard Model and the MSSM}},}\ }}\href {\doibase
  10.1016/j.cpc.2013.02.006} {\bibfield  {journal} {\bibinfo  {journal}
  {Comput. Phys. Commun.}\ }\textbf {\bibinfo {volume} {184}},\ \bibinfo
  {pages} {1605} (\bibinfo {year} {2013})},\ \Eprint
  {http://arxiv.org/abs/1212.3249} {arXiv:1212.3249 [hep-ph]} \BibitemShut
  {NoStop}%
\bibitem [{\citenamefont {Eriksson}\ \emph {et~al.}(2010)\citenamefont
  {Eriksson}, \citenamefont {Rathsman},\ and\ \citenamefont
  {Stal}}]{Eriksson:2009ws}%
  \BibitemOpen
  \bibfield  {author} {\bibinfo {author} {\bibfnamefont {D.}~\bibnamefont
  {Eriksson}}, \bibinfo {author} {\bibfnamefont {J.}~\bibnamefont {Rathsman}},
  \ and\ \bibinfo {author} {\bibfnamefont {O.}~\bibnamefont {Stal}},\
  }\bibfield  {title} {\emph {\enquote {\bibinfo {title} {{2HDMC:
  Two-Higgs-Doublet Model Calculator Physics and Manual}},}\ }}\href {\doibase
  10.1016/j.cpc.2009.09.011} {\bibfield  {journal} {\bibinfo  {journal}
  {Comput. Phys. Commun.}\ }\textbf {\bibinfo {volume} {181}},\ \bibinfo
  {pages} {189} (\bibinfo {year} {2010})},\ \Eprint
  {http://arxiv.org/abs/0902.0851} {arXiv:0902.0851 [hep-ph]} \BibitemShut
  {NoStop}%
\bibitem [{\citenamefont {Grimus}\ \emph
  {et~al.}(2008{\natexlab{a}})\citenamefont {Grimus}, \citenamefont {Lavoura},
  \citenamefont {Ogreid},\ and\ \citenamefont {Osland}}]{Grimus:2008nb}%
  \BibitemOpen
  \bibfield  {author} {\bibinfo {author} {\bibfnamefont {W.}~\bibnamefont
  {Grimus}}, \bibinfo {author} {\bibfnamefont {L.}~\bibnamefont {Lavoura}},
  \bibinfo {author} {\bibfnamefont {O.}~\bibnamefont {Ogreid}}, \ and\ \bibinfo
  {author} {\bibfnamefont {P.}~\bibnamefont {Osland}},\ }\bibfield  {title}
  {\emph {\enquote {\bibinfo {title} {{The Oblique parameters in
  multi-Higgs-doublet models}},}\ }}\href {\doibase
  10.1016/j.nuclphysb.2008.04.019} {\bibfield  {journal} {\bibinfo  {journal}
  {Nucl. Phys. B}\ }\textbf {\bibinfo {volume} {801}},\ \bibinfo {pages} {81}
  (\bibinfo {year} {2008}{\natexlab{a}})},\ \Eprint
  {http://arxiv.org/abs/0802.4353} {arXiv:0802.4353 [hep-ph]} \BibitemShut
  {NoStop}%
\bibitem [{\citenamefont {Grimus}\ \emph
  {et~al.}(2008{\natexlab{b}})\citenamefont {Grimus}, \citenamefont {Lavoura},
  \citenamefont {Ogreid},\ and\ \citenamefont {Osland}}]{Grimus:2007if}%
  \BibitemOpen
  \bibfield  {author} {\bibinfo {author} {\bibfnamefont {W.}~\bibnamefont
  {Grimus}}, \bibinfo {author} {\bibfnamefont {L.}~\bibnamefont {Lavoura}},
  \bibinfo {author} {\bibfnamefont {O.}~\bibnamefont {Ogreid}}, \ and\ \bibinfo
  {author} {\bibfnamefont {P.}~\bibnamefont {Osland}},\ }\bibfield  {title}
  {\emph {\enquote {\bibinfo {title} {{A Precision constraint on
  multi-Higgs-doublet models}},}\ }}\href {\doibase
  10.1088/0954-3899/35/7/075001} {\bibfield  {journal} {\bibinfo  {journal} {J.
  Phys. G}\ }\textbf {\bibinfo {volume} {35}},\ \bibinfo {pages} {075001}
  (\bibinfo {year} {2008}{\natexlab{b}})},\ \Eprint
  {http://arxiv.org/abs/0711.4022} {arXiv:0711.4022 [hep-ph]} \BibitemShut
  {NoStop}%
\bibitem [{\citenamefont {Veltman}(1977)}]{Veltman:1977kh}%
  \BibitemOpen
  \bibfield  {author} {\bibinfo {author} {\bibfnamefont {M.}~\bibnamefont
  {Veltman}},\ }\bibfield  {title} {\emph {\enquote {\bibinfo {title} {Limit on
  mass differences in the weinberg model},}\ }}\href {\doibase
  https://doi.org/10.1016/0550-3213(77)90342-X} {\bibfield  {journal} {\bibinfo
   {journal} {Nuclear Physics B}\ }\textbf {\bibinfo {volume} {123}},\ \bibinfo
  {pages} {89} (\bibinfo {year} {1977})}\BibitemShut {NoStop}%
\bibitem [{\citenamefont {Ginzburg}\ and\ \citenamefont
  {Ivanov}(2005)}]{Ginzburg2005}%
  \BibitemOpen
  \bibfield  {author} {\bibinfo {author} {\bibfnamefont {I.~F.}\ \bibnamefont
  {Ginzburg}}\ and\ \bibinfo {author} {\bibfnamefont {I.~P.}\ \bibnamefont
  {Ivanov}},\ }\bibfield  {title} {\emph {\enquote {\bibinfo {title}
  {{Tree-level unitarity constraints in the most general two Higgs doublet
  model}},}\ }}\href {\doibase 10.1103/PhysRevD.72.115010} {\bibfield
  {journal} {\bibinfo  {journal} {Physical Review D - Particles, Fields,
  Gravitation and Cosmology}\ }\textbf {\bibinfo {volume} {72}},\ \bibinfo
  {pages} {1} (\bibinfo {year} {2005})},\ \Eprint
  {http://arxiv.org/abs/0508020} {arXiv:0508020 [hep-ph]} \BibitemShut
  {NoStop}%
\bibitem [{\citenamefont {Kanemura}\ \emph {et~al.}(1993)\citenamefont
  {Kanemura}, \citenamefont {Kubota},\ and\ \citenamefont
  {Takasugi}}]{Kanemura1993}%
  \BibitemOpen
  \bibfield  {author} {\bibinfo {author} {\bibfnamefont {S.}~\bibnamefont
  {Kanemura}}, \bibinfo {author} {\bibfnamefont {T.}~\bibnamefont {Kubota}}, \
  and\ \bibinfo {author} {\bibfnamefont {E.}~\bibnamefont {Takasugi}},\
  }\bibfield  {title} {\emph {\enquote {\bibinfo {title} {{Lee-Quigg-Thacker
  bounds for Higgs boson masses in a two-doublet model}},}\ }}\href {\doibase
  10.1016/0370-2693(93)91205-2} {\bibfield  {journal} {\bibinfo  {journal}
  {Physics Letters B}\ }\textbf {\bibinfo {volume} {313}},\ \bibinfo {pages}
  {155} (\bibinfo {year} {1993})},\ \Eprint {http://arxiv.org/abs/9303263}
  {arXiv:9303263 [hep-ph]} \BibitemShut {NoStop}%
\bibitem [{\citenamefont {Deshpande}\ and\ \citenamefont
  {Ma}(1978)}]{Deshpande1978}%
  \BibitemOpen
  \bibfield  {author} {\bibinfo {author} {\bibfnamefont {N.~G.}\ \bibnamefont
  {Deshpande}}\ and\ \bibinfo {author} {\bibfnamefont {E.}~\bibnamefont {Ma}},\
  }\bibfield  {title} {\emph {\enquote {\bibinfo {title} {{Pattern of symmetry
  breaking with two Higgs doublets}},}\ }}\href {\doibase
  10.1103/PhysRevD.18.2574} {\bibfield  {journal} {\bibinfo  {journal} {Phys.
  Rev. D}\ }\textbf {\bibinfo {volume} {18}},\ \bibinfo {pages} {2574}
  (\bibinfo {year} {1978})}\BibitemShut {NoStop}%
\bibitem [{\citenamefont {Bernon}\ \emph {et~al.}(2016)\citenamefont {Bernon},
  \citenamefont {Gunion}, \citenamefont {Haber}, \citenamefont {Jiang},\ and\
  \citenamefont {Kraml}}]{Bernon:2015wef}%
  \BibitemOpen
  \bibfield  {author} {\bibinfo {author} {\bibfnamefont {J.}~\bibnamefont
  {Bernon}}, \bibinfo {author} {\bibfnamefont {J.~F.}\ \bibnamefont {Gunion}},
  \bibinfo {author} {\bibfnamefont {H.~E.}\ \bibnamefont {Haber}}, \bibinfo
  {author} {\bibfnamefont {Y.}~\bibnamefont {Jiang}}, \ and\ \bibinfo {author}
  {\bibfnamefont {S.}~\bibnamefont {Kraml}},\ }\bibfield  {title} {\emph
  {\enquote {\bibinfo {title} {{Scrutinizing the alignment limit in
  two-Higgs-doublet models. II. m$_H$=125 GeV}},}\ }}\href {\doibase
  10.1103/PhysRevD.93.035027} {\bibfield  {journal} {\bibinfo  {journal} {Phys.
  Rev. D}\ }\textbf {\bibinfo {volume} {93}},\ \bibinfo {pages} {035027}
  (\bibinfo {year} {2016})},\ \Eprint {http://arxiv.org/abs/1511.03682}
  {arXiv:1511.03682 [hep-ph]} \BibitemShut {NoStop}%
\bibitem [{\citenamefont {{ATLAS Collaboration}}(2018)}]{ATLAS_AZh}%
  \BibitemOpen
  \bibfield  {author} {\bibinfo {author} {\bibnamefont {{ATLAS
  Collaboration}}},\ }\bibfield  {title} {\emph {\enquote {\bibinfo {title}
  {{Search for heavy resonances decaying into a $W$ or $Z$ boson and a Higgs
  boson in final states with leptons and $b$-jets in 36 fb$^{-1}$ of $\sqrt s =
  13$ TeV $pp$ collisions with the ATLAS detector}},}\ }}\href {\doibase
  10.1007/JHEP11(2018)051, 10.1007/JHEP03(2018)174} {\bibfield  {journal}
  {\bibinfo  {journal} {JHEP}\ }\textbf {\bibinfo {volume} {03}},\ \bibinfo
  {pages} {174} (\bibinfo {year} {2018})},\ \bibinfo {note} {[Erratum:
  JHEP11,051(2018)]}\BibitemShut {NoStop}%
\end{thebibliography}%


%

\end{document}